\documentclass{aa}  

\usepackage{graphicx}
\usepackage{txfonts}
\usepackage[bookmarksopen=false,bookmarksnumbered=true,breaklinks=true,colorlinks=true,linkcolor=blue,citecolor=blue]{hyperref} 
\usepackage{natbib}
\newcommand{\postref}[1]{{#1}}
\newcommand{\mpostref}[1]{{#1}}

\begin{document}
%
    \title{The wind-driven halo in high-contrast images I: analysis from the focal plane images of SPHERE}
    \author{
      	F.~Cantalloube\inst{1} 
        \and O.~J.~D.~Farley\inst{2}
        \and J.~Milli\inst{3,4}
        \and N.~Bharmal\inst{2}
        \and W.~Brandner\inst{1}
        \and C.~Correia\inst{5,6}
        \and K.~Dohlen\inst{5}
        \and Th.~Henning\inst{1}
        \and J.~Osborn\inst{2}
        \and E.~Por\inst{7}
       \and M.~Su\'arez~Valles\inst{8}
        \and A.~Vigan\inst{5}
	  		}
	  
   \institute{
	    Max Planck Institute for Astronomy, K\"{o}nigstuhl 17, D-69117 Heidelberg, Germany \email{cantalloube@mpia.de}    
	    \and Centre for Advanced Instrumentation (CfAI), Department of Physics, Durham University, Durham DH1 3LE, UK
	     	\and European Southern Observatory (ESO), Alonso de C\'{o}rdova 3107, Vitacura, Casilla 19001, Santiago, Chile
	    	\and Universit\'e Grenoble Alpes, CNRS, IPAG, F-38000 Grenoble, France 
		\and Aix Marseille Univ, CNRS, CNES, LAM, Marseille, France
		\and W. M. Keck Observatory, 65-1120 Mamalahoa Highway, Kamuela, HI 96743	    
	    	\and Leiden Observatory, Leiden University, PO Box 9513, 2300 RA Leiden, The Netherlands
	  	\and European Southern Observatory (ESO), Karl-Schwarzschild-Str. 2, 85748 Garching, Germany
             	}

   \date{Received ; accepted }

  \abstract
{
The wind driven halo is a feature observed within the images delivered by the latest generation of ground-based instruments equipped with an extreme adaptive optics system and a coronagraphic device, such as SPHERE at the VLT. This signature appears when the atmospheric turbulence conditions are varying faster than the adaptive optics loop can correct. 
The wind driven halo shows as a radial extension of the point spread function along a distinct direction (sometimes referred to as the butterfly pattern). 
When present, it significantly limits the contrast capabilities of the instrument and prevents the extraction of signals at close separation or extended signals such as circumstellar disks. 
This limitation is consequential because it contaminates the data a substantial fraction of the time: about $30\%$ of the data produced by the VLT/SPHERE instrument are affected by the wind driven halo. 
}
{
This paper reviews the causes of the wind driven halo and presents a method to analyze its contribution directly from the scientific images. Its effect on the raw contrast and on the final contrast after post-processing is demonstrated. 
}
{
We used simulations and on-sky SPHERE data to verify that the parameters extracted with our method are capable of describing the wind driven halo present in the images. We studied the temporal, spatial and spectral variation of these parameters to point out its deleterious effect on the final contrast.
}
{
The data driven analysis we propose does provide information to accurately describe the wind driven halo contribution in the images. This analysis justifies why this is a fundamental limitation to the final contrast performance reached. 
}
{
With the established procedure, we will analyze a large sample of data delivered by SPHERE in order to propose, in the future, post-processing techniques tailored to remove the wind driven halo. 
}


   \maketitle

\section{Introduction}
Thanks to the latest generation of instruments dedicated to exoplanet and circumstellar disk imaging, the last five years have witnessed a huge step in high-contrast imaging (HCI) of the close environment of nearby stars. 
To detect the light emitted by young Jupiter-like companions in the near infrared that are orbiting at a few astronomical units ($au$) from their host star, itself located at a few tens of parsecs from Earth, it is required to reach a contrast better than $10^{-5}$ at an angular separation of $500$ milliarcseconds ($\mathrm{mas}$) from the star. 
By equipping 8-m class telescopes with dedicated instruments combining extreme adaptive optics systems (AO) using high density deformable mirrors (DM) with specific coronagraphs and advanced post-processing techniques, instruments such as VLT/SPHERE \citep{Beuzit2019}, Gemini/GPI \citep{Macintosh2008gpi} and Subaru/\postref{SCExAO} \citep{Jovanovic2015scexao} successfully addressed this challenge. 
However, after achieving such high resolution and contrast, new limitations are now showing up in the focal plane images that were not visible with the first generation of HCI instruments such as Gemini/NICI \citep{Artigau2008}, VLT/NaCo \citep{Rousset2003} or Keck/NIRC2 \citep{McLean2000}. 

\begin{figure}
\resizebox{\hsize}{!}{\includegraphics{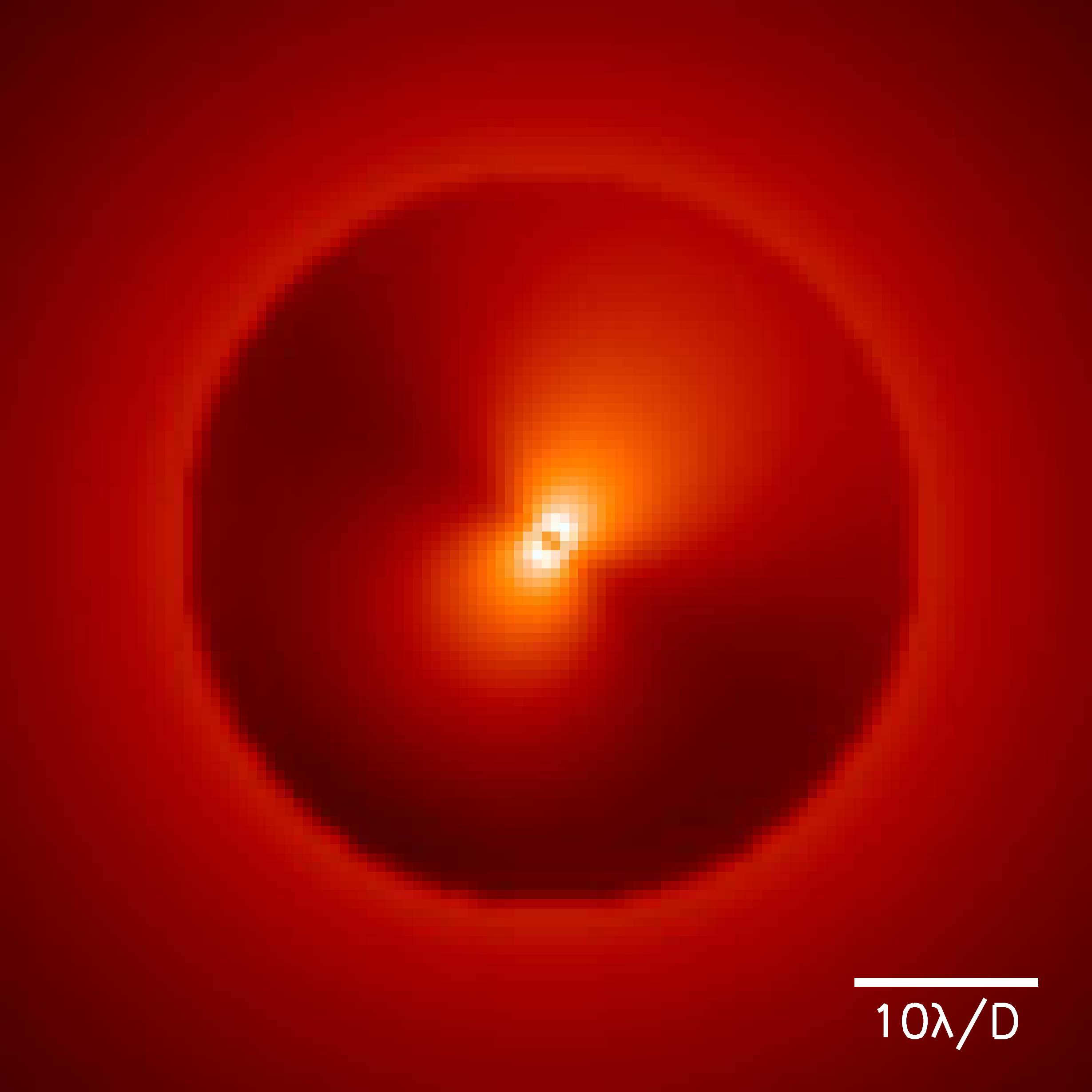}\includegraphics{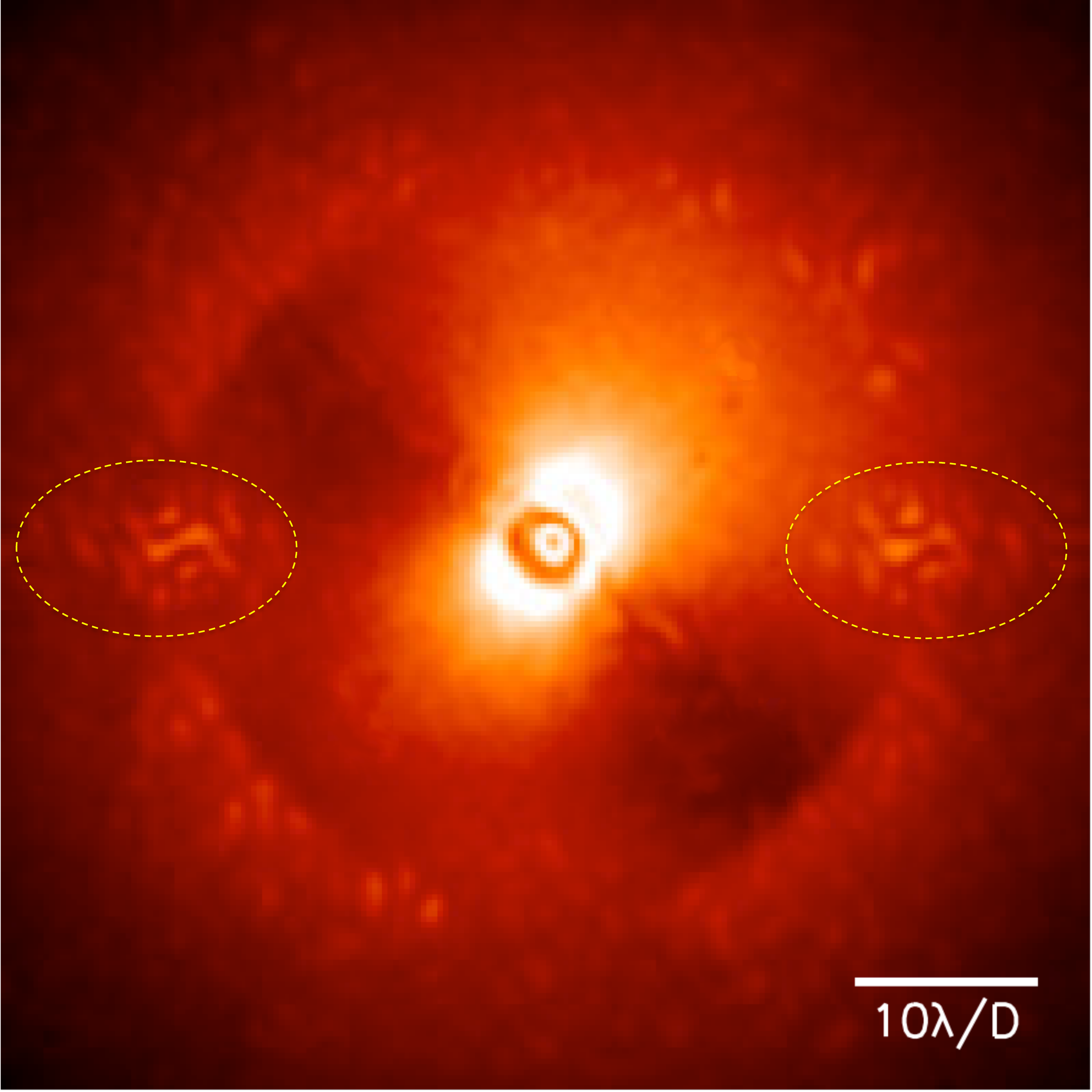}}
\caption{Coronagraphic focal plane images showing the wind driven halo. Left: Simulation of a perfect post-AO coronagraphic image of infinite exposure using an analytic AO tool (accounting only for fitting and servolag errors). Right: One exposure obtained with SPHERE-IRDIS (H2 band). Both images are in logarithmic scale to emphasize the WDH. \postref{The two regions encircled with yellow dotted lines are artefacts due to the deformable mirror manufacturing technique.}}
\label{Fig-wdh}
\end{figure}

The scientific region of interest is the close vicinity of the star (below $500~\mathrm{mas}$) where the detection of exoplanets is crucial to reject one or the other planet formation scenario as most giant planets are expected to be found in this region \postref{\citep{Chauvin2018,Nielsen2019}}, and where circumstellar disks are sometimes expected from the host \postref{stars} infrared excess. 
With the latest generation of HCI instruments, the main limitations that particularly affect those inner regions by provoking leakages of the starlight from the focal plane mask element of the coronagraph are (1) the quasi-statics non-common path aberrations \citep[NCPA,][]{Guyon2005,Fusco2006saxo}, which are differential aberrations between the AO arm and the science arm that are either not seen by the AO or that are corrected whereas not present in the science arm, (2) the low wind effect \citep[LWE,][]{Sauvage2016lwe,Milli2018lwe}, inducing differential piston and tip-tilt errors between the fragments of the pupil, (3) the low order residuals (LOR), \postref{such as residual tip-tilt}, which can be either due to atmospheric residuals, mechanical low frequency vibrations (about $10~\mathrm{Hz}$) induced by the telescope pointing \citep{Lozi2018SPIE} or atmospheric dispersion residuals \citep{pathak2016adc1}, and (4) the wind driven halo \citep[WDH,][]{Cantalloube2018} due to the atmospheric turbulence evolution being faster than the AO correction timescale (shown in Fig.~\ref{Fig-wdh} and described in this paper). Current post-processing techniques fail to overcome these limitations and the final contrast can decrease by a factor $20$ at separation between $200$ and $500~\mathrm{mas}$ \citep[e.~g.][for the LWE]{Milli2018lwe}. For more details about these various contributions, \cite{CantalloubeMsgr} present a review of the contrast limitations observed in the VLT/SPHERE images and \cite{Mouillet2018SPIE} present a review of the impact of the AO performance on the SPHERE images.

The wind driven halo originates from one of the AO error terms, namely the AO servolag (or temporal bandwidth) error, which is due to the finite and time-delayed nature of the AO correction. 
Astronomical AO-systems run in closed loop so that the wavefront sensor (WFS) sees the residual phase after the AO-correction, and therefore the command sent to the deformable mirror (DM) is relative to the previous correction. 
As there is some time delay between the WFS measurement and the setting up of the DM, if the atmospheric turbulence has varied significantly between the measure and the applied correction, the AO servolag error becomes consequential. 
For a fixed AO delay, the AO servolag error therefore depends on the turbulence coherence time $\tau_0$ itself dependent upon the seeing and the effective wind velocity at the telescope pupil. 
As a consequence of this servolag error, the AO-corrected phase shows strong low order residuals along the effective wind direction, which result in a shattering of the point spread function (PSF) along the effective wind direction. \postref{For a long exposure time (from $1$ to $60~\mathrm{seconds}$ for observations using SPHERE), these AO residual speckles add-up to form a smooth halo, the wind driven halo}, at a contrast typically below $10^{-3}$. 
In the context of HCI, when using a coronagraph to suppress the coherent peak of the starlight, a raw contrast \postref{of about} $10^{-3}$ is reached and this feature becomes visible. 
Moreover, we recently found that the temporally delayed AO residual phase interferes with amplitude errors, creating an asymmetry of the WDH in its radial direction \citep{Cantalloube2018, Madurowicz2019}: the more the amplitude error and the delayed phase error are correlated, the less the asymmetry. 
Figure~\ref{Fig-wdh} shows this wind driven halo contribution in a coronagraphic image from the VLT/SPHERE-IRDIS instrument \citep{Dohlen2008}, in the case of a simulation (left, infinite exposure with a perfect coronagraph) and an on-sky image (right).

Since a significant part of the images obtained with SPHERE are affected by the WDH, ultimately, we would like to reconstruct this effect to apply correction to existing data acquired during the 5 years of SPHERE operations. For point source detection, we can apply a spatial high-pass filter to the images but for disk imaging, this removes most of the object information as the disk signals are mainly spread within low spatial frequencies. We therefore decided to characterize finely this WDH signature in the view of developing more specific post-processing techniques to remove the WDH from the images. In addition, to prepare for the future generation of high-contrast instruments and to optimize the operation of such instruments, this paper presents a complete review of the parameters at stake, in terms of turbulence profiling, AO control and post-processing techniques.

In the following, we first review the physical origin of the WDH to highlight on which parameters it depends and show its effect on the raw contrast (Sect.~\ref{sec-WDH}). We then propose a method and metrics to analyze its contribution directly from the focal plane images (Sect.~\ref{sec-anaWDH}).
\postref{We then apply this procedure on on-sky SPHERE images to highlight the impact of the WDH on the contrast after post-processing}, by studying its typical spatial, temporal and spectral variations (Sect.~\ref{sec-PP}). From these analysis, we conclude that the current post-processing techniques based on differential imaging are not capable of fully removing the wind driven halo and consequently the contrast performance is decreased by an order of magnitude in the AO corrected area. 

\section{Origin and consequences of the wind driven halo}
\label{sec-WDH}

In the following, we detail the temporal aspect of the AO-loop (Sect.~\ref{ssec-aolag}) and of the atmospheric turbulence (Sect.~\ref{ssec-turblag}) and we specify how it affects the point spread function (Sect.~\ref{ssec-psflag}) and finally how it affects the raw contrast in the specific case of coronagraphic imaging (Sect.~\ref{ssec-contlag}).  

\subsection{The adaptive optics temporal lag}
\label{ssec-aolag}
A classical on-axis single conjugated AO system is composed of three main components: (i) the wavefront sensor (WFS) analysing the incoming phase distortion, (ii) the real-time computer (RTC) \postref{calculating}, from the WFS measurement, the command to be \postref{sent} to the phase corrector and (iii) the deformable mirror (DM) correcting for the phase distortion. The adaptive optics system of the VLT/SPHERE instrument, SAXO, is constituted of a $40 \times 40$ sub-aperture spatially-filtered Shack-Hartmann WFS \citep{Fusco2006saxo} using an EMCCD detector \citep{Sauvage2014wfs}; the RTC is the ESO-provided SPARTA architecture \citep{Valles2012sparta2,Petit2014control}; and the correction is in two stages thanks to one tip-tilt mirror and a $41 \times 41$-actuator high-order DM \citep[HODM,][]{Sinquin2008}. For a full description of the SAXO system, see \cite{Fusco2006saxo}.

\begin{figure*}
\resizebox{\hsize}{!}{\includegraphics{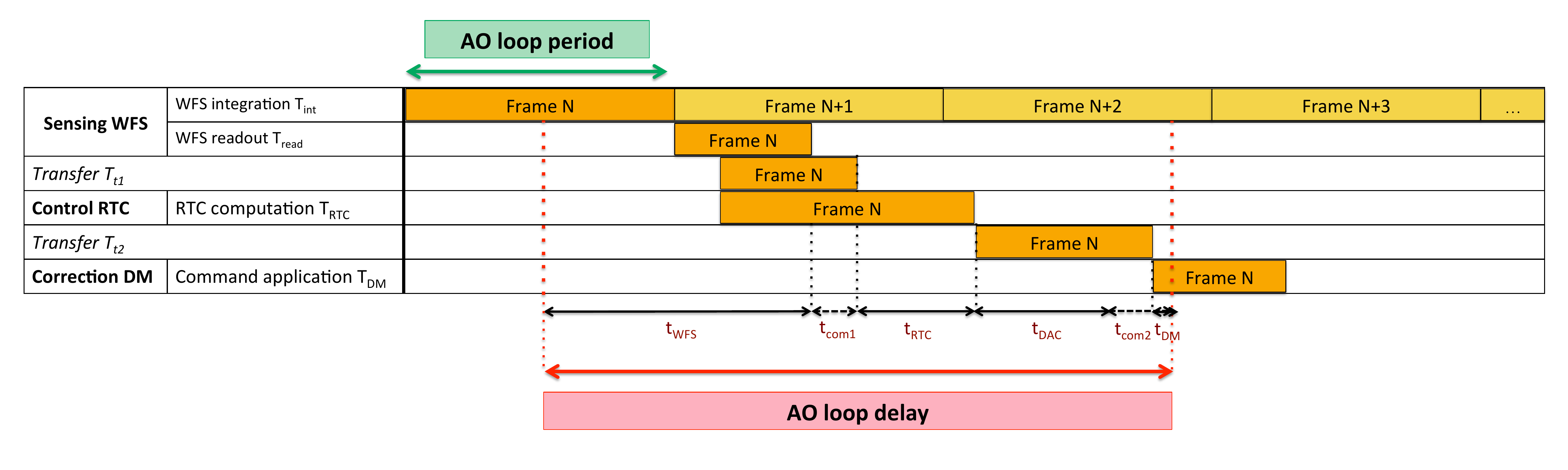}}
\caption{Typical AO chronogram summarizing the different sequences of an AO-loop using a CCD WFS as on SPHERE-SAXO (arrow lengths are not to scale). One frame is taken every AO loop period (green box) and the AO loop delay (red box) consists in about $2.2$ frames delay, starting in the middle of the first frame ($T_{int}/2$) until the DM is effectively in the corresponding shape.}
\label{Fig-chrono}
\end{figure*}

The different steps occurring during an AO closed loop run are summarized in the chronogram presented \postref{in} Fig.~\ref{Fig-chrono} \citep[adapted from different schemes from the literature, including][]{petit2008first}. 
From the first photon reaching the WFS detector to the DM being \postref{set}, it proceeds as:
\begin{enumerate}
    \item The sequence of events at the WFS follows $T_{WFS} = T_{int} + T_{read}$, where $T_{int}$ is a tunable integration time (charge collection in pixel well of the CCD) and $T_{read}$ is the fixed readout time (depending on the WFS detector technology under use). For SPHERE, $T_{read} = 725~\mathrm{\mu s}$, including all the necessary operations needed to complete the image readout. By construction it is such that $T_{int} \geq T_{read}$. The minimum interval between the first pixel being integrated of two successive frames, defining the maximum frame rate, is therefore $1/T_{read} = 1/725~\mathrm{\mu s} = 1380~\mathrm{Hz}$. 
    \item The RTC computing time $T_{RTC}$, is the time the RTC takes to carry out the processing for a given loop cycle. The RTC starts when the first pixel is received from the WFS (it continues in parallel to the readout of the WFS) and finishes when the last command is sent to the DM. On SPHERE it has been measured as $T_{RTC} = 734~\mathrm{\mu s}$.
    \item The DM settling time $T_{DM}$, is the time the DM takes to \postref{reach the requested shape} after receiving the first command from the RTC. It depends on the rise time of the actuators $T_{rise}$, fixed by the DM technology under use. For SPHERE, $T_{DM}$ is very small compared to all other times involved and can be neglected \postref{\citep[ranging from $15$ to $20~\mathrm{kHz}$,][]{Sinquin2008}}. 
\end{enumerate}
In between these three main parts, there are also fixed transfer times (gathered in $T_{t}$) from the WFS to the RTC and from the RTC to the DM. 
In terms of AO control, the frame rate for the WFS defines the AO loop frequency (sometimes also referred to as the AO loop rate). 

The AO loop delay, $\tau_{AO}$, is a pure delay defined as the addition of the equivalent delays (to a first order approximation) from the various processes involved between taking a measurement of the atmospheric disturbance via the WFS and commanding the DM accordingly: the WFS delay ($t_{WFS}$), the RTC delay ($t_{RTC}$), the digital to analog conversion delay at the DM amplifier ($t_{DAC}$), the DM positioning delay ($t_{DM}$) and at last an overall data transfer delay ($t_{com}$). At low running frequency, each term can be approximated as follow: \begin{itemize}
    \item $t_{WFS}$ is approximated by the sum of the readout time $T_{read}$ and half the integration time $T_{int}/2$;
    \item $t_{RTC}$ is the RTC latency, the time between the reception of the last pixel from the WFS to the last DM command data sent, measured in-lab as $t_{RTC} = 80~\mathrm{\mu s}$;
    \item $t_{DAC}$ is approximated by $T_{int}/2$ (when assuming that, at first order, the response to an impulse is a pulse of duration $T_{int}$);
    \item $t_{DM}$ is approximated as half the rise-time of the DM, $T_{rise}/2$;
    \item $t_{com}$ is the overall data transfer delay and has been fitted empirically on SPHERE, by measuring the closed loop transfer function, to be such as $t_{DM}+t_{com} = 35~\mathrm{\mu s}$.
\end{itemize}
For SAXO, the main contributors are therefore the integration time $T_{int}$, then the RTC latency $t_{RTC}$. In this framework, we consider the WFS as a low-pass filter that takes an average of the atmosphere during the measurement time $T_{int}$. As a whole, the AO loop delay for SPHERE is $1.56~\mathrm{ms}$, corresponding to about $2.2$ loop cycles
when running at $1380~\mathrm{Hz}$. In practice, according to the latest tests performed on SPHERE (in December 2018), the measurement of frame number $n$ is mainly affected by the command number $n-2$ (by $84\%$) and less affected by command number $n-3$ (by $16\%$). 

As a consequence, as soon as the atmospheric turbulence evolves faster than $2.17~\mathrm{ms}$ (corresponding to three frames at $1380~\mathrm{Hz}$), the AO servolag error appears in SPHERE\footnote{On fainter stars, SPHERE SAXO runs slower, thus this statement is only strictly correct for bright stars with SAXO running at fastest/optimal loop frequency.}, which affects the starlight distribution in focal plane. The trade off between the AO loop delay and the temporal evolution of the atmospheric turbulence is the critical parameter \postref{provoking} the WDH in high-contrast images. 

In the following, we use this temporal description of SAXO to simulate the AO residual phases that are used to produce the SPHERE-like simulated coronagraphic images and to discuss the effect of the atmospheric turbulence evolution. 


\subsection{The atmospheric turbulence temporal variation}
\label{ssec-turblag}
At a given instant, the atmospheric turbulence state can be represented as a 2-dimensional phase screen reaching the telescope aperture. 
This phase screen can be described by its power spectral density (PSD) along models such as the widely used stratified von Karman model \citep{conan2000modelisation}. 
\postref{This model} is parametrized by the Fried parameter ($r_0$, the typical spatial extension of the turbulence cells) and the outer scale ($L_0$, the largest size of the turbulence cells). 
During the observation sequence, this phase screen evolves in two fashions: by translation (the flow) and by the evolution of the turbulent cells shape and spatial distribution (the boiling). In a multilayer description, the flow is associated with the variations of the wind speed and direction of each layer and therefore mainly affects low spatial frequencies variations. The boiling is associated to a change in the mixing of the different layers and therefore affects high spatial frequencies variations. 
It has been empirically shown that the phase screen autocorrelation decays linearly with time over typical timescales \postref{longer than} $25~\mathrm{ms}$ 
\citep[see the studies][for three different sites, Cerro Pachon, Mauna Kea and Albuquerque]{Guesalaga2014,Poyneer2006,schock2000}. 
According to the number of subapertures of the WFS per telescope diameter ($20~\mathrm{cm}$ sampling for SAXO) and the AO-loop delay ($1.56~\mathrm{ms}$ for SPHERE-SAXO), the effect of boiling can be ignored for SPHERE-like instruments. In such case, we can work under the frozen flow assumption\footnote{See \cite{Bharmal2015} for a review on the validation of the Taylor hypothesis in AO astronomy.} \citep[the so-called Taylor hypothesis,][]{Taylor1938}, stating that the temporal evolution of the phase screen is largely dominated by translation following the projected wind speed and direction (the so-called effective wind velocity). Moreover, boiling will cause an isotropic starlight leakage in the coronagraphic image, that is therefore not linked to the wind driven halo. 

\begin{figure}
\resizebox{\hsize}{!}{\includegraphics{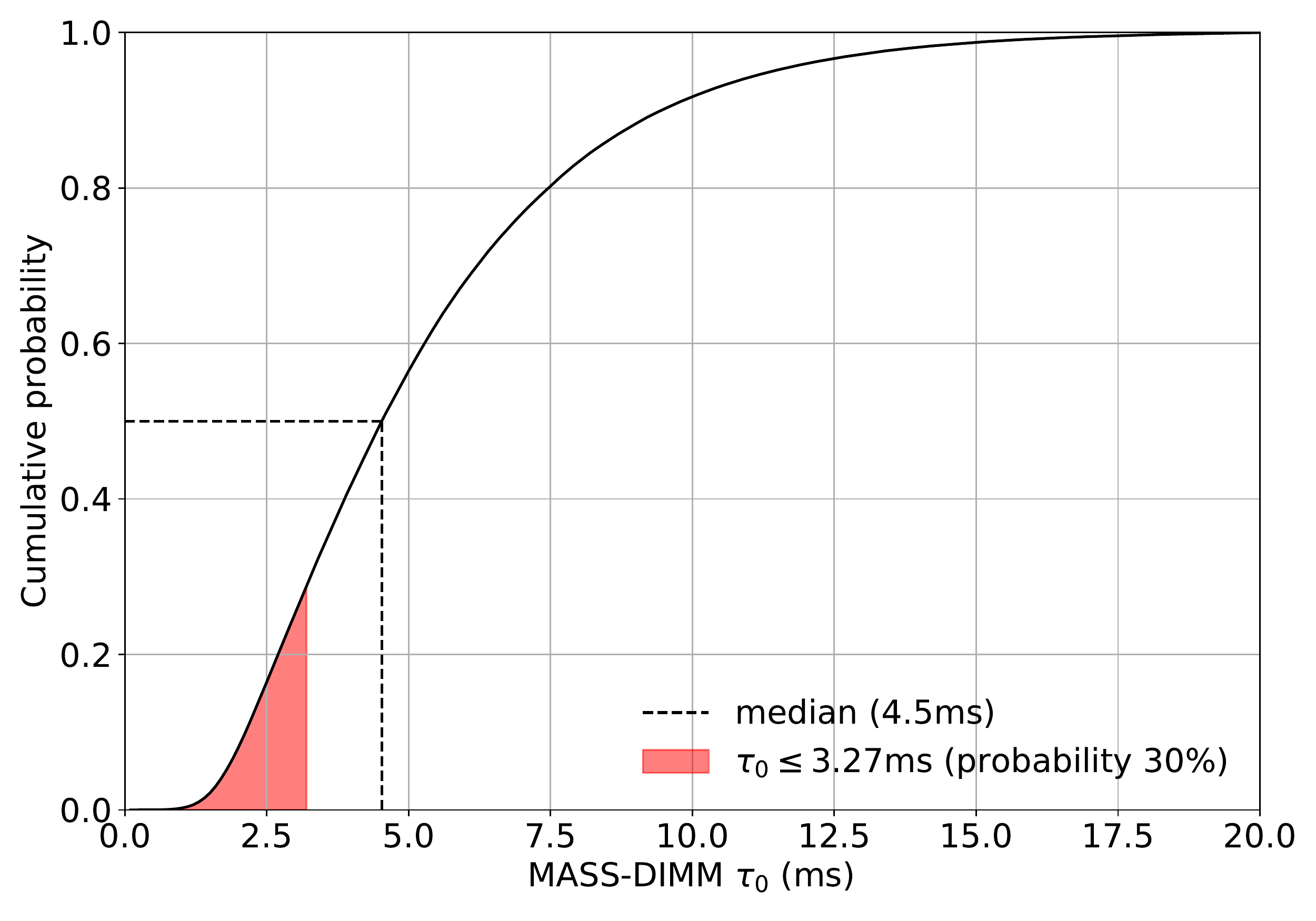}}
\caption{Cumulative histogram of the turbulence coherence time $\tau_0$ value at the zenith and at $500~\mathrm{nm}$ over 3 years of MASS-DIMM measurements (from 2016 to 2019). This figure shows that the median $\tau_0$ value is $4.5~\mathrm{ms}$ (black dashed lines).}
\label{Fig-occ}
\end{figure}

Under the frozen flow hypothesis, the temporal variation of the atmospheric turbulence is parametrized by the turbulence coherence time, $\tau_0$, analytically\footnote{From MASS-DIMM turbulence profiler \citep{Kornilov2007massdimm} measurements taken at Paranal observatory, the distribution of the coherence time as a function of the seeing is shown in  Fig.~\ref{Fig-coude}. The median value of the measured $\tau_0$ is $4.5~\mathrm{ms}$ at $500~\mathrm{nm}$ and at the zenith.} defined as \citep{Roddier1981,Roddier1982,hardy1998ao}: 
\begin{equation}
 \tau_{0} = 0.314 \frac{r_0}{v_{eff}},   
 \label{eq:tau0}
\end{equation}
where $r_0$ is the Fried parameter (dependent upon the wavelength of observation and the zenith angle) and $v_{eff}$ is the effective wind velocity, itself defined as $v_{eff} =\left[ \frac{\int_\infty C_{n}^2(h).v(h)^{5/3} dh}{\int_\infty C_{n}^2(h) dh} \right]^{3/5}$, with $v(h)$ the wind velocity profile with altitude \postref{$h$,} and $C_{n}^2(h)$ the refractive index structure constant profile with altitude. The turbulence coherence time characterizes the time interval for which the temporal fluctuations of the turbulent phase are equal to $1~\mathrm{rad^2}$. If $\tau_0$ equals $\tau_{AO}$ (at the sensing wavelength), it means that between the measurement of the incoming phase and the DM \postref{reaching the requested shape}, the actual phase evolved of $1~\mathrm{rad^2}$, that is to say, its Strehl ratio decreased by $63\%$ due to the \postref{servolag} error (under the small phase approximation, \postref{which is} valid during AO correction). 
The cumulative histogram of the coherence time values over Paranal observatory (at zenith and at $500~\mathrm{nm}$) during three years of MASS-DIMM \citep[Multi-Aperture Scintillation Sensor - Differential Image Motion Monitor,][]{Kornilov2007massdimm} measurements is presented in Fig.~\ref{Fig-occ} and shows a steep curve at short $\tau_0$ (below the median of $4.5~\mathrm{ms}$). It is not possible to obtain a direct trade-off value that compares the AO delay and the turbulence coherence time to state when the WDH appears in the images. In the following, we establish a rule of thumb, based on simulations, to estimate a typical $\tau_0$ value below which the WDH dominates in the image. In a next paper, we will apply our WDH analysis procedure to the SPHERE-SHINE guaranteed time survey \citep{Chauvin2017SHINE} data to extract a realistic occurrence rate of the WDH. 

\begin{figure}
\resizebox{\hsize}{!}{\includegraphics{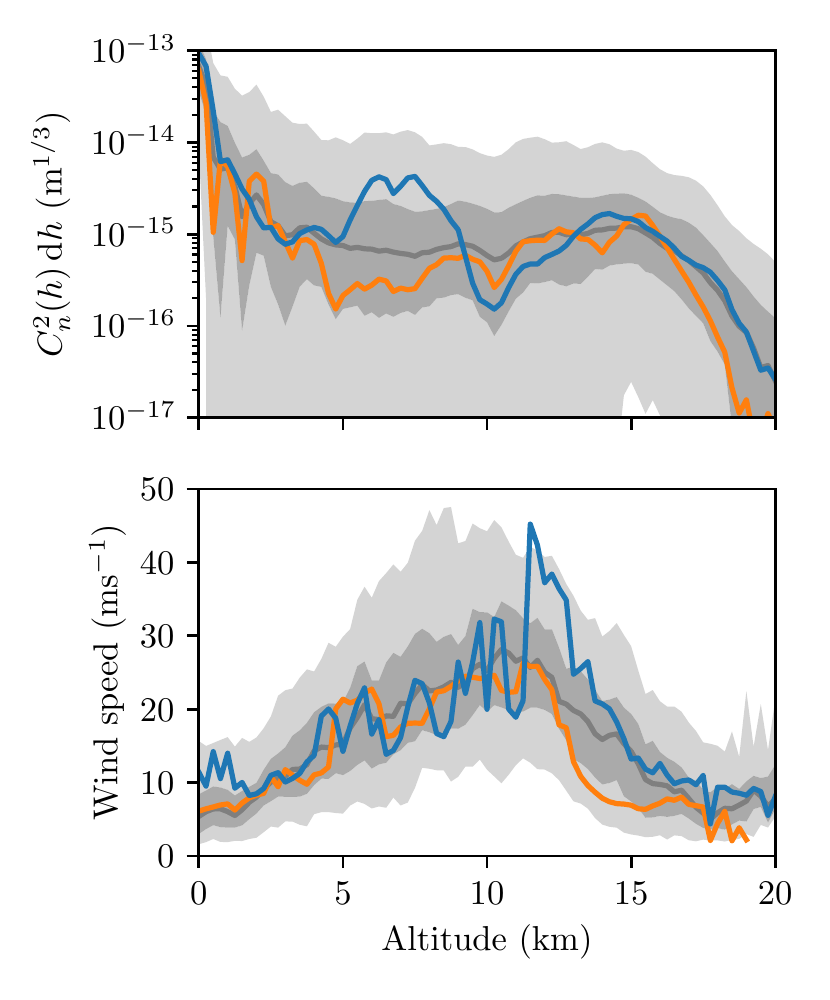}}
\caption{Turbulence strength and wind speed profiles with altitude measured at Paranal observatory during the 2018 Stereo-SCIDAR campaign. 
In both figures, the lighter shaded region contains $90\%$ of the data, the darker shaded region $50\%$ and the solid line is the median. 
Top: Integrated refractive index structure constant between two altitude limits, $\int_{h_2}^{h_1} C_n^2 \; dh$, as a function of the altitude $h$.
Bottom: Wind speed as a function of the altitude $h$, with blue solid line corresponding to the median in winter and orange solid line to the median in summer.}
\label{Fig-jetstream}
\end{figure}

\begin{figure}
\resizebox{\hsize}{!}{\includegraphics{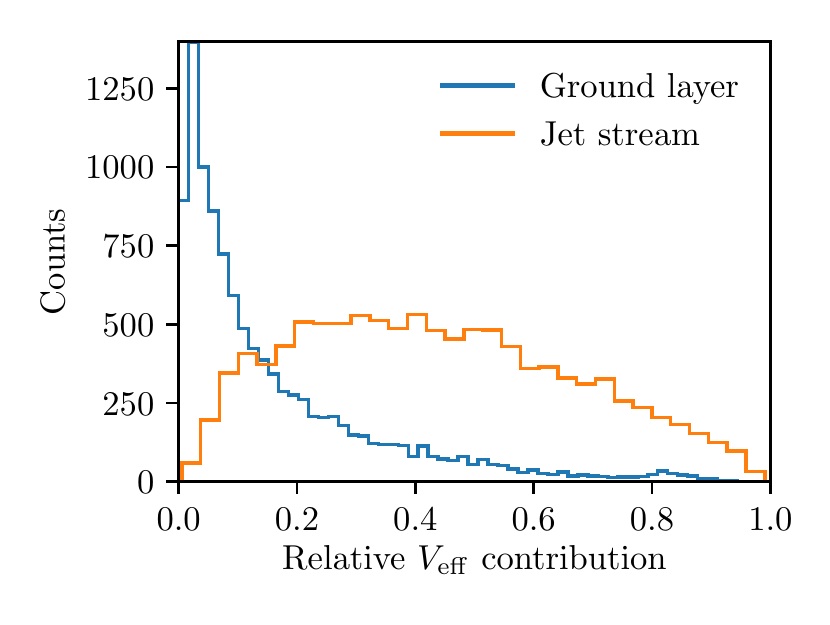}}
\caption{Histogram of the contribution to $v_{eff}$ for the ground layer (below $1~\mathrm{km}$ altitude, blue line) and the jet stream layer ($8$ to $14~\mathrm{km}$ altitude, orange line), extracted from the 2018 Stereo-SCIDAR measurements at Paranal observatory. 
}
\label{Fig-jsculprit}
\end{figure}

For a given AO system, the important external parameters playing a role in the servolag error expression are therefore the seeing and the effective wind velocity, that is to say the balance between strong turbulent layers and high wind speed layers.  
As shown on the $C_{n}^2(h)$ profile, extracted from the Stereo-SCIDAR \citep[SCIntillation Detection And Ranging,][]{VerninRoddier1973, shepherd2013stereo} measurements 2018 campaign \citep{osborn2018scidar} at Paranal observatory, presented in Fig.~\ref{Fig-jetstream} (top), the strongest turbulence layers are close to the ground layer. As shown on the wind speed profile $v(h)$ presented in Fig.~\ref{Fig-jetstream} (bottom), the fastest wind occurs at the jet stream layer, located at about $12~\mathrm{km}$ above sea level ($200~\mathrm{mbar}$) and whose wind speed can go up to $50~\mathrm{m/s}$, depending on the latitude and the season\footnote{In the southern hemisphere, the subtropical jet stream showing up from March to December varies from about $20~\mathrm{m/s}$ during summer, up to about $50~\mathrm{m/s}$ during winter \citep{Gallego2005}. The jet stream is less prominent on other astronomical sites such as Mauna Kea \citep{Sarazin2003}.}. 
Notably, before the installation of the MASS, the effective wind velocity used to estimate the coherence time was empirically computed by \cite{Sarazin2002} as $v_{eff} \simeq \mathrm{max}(v_{ground},0.4v_{jet-stream})$, showing the impact of the jet stream and how it affects the astronomical data \citep{Masciadri2013MOSE}. 
In order to highlight the predominance of the jet stream layer to the effective wind speed, and therefore to the WDH, Fig.~\ref{Fig-jsculprit} shows the relative contribution of the ground layer (below $1~\mathrm{km}$) and of the jet stream layer (between $8$ to $14~\mathrm{km}$) to $v_{eff}$, extracted from the Stereo-SCIDAR data. The median contribution of the ground layer to $v_{eff}$ is about $10\%$ while the median contribution of the jet stream layer to $v_{eff}$ is about  $40\%$. In addition, by comparing the contribution from the ground layer and the contribution from the jet stream layer for each individual profile, we observe that for about 80\% of the profiles, the jet stream layer has a higher contribution to $v_{eff}$.
By correlating the observed WDH direction within high-contrast images from GPI, \cite{Madurowicz2018SPIE} also show that the jet stream layer is indeed the main responsible for the apparition of the WDH in HCI data. 

\begin{figure}
\resizebox{\hsize}{!}{\includegraphics{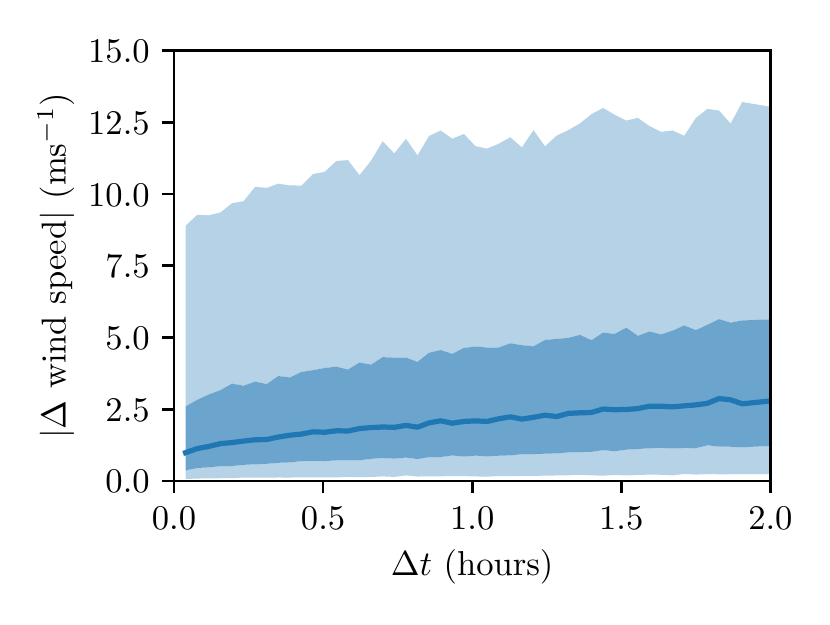}}
\resizebox{\hsize}{!}{\includegraphics{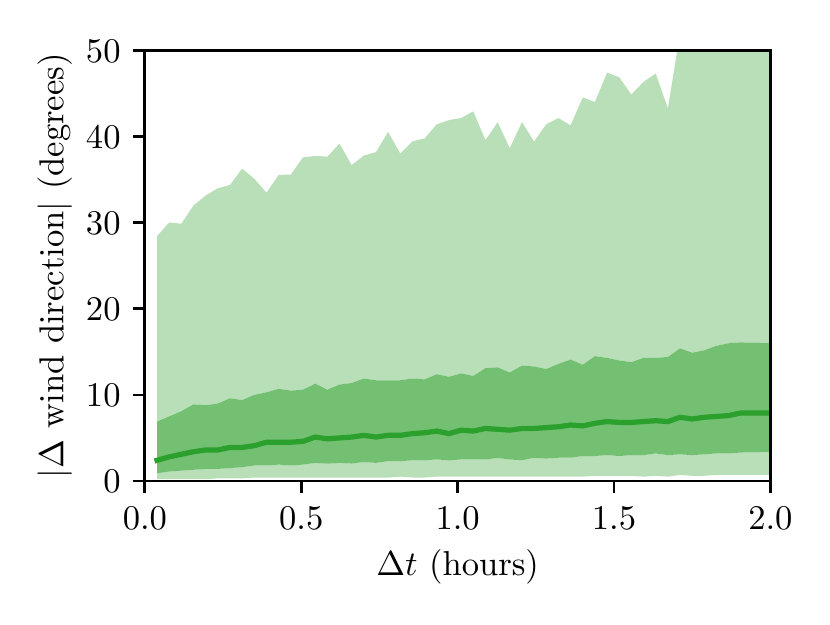}}
\caption{Typical temporal evolution of the wind speed (top) and wind direction (bottom) within the $8\mathrm{-}12~\mathrm{km}$ altitude region, extracted from the 2018 Stereo-SCIDAR campaign at Paranal observatory. In every figure, the lighter shaded region contains $90\%$ of the data, the darker shaded region $50\%$ and the solid line is the median.}
\label{Fig-evol}
\end{figure}

From the Stereo-SCIDAR measurements, we also computed the typical temporal evolution of the wind speed (Fig.~\ref{Fig-evol}, top) and the wind direction (Fig.~\ref{Fig-evol}, bottom) at the jet stream layer, during two hours, the typical time of an observing sequence with SPHERE. The latter is the distribution of the absolute value of the change in wind speed and direction with some time lag across the entire data set. It shows that the wind speed at the jet stream layer rarely remains stable and is more likely to vary of up to typically $5~\mathrm{m/s}$ over an hour while the change in wind direction is likely to remain below $10^\circ$.

As reminded in the introduction, the asymmetry of the WDH is due to the scintillation. For a given AO delay and turbulence coherence time, the asymmetry of the wind driven halo increases with the scintillation \citep{Cantalloube2018}. 
Scintillation is due to the effect of long distance propagation that transforms phase aberrations into amplitude aberrations. The scintillation is dependent upon the propagation length $z$, the strength of the atmospheric turbulence $C_n^2(h)$, the diameter and structure of the telescope aperture $D_{tel}$, the wavelength of observation\footnote{For large telescopes of the 8-m class, the scintillation is almost achromatic.} $\lambda$, the exposure time $\tau_{exp}$, and the wind speed $w(z)$ at and above the tropopause. 
Under the frozen-flow hypothesis, the scintillation index (variance of the flux fluctuation through the telescope pupil) can be expressed as $\sigma^2_{s} = 10.66 \, D_{tel}^{-4/3} \, \tau_{exp}^{-1} \int_{A} C_n^2(z) \, \frac{z^2}{w(z)} \, dz$, where $A$ is the telescope aperture filter \citep{kornilov2011}.
The temporal variation of the scintillation as been measured with the Multi-Aperture Scintillation Sensor instrument \citep[MASS,][]{kornilov2003mass} measuring the scintillation index during various campaigns on different sites\footnote{The measured scintillation index from \cite{Kornilov2012} has comparable values on all site but is lower at Mauna Kea site because of its higher altitude and lower turbulence in the upper atmosphere.} over 8-years, from 2004 to 2012, published in \cite{Kornilov2012}: the authors show that the scintillation index has seasonal variations depending on the latitude of the observatory (being higher during winter), directly related to the wind speed seasonal variations in the upper atmosphere. In addition, they observed that the power of the scintillation is quite stable over $1~\mathrm{h}$ timescales at Paranal observatory\footnote{According to the study led in \cite{Kornilov2012}, the temporal behavior of the scintillation is similar among the 11 surveyed sites.}, directly related to the jet stream speed variations, which are also shown in Fig.~\ref{Fig-evol} (top). This indicates again the importance of the jet stream layer on the WDH signature. 

In the following, all these aspects are taken into account \postref{to simulate} SPHERE-like images.

\subsection{The AO servolag error consequences in the images}
\label{ssec-psflag}
The spatial variance of the AO residual phase in the pupil due to the AO servolag error varies as:
\begin{equation}
\sigma^2_{servo} = \left(\frac{\tau_{AO}}{\tau_0}\right)^{5/3}.
\label{eq:servolag}
\end{equation}
For a single conjugated AO system, the power spectral density (PSD) of the residual phase due to the AO servolag error, showing the distribution of the averaged power of the phase fluctuations over the spatial frequencies $k$, can be expressed as \postref{\citep{rigaut1998analytical,Cantalloube2018}}:
\begin{equation}
\begin{split}
\mathrm{PSD}_{servo} (k) = & \; 0.023 \; k^{-11/3} \int_\infty r_0(h)^{5/3} \; \Bigl[ \sin_c^2(k.v(h).T_{int}) \\
& + 1 - 2 \cos (2 \pi.k.v(h).\tau_{AO}) \sin_c(k.v(h).T_{int}) \Bigr] \\
& \times (1-sin(2 \pi.h/h_T) \pm 2 \pi.k.T_{int}.v(h)) \; dh,
\end{split}
\label{eq:servodsp}
\end{equation}
with  $h_T = 2/ (\lambda k^2)$ the Talbot length \citep{Antichi2011}.
The last \postref{line} of the expression accounts for the asymmetry due to the interaction with the \postref{atmospheric} amplitude error. In the following, the simulated images (as in Fig.~\ref{Fig-wdh}, left) are made from residual phase screens produced by an analytical AO simulator \citep{Jolissaint2006} using this updated expression of the AO servolag error \postref{(Eq.~\ref{eq:servodsp}). This updated expression accounts for the interaction of the servolag with the scintillation, therefore generating} the asymmetry \postref{(it includes both the amplitude and the phase of the electric field)}. \postref{Note that for} small \postref{residual errors (high Strehl ratio)}, the PSD is an approximation of the smooth structures of the PSF (such as the WDH).

\begin{figure}
\resizebox{\hsize}{!}{\includegraphics{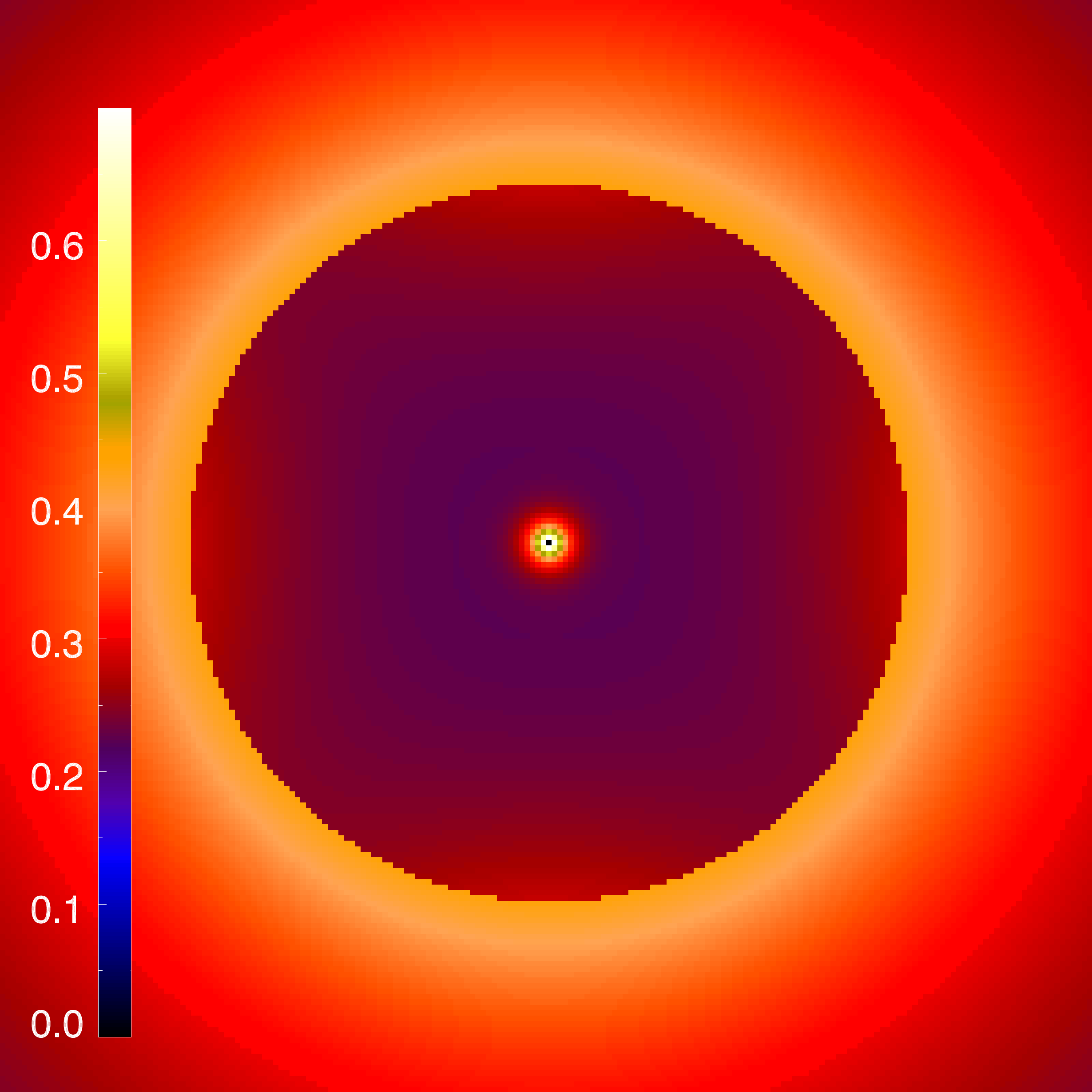}\includegraphics{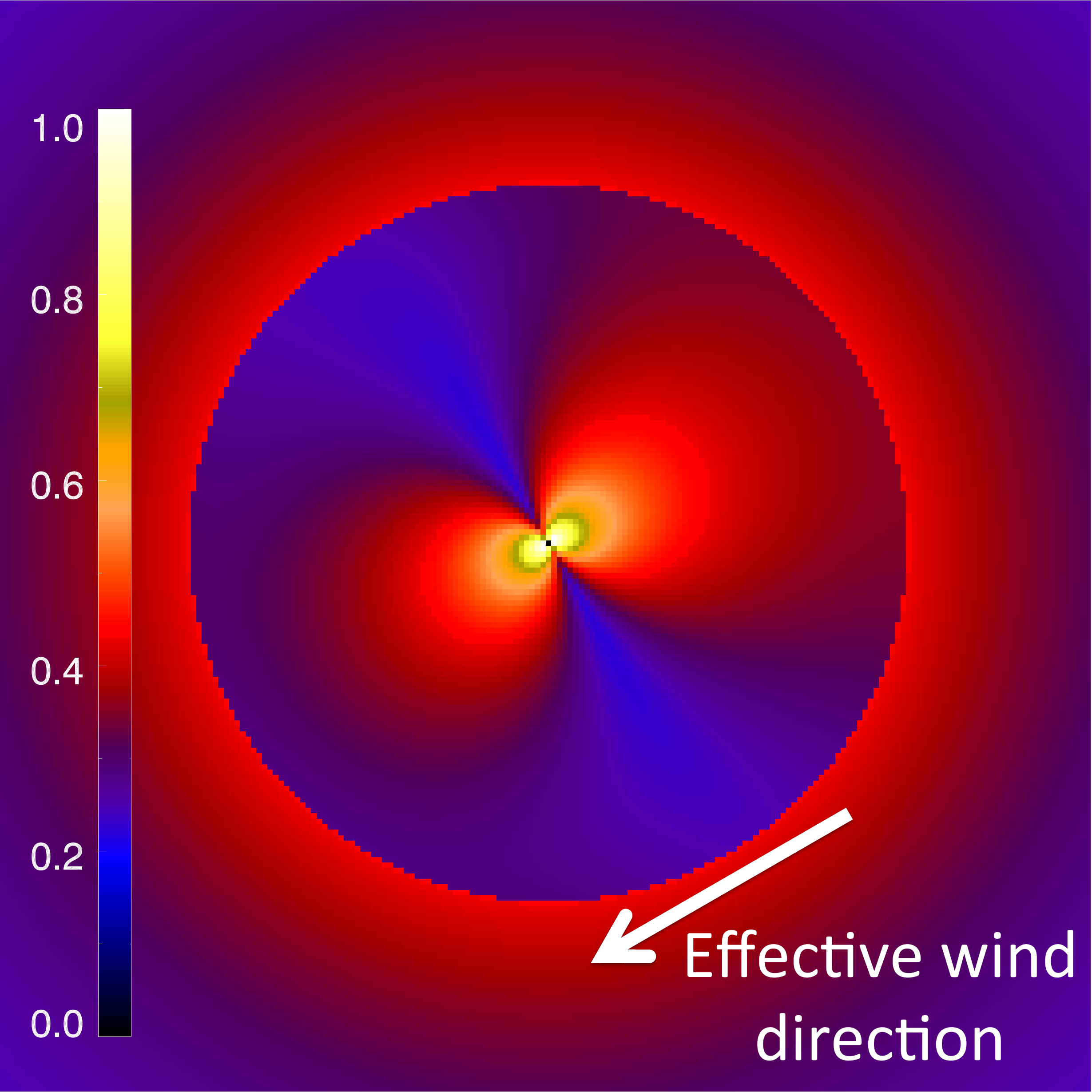}}
\caption{Power spectral densities of the AO residual phase simulated for a SAXO-like system. Left: PSD without the servolag error. Right: PSD including the servolag error and its \postref{interference} with amplitude errors, \postref{for an effective wind along a direction of $30^\circ$ (white arrow)}. 
In order to highlight the \postref{servolag} contribution, the colorbar is normalized to the maximum of the \postref{right} image.}
\label{fig-psd}
\end{figure}

\postref{Simulations of the two-dimensional AO residual phase PSD, with and without the servolag error, are shown in Fig.~\ref{fig-psd} (right and left respectively)}. The residual phase PSD due to the AO servolag error shows a low spatial frequency structure along the effective wind direction \postref{(white arrow on Fig.~\ref{fig-psd}, right)} decreasing radially (until the AO correction radius) and shows no power in the direction perpendicular to the effective wind direction. In addition, due to the interference term with amplitude error, one wing of the AO servolag signature is smaller than the other \postref{(in the opposite direction of the wind)}. 

By comparing the simulated PSDs with and without the servolag error, we can conclude that, for a SAXO-like AO system working under the typical median $\tau_0$ of Paranal observatory, about  $69\%$ of the starlight within the AO corrected zone \postref{is} scattered outside of the coherent peak due to the servolag error. \postref{The remainder of the scattered light is provoked by} other typical AO errors (aliasing, chromaticity, anisoplanetism and noise propagation, but excluding NCPA or other exogenous errors). 

\subsection{Effect of the wind driven halo on the raw contrast}
\label{ssec-contlag}
In this section, we analyze how the WDH affects the raw contrast, that is to say the contrast obtained in an image after the AO correction with a coronagraph, but before the application of any post-processing technique. In practice it is computed as the \postref{mean} radial profile of the coronagraphic image normalized by the maximum of the non-coronagraphic image.

\begin{figure}
\resizebox{\hsize}{!}{\includegraphics[trim={0 0 0 1cm, clip}]{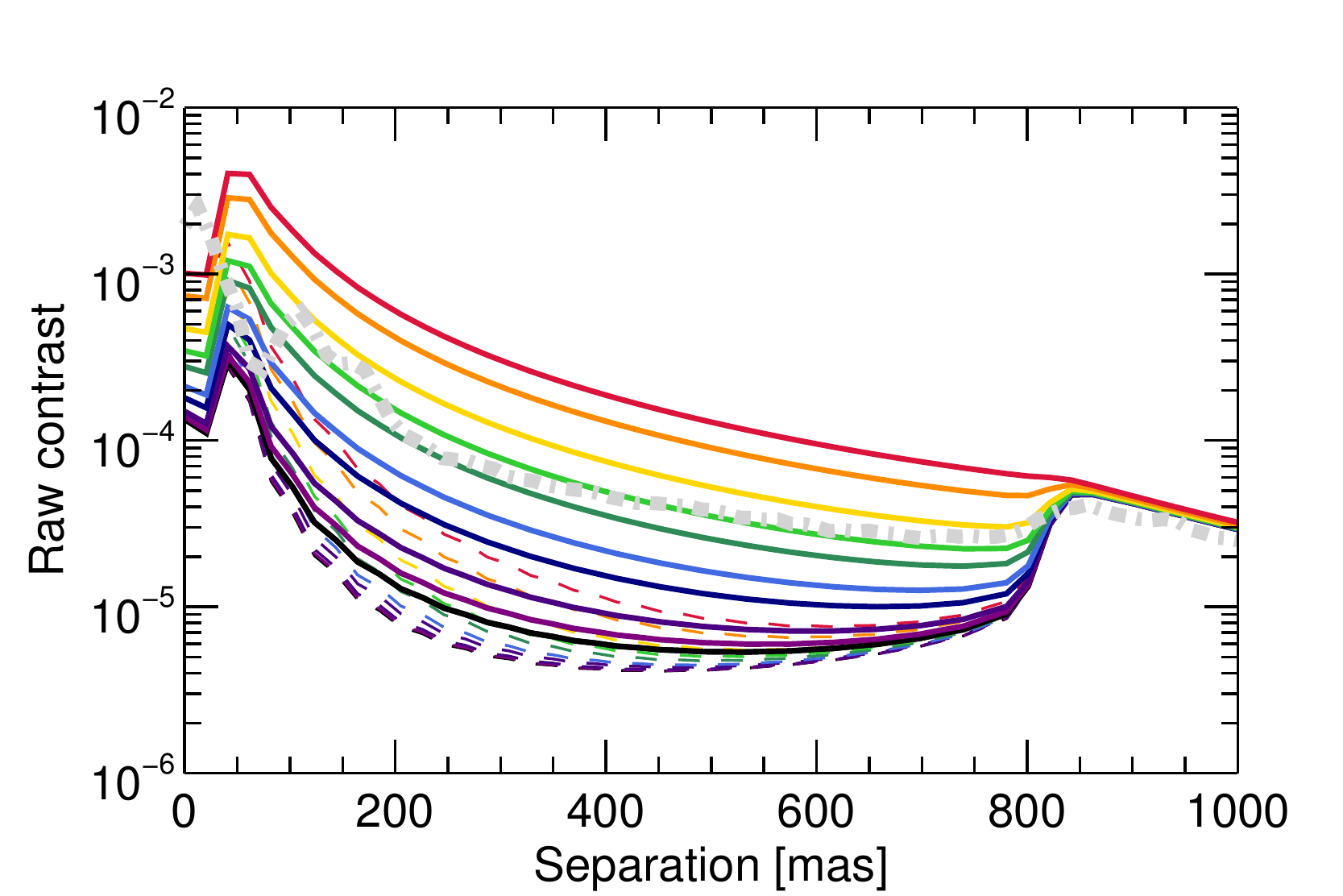}}
\caption{Illustration of the raw contrast as a function of separation to the star for various $\tau_0$, ranging from $12.5~\mathrm{ms}$ (black) to $1.25~\mathrm{ms}$ (red). From bottom to top, the curves are for turbulence coherence times of respectively: $12.5$, $10$, $7.5$, $5.0$, $4.0$, $3.0$, $2.5$, $2.0$, $1.5$ and $1.25~\mathrm{ms}$. The solid lines are along the wind direction whereas the dashed lines are along its perpendicular direction. For comparison, the grey dash-dotted line shows the raw contrast \postref{(azimuthal mean of the image)} of a SPHERE-IRDIS H2-band (centered on $1.59~\mathrm{\mu m}$) image under median seeing, airmass, and long $\tau_0 \sim 9~\mathrm{ms}$.}
\label{Fig-prof}
\end{figure}

To highlight the contribution of the WDH in high-contrast images, we simulated AO-corrected (using a SAXO-like system as  described in Sect.~\ref{ssec-aolag}) and ideal coronagraphic \citep{cavarroc2006fundamental, Sauvage2010} infinite exposure images in H-band ($1.6~\mathrm{\mu m}$), producing Fig.~\ref{Fig-wdh} (left). We used a single layer atmospheric model, moving along one given direction, varying only the wind velocity, under the median seeing conditions at Paranal observatory. 
To assess the impact of the WDH on the raw contrast, Fig.~\ref{Fig-prof} shows the radial profiles of simulated coronagraphic images for various $\tau_0$, along the wind direction (solid lines, where the raw contrast is highly impacted) and along its perpendicular direction (dashed lines, where the raw contrast is less impacted), under median seeing condition ($r_0 = 13.8~\mathrm{cm}$, according to MASS-DIMM measurements) and median airmass ($\mpostref{a =} 1.15$). 

\begin{figure}
\resizebox{\hsize}{!}{\includegraphics{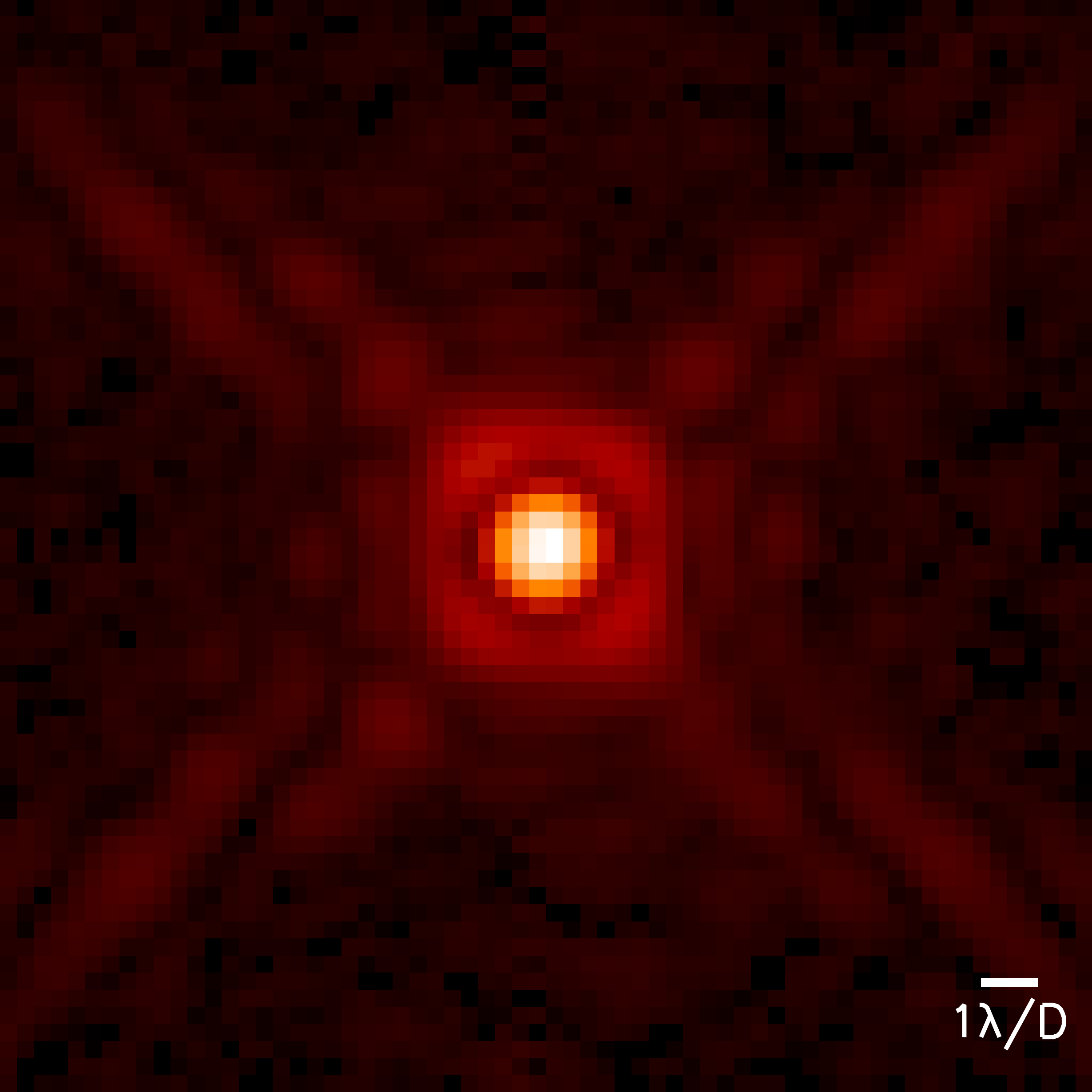}\includegraphics{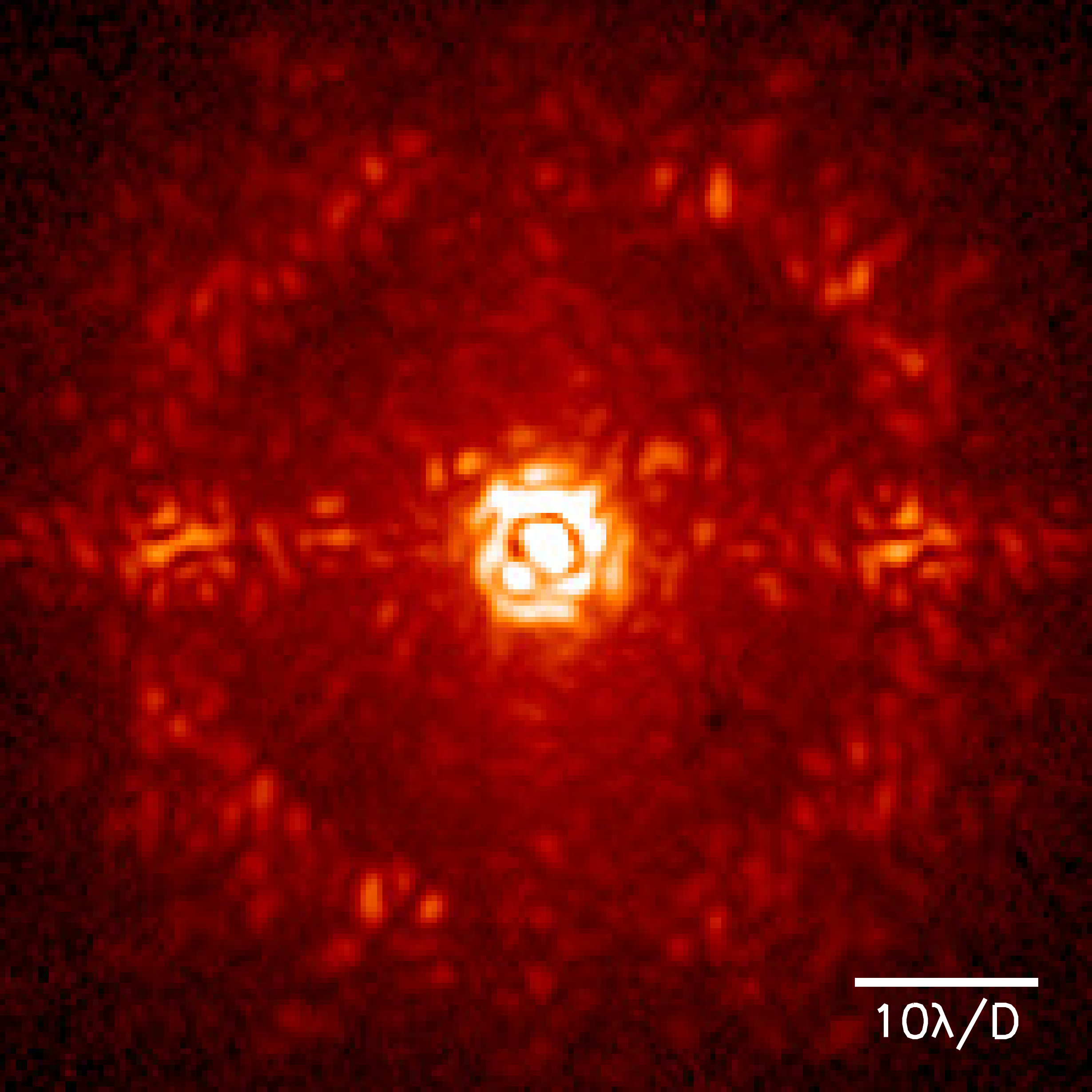}}
\resizebox{\hsize}{!}{\includegraphics{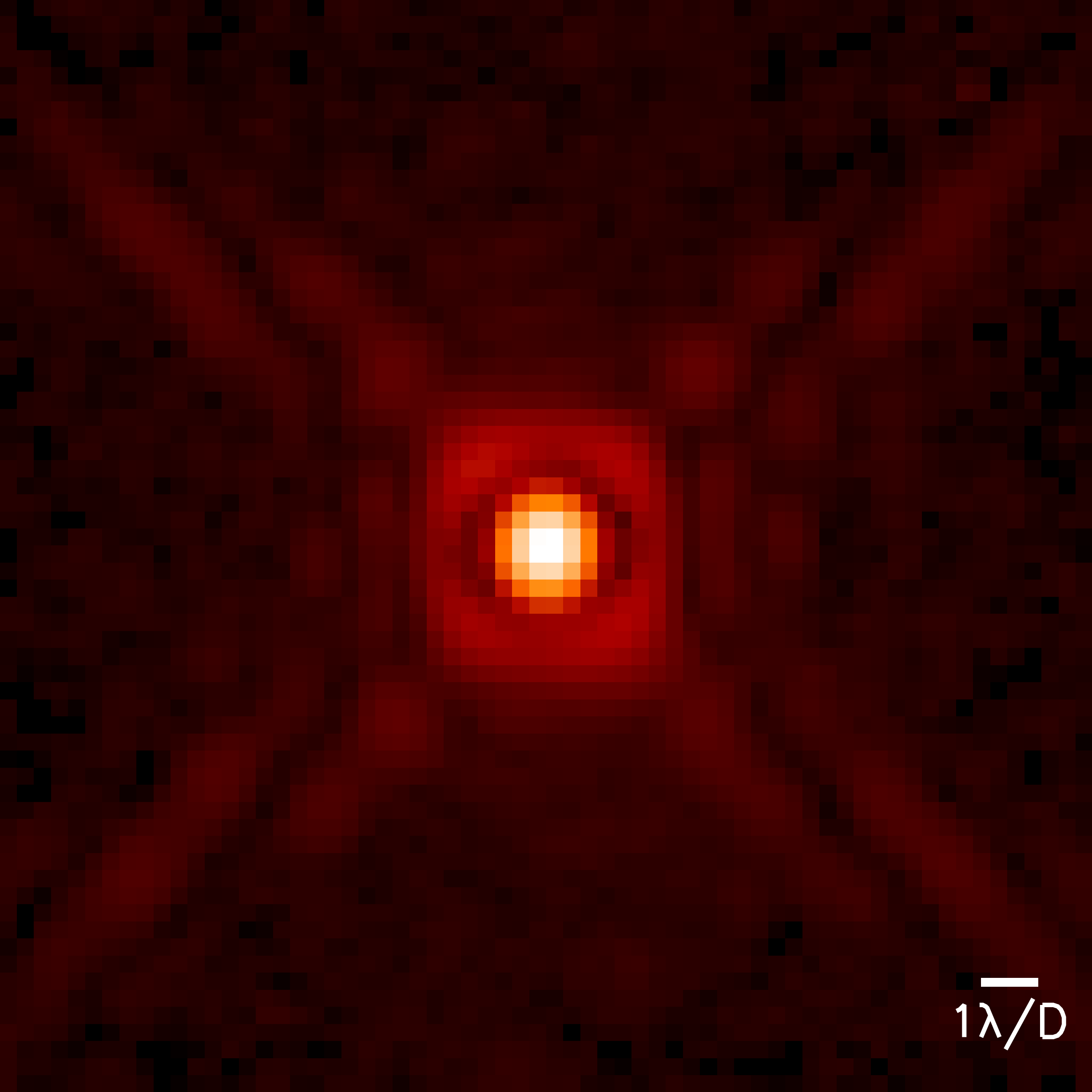}\includegraphics{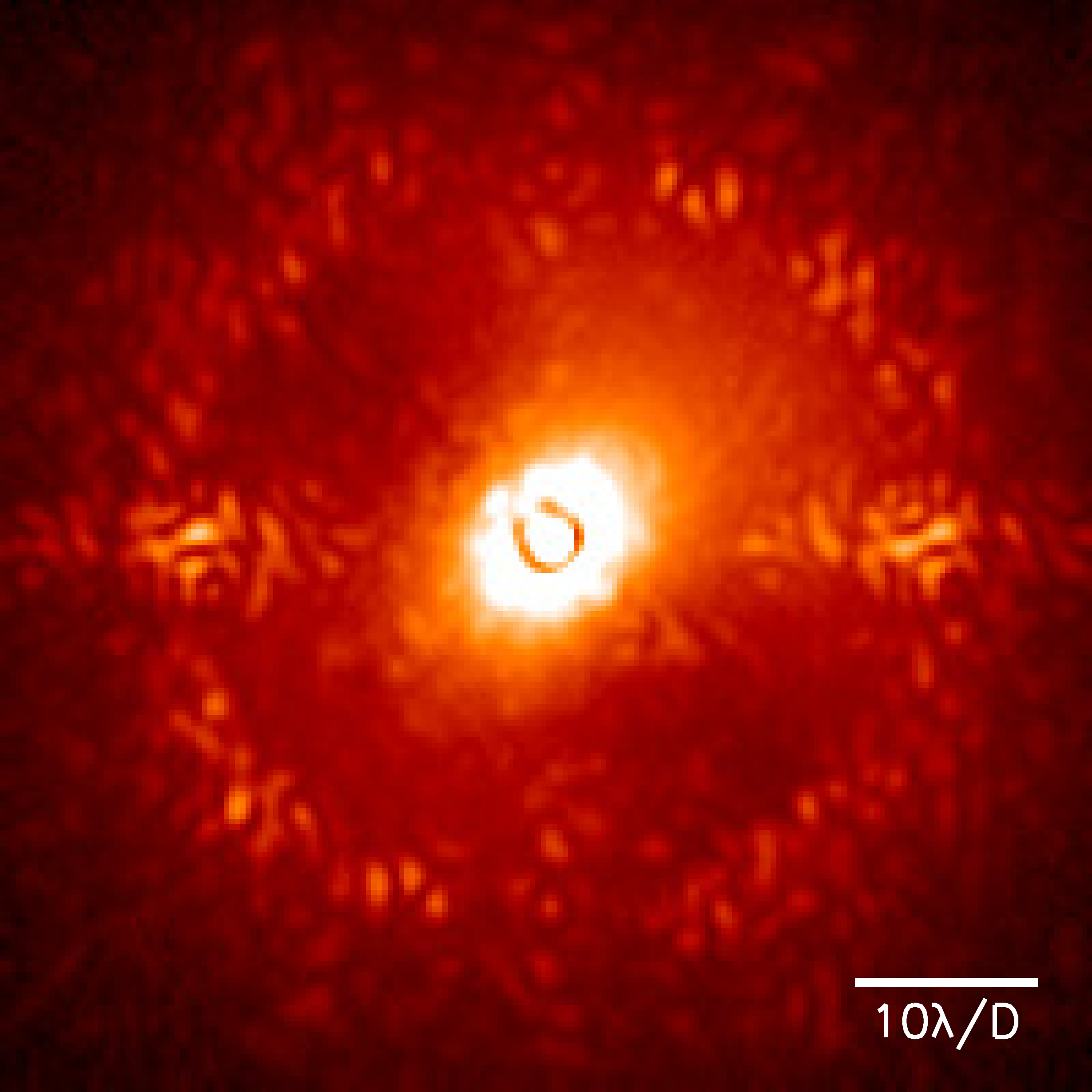}}
\caption{\postref{On-sky images from the SPHERE-IRDIS instrument in H2-band, without (top) and with WDH (bottom). 
Left: Non coronagraphic images. Right: Corresponding coronagraphic images. The top images are taken under a very long coherence time $\tau_0 \sim 9~\mathrm{ms}$.}}
\label{Fig-coro}
\end{figure}

\postref{We compared these simulations with the raw contrast of \postref{a} SPHERE image taken under very good observing conditions (median seeing, $\tau_0 > 9~\mathrm{ms}$), in which no WDH is visible (Fig.~\ref{Fig-prof}, grey dash-dotted line and Fig.~\ref{Fig-coro}, top-right). At a separation of $300~\mathrm{mas}$ (where the raw contrast without servolag error reaches a plateau, and way beyond the influence of the coronagraph inner working angle), the raw contrast reached in the H-band is about $7.10^{-5}$ \citep[see also][]{Vigan2015sirius}.} 
\postref{In this on-sky image, the raw contrast is limited by the presence of speckles due to NCPA. As NCPA are always present and are the main limitation at $300~\mathrm{mas}$, this is the ultimate raw contrast we can reach under good observing conditions \citep{Vigan2019zelda}.} 
\postref{From Fig.~\ref{Fig-prof}, the measured contrast curve (grey dash-dotted line), which was taken with a coherence time of $9~\mathrm{ms}$, is above that expected for the system with a $3~\mathrm{ms}$ coherence time (green line). As such we expect the WDH to show up from a coherence time below $3~\mathrm{ms}$.} 
Reported on the cumulative histogram in Fig.~\ref{Fig-occ}, it yields an occurrence rate of WDH of about $30\%$ after correcting for the median airmass of $1.13$ (corresponding to a zenith angle of $27.7^\circ$) as measured for SPHERE. \postref{This value is an approximate value to motivate the present work, and will be further refined by a statistical study of the SPHERE data acquired during its 5 years of operation}.

\begin{figure}
\resizebox{\hsize}{!}{\includegraphics[trim={0 0 0 0.3cm, clip}]{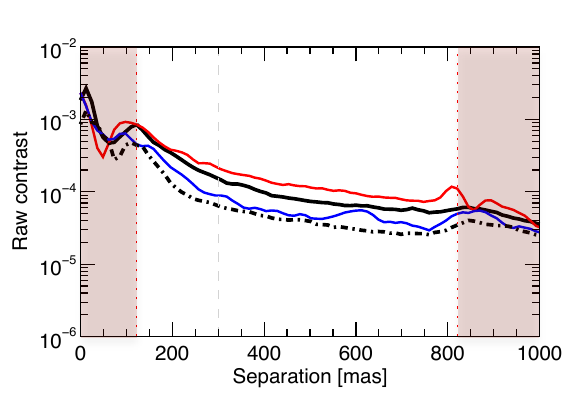}}
\caption{\postref{Raw contrast as a function of the separation to the star for the two images with (solid lines) and without (dotted-dash line) WDH presented in Fig.~\ref{Fig-coro}. The red line is along the WDH, the blue line in the perpendicular direction and the black lines are the azimuthal median. The two red shaded areas correspond to the region affected by the coronagraph (left) and outside the AO correction zone (right).}}
\label{Fig-cprof}
\end{figure}

The WDH has a high contrast and is only unveiled thanks to the use of a coronagraph, \postref{as highlighted in Fig.~\ref{Fig-coro}, where no difference can be seen in the two non-coronagraphic images (left), while the WDH is very bright in the coronagraphic image under short coherence time (bottom right)}. \postref{The corresponding raw contrast profiles are shown on Fig.~\ref{Fig-cprof}, showing the impact of the WDH on real data}. Hence this effect appears as an important limitation only for the latest generation of HCI instrument such as VLT/SPHERE, Gemini/GPI and Subaru/SCExAO using both extreme-AO and advanced coronagraph technology. The careful analyses of the WDH presented in this paper is thus motivated by the strong impact of the WDH on the achieved contrast with these high contrast instruments.

\section{Analysis of the wind driven halo in the focal plane images of SPHERE}
\label{sec-anaWDH}
In this section, we present a method to analyze the WDH by deriving the three properties used to describe it: its direction (Sect.~\ref{ssec-dir}), its intensity in the focal plane image (Sect.~\ref{ssec-str}) and its asymmetry (Sect.~\ref{ssec-asy}). The analysis are conducted directly within the focal plane images as the results are more reliable in the context of high-contrast performance than using external data such as AO-telemetry or turbulence profiling. 
\postref{In addition to the analysis presented in this paper on how the WDH affects the contrast after post-processing, we intend to use this WDH analysis procedure for two studies: (i) making a statistical analysis of how the SPHERE data are affected by the WDH and correlate it with the AO telemetry and profiling data and (ii) estimate the WDH to remove it from the data.}

\begin{figure*}
\centering
\resizebox{\hsize}{!}{\includegraphics{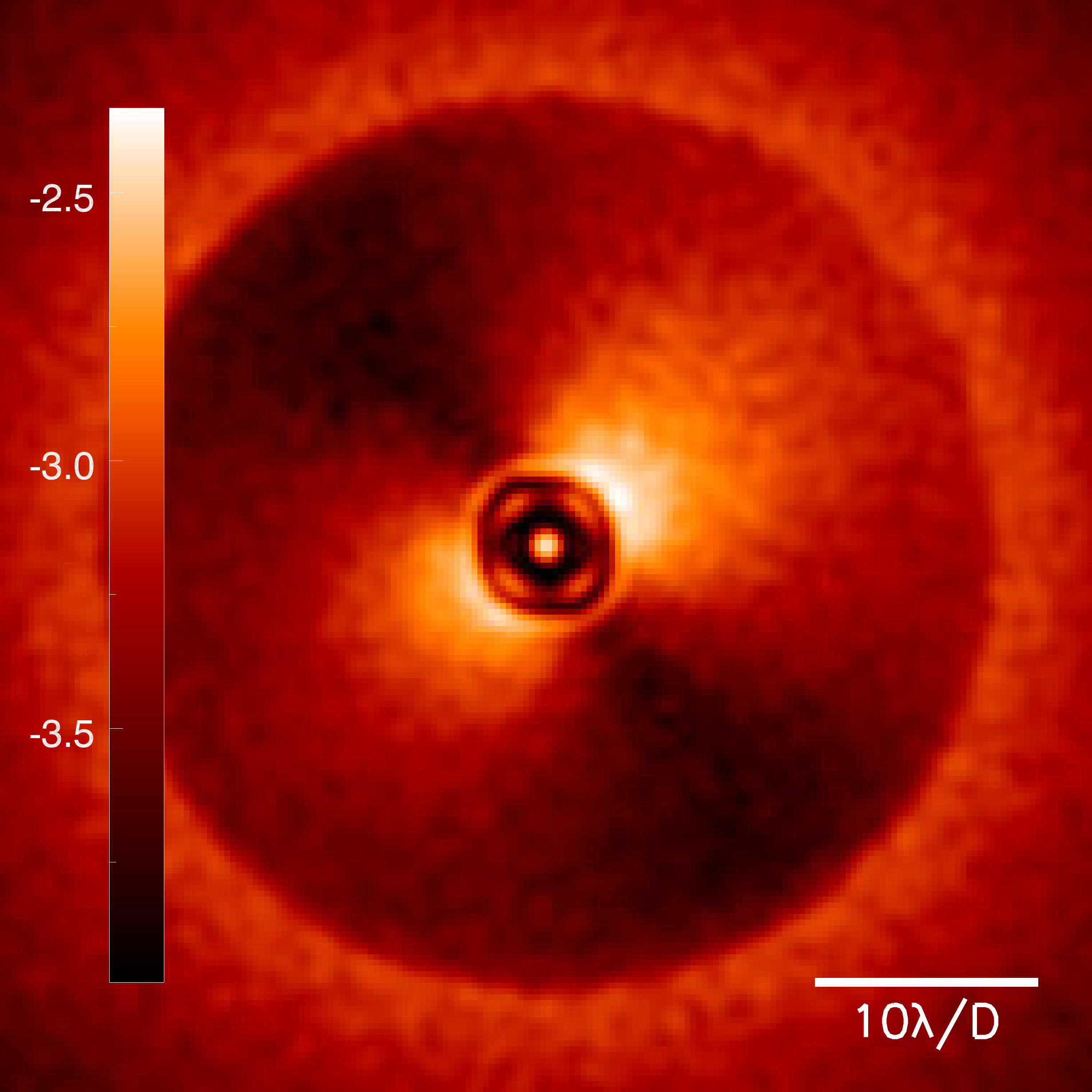}
\includegraphics{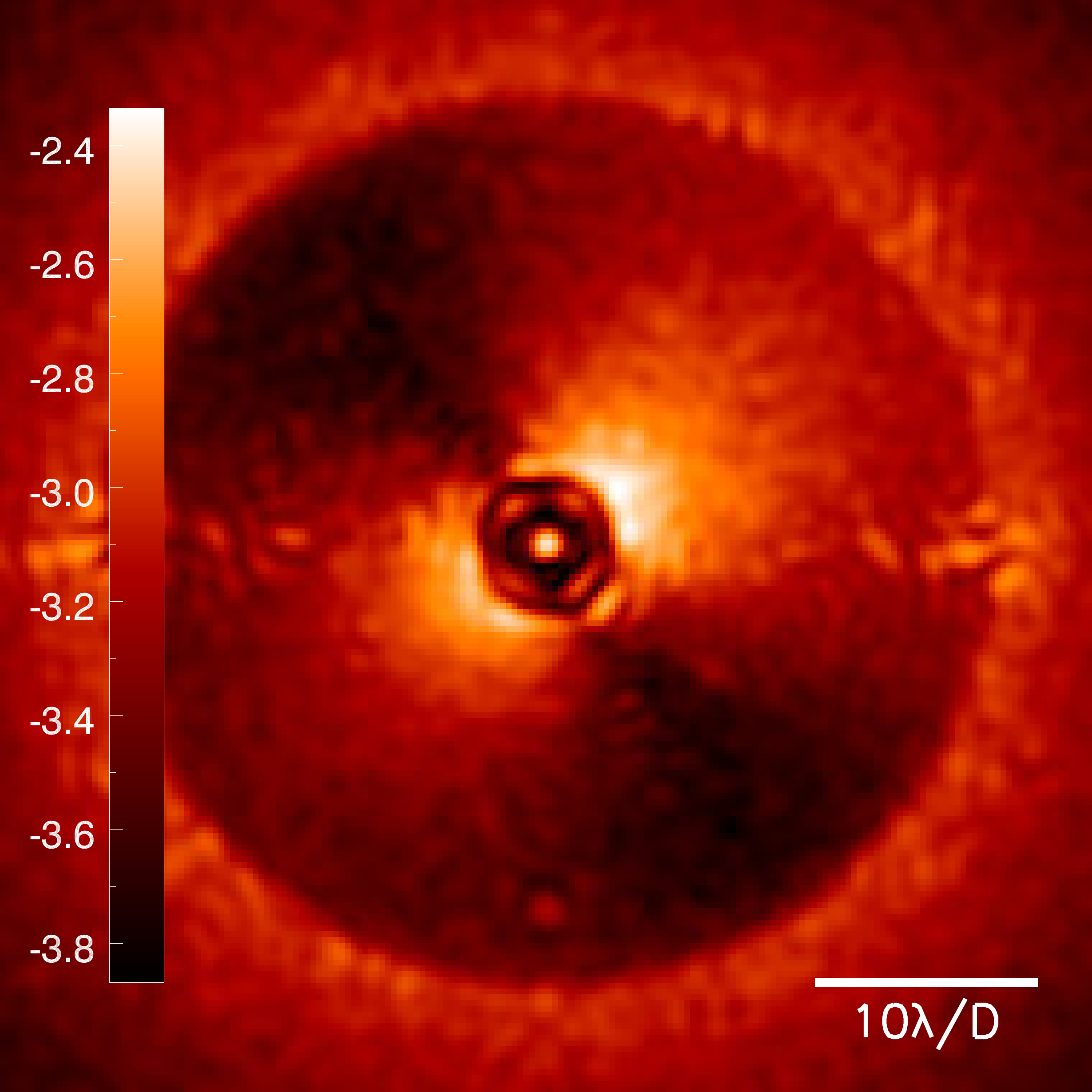}
\includegraphics{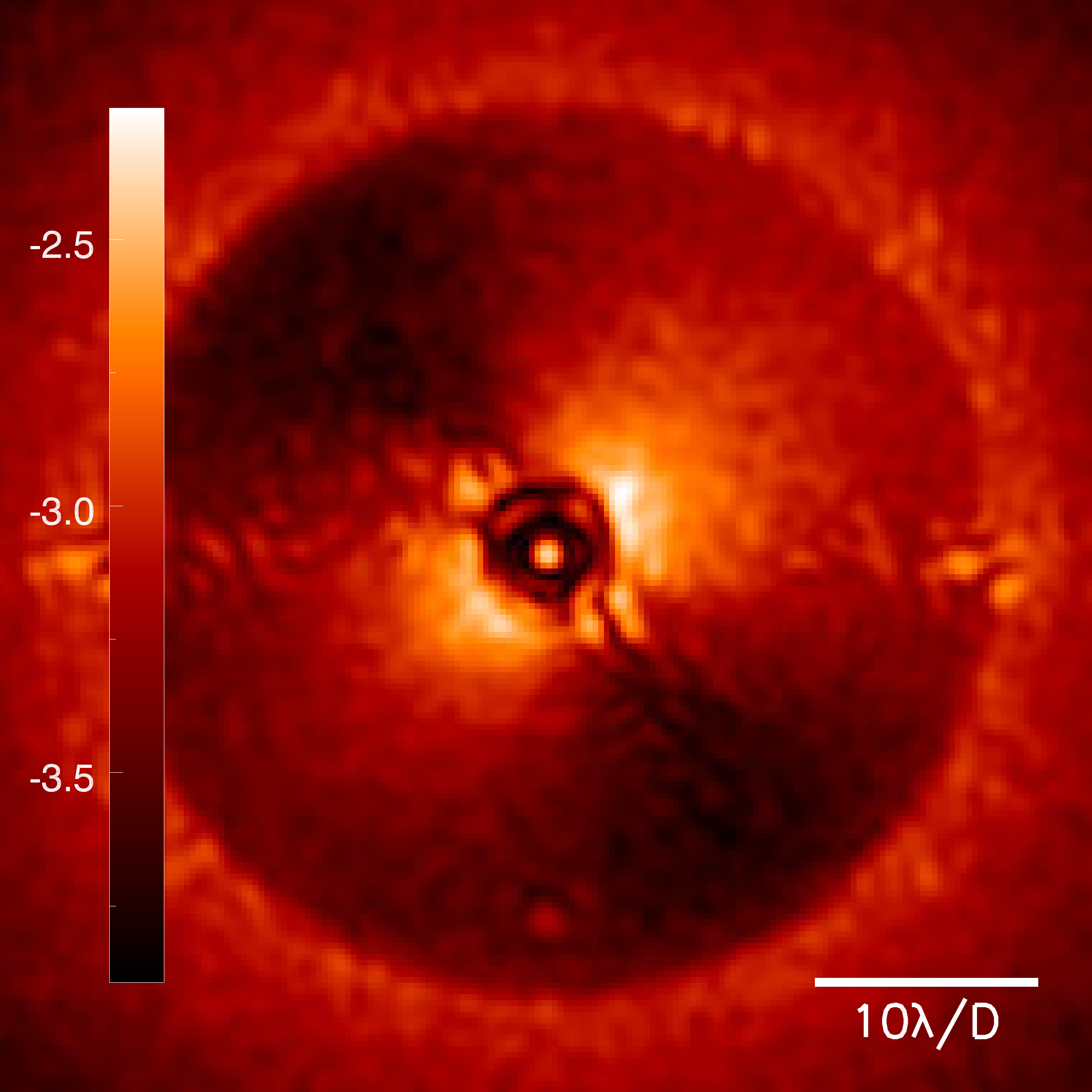}
\includegraphics{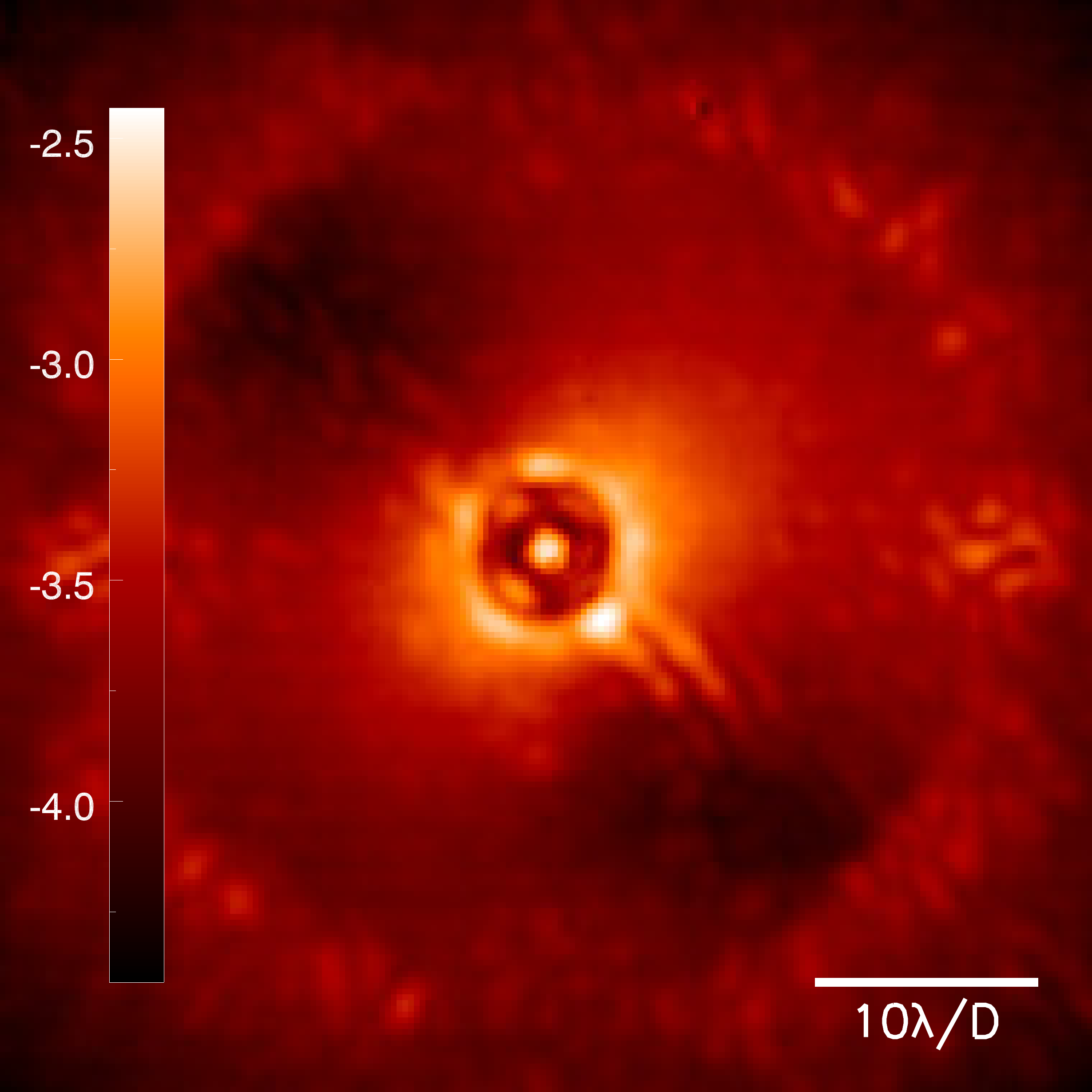}}
\caption{High contrast images (at the shortest wavelength, $0.966~\mathrm{\mu m}$) used to verify the WDH analysis procedure (logarithmic scale). 
Left: Simulation using only the fitting and AO-servolag errors. 
Middle-left: Simulation including additionally NCPA upstream (given by the ZELDA on-sky estimate) and downstream the APLC focal plane mask. 
Middle-right: Simulation including additionally LOR and LWE. 
Right: On-sky image of the star 51 Eri taken with the VLT/SPHERE-IFS instrument.}
\label{Fig-imgall}
\end{figure*}

To illustrate and verify our approach, we use four types of multispectral coronagraphic images, following the integral field spectrograph \citep[IFS,][]{Antichi2009} of SPHERE in the YH-band (from $0.96$ to $1.66~\mathrm{\mu}m$, with a spectral resolution $R\sim30$):
\begin{itemize}
\item Case~1: Simulated images obtained as a temporal stack of short exposures, containing only the AO-residuals due to fitting and servolag errors (produced using our analytical AO simulator) and using an apodized Lyot coronagraph \citep[APLC,][]{Soummer2011,martinez2009aplc} as the one used on SPHERE (see Fig.~\ref{Fig-imgall}, left);
\item Case~2: Simulated images like before, additionally including NCPA upstream and downstream the coronagraph focal plane mask. In order to have realistic simulations, close to the images actually obtained with SPHERE, we used the upstream phase estimated by the ZELDA mask \citep{NDiaye2013, Vigan2019zelda} during the latest tests conducted on SPHERE (see Fig.~\ref{Fig-imgall}, middle-left);
\item Case~3: Simulated image like before but additionally including a small amount of LOR ($5~\mathrm{mas}$ tip-tilt) and of LWE ($5~\mathrm{mas}$) in random direction for each pupil fragment separated by the spider of the telescope (see Fig.~\ref{Fig-imgall}, middle-right);
\item Case~4: On-sky VLT/SPHERE-IFS data cube of the 51~Eri star taken during the SPHERE-SHINE guaranteed time survey \citep{Chauvin2017SHINE} and described in \cite{Samland2017} and \cite{Maire201951eri} (see Fig.~\ref{Fig-imgall}, right).
\end{itemize}
For the three sets of simulated images, $200$ short exposures are stacked to obtain a long-exposure image. \postref{The injected wind direction is $125$ degrees.} For each multispectral cube, Fig.~\ref{Fig-imgall} shows only the shortest wavelength of the cube, at $0.966~\mathrm{\mu m}$. 

\subsection{Extracting the WDH contribution in the image}
In the focal plane image, the WDH extends from the center to the edge of the AO correction zone, as shown in Fig.~\ref{fig-psd}. Excluding the inner working angle of the coronagraph, the WDH is a low spatial frequency feature, whose intensity is dependent upon the atmospheric turbulence temporal variation (Sect.~\ref{ssec-turblag}), the AO \postref{servolag error} (Sect.~\ref{ssec-aolag}) and the image exposure time (DIT). In the case of SPHERE, the AO delay is fixed and the DIT are long enough so that the WDH shows as a smooth structure. Therefore, one simple way to separate the WDH contribution from the other starlight residuals that are mainly high-spatial frequency speckles (originating either from residual atmospheric turbulence or from NCPA) is to spatially filter the data in the Fourier domain to keep only its low frequency content. To perform the filtering, a Hamming window is a good compromise to avoid Gibbs effects (being a continuous function) while being steep enough to separate the frequencies at the user-specified cutoff frequency. The estimated WDH is the low-pass filtered image on which we apply an annular binary mask to cover the coronagraph signature (below $2\lambda/D$) and the seeing-limited area (beyond $20\lambda/D$). As SPHERE images show two artefacts on its correction ring, which are due to the DM square grid (see Fig.~\ref{Fig-wdh} \postref{right image, encircled with yellow dashed line}), these two spots are also masked.

\begin{figure*}
\centering
\resizebox{\hsize}{!}{\includegraphics{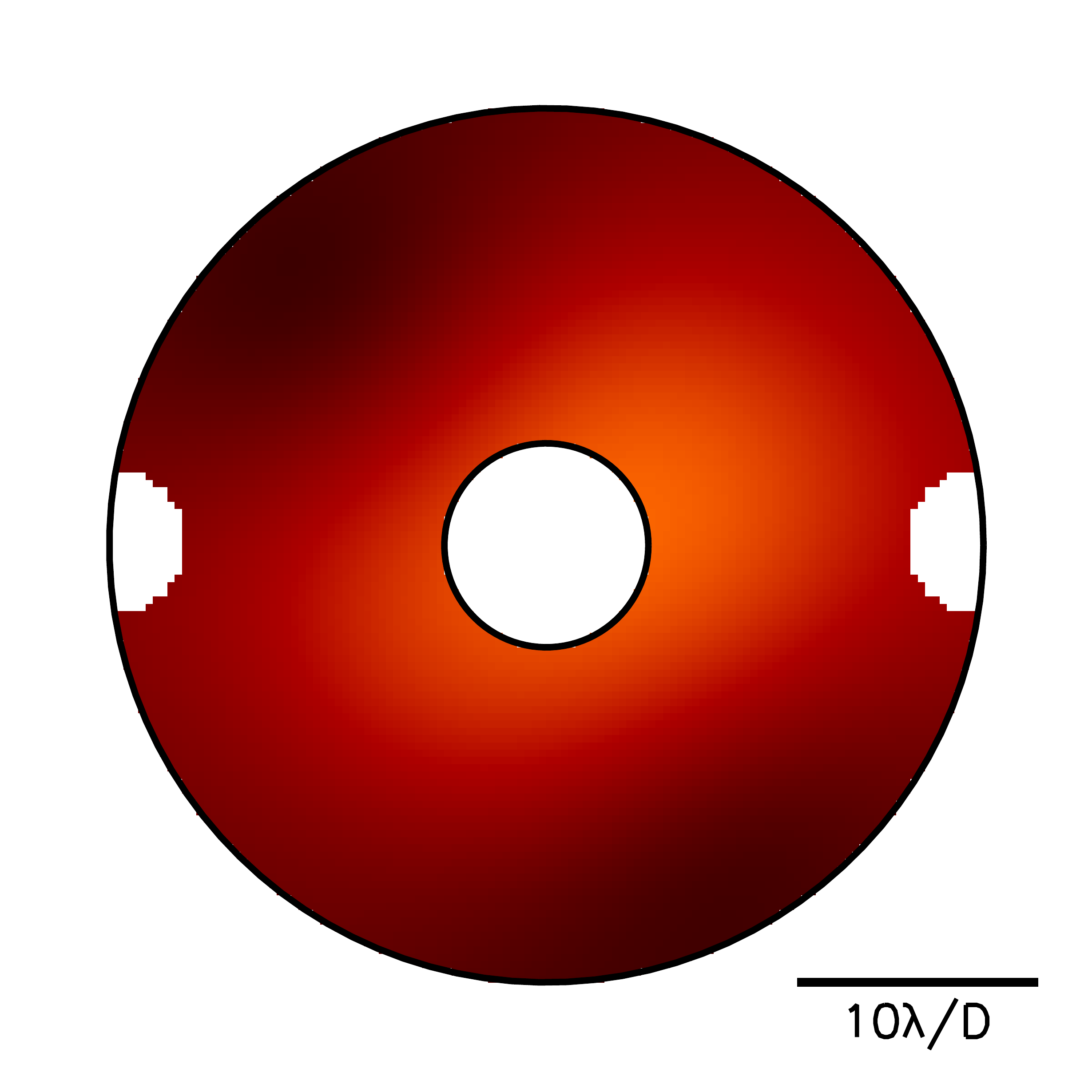}
\includegraphics{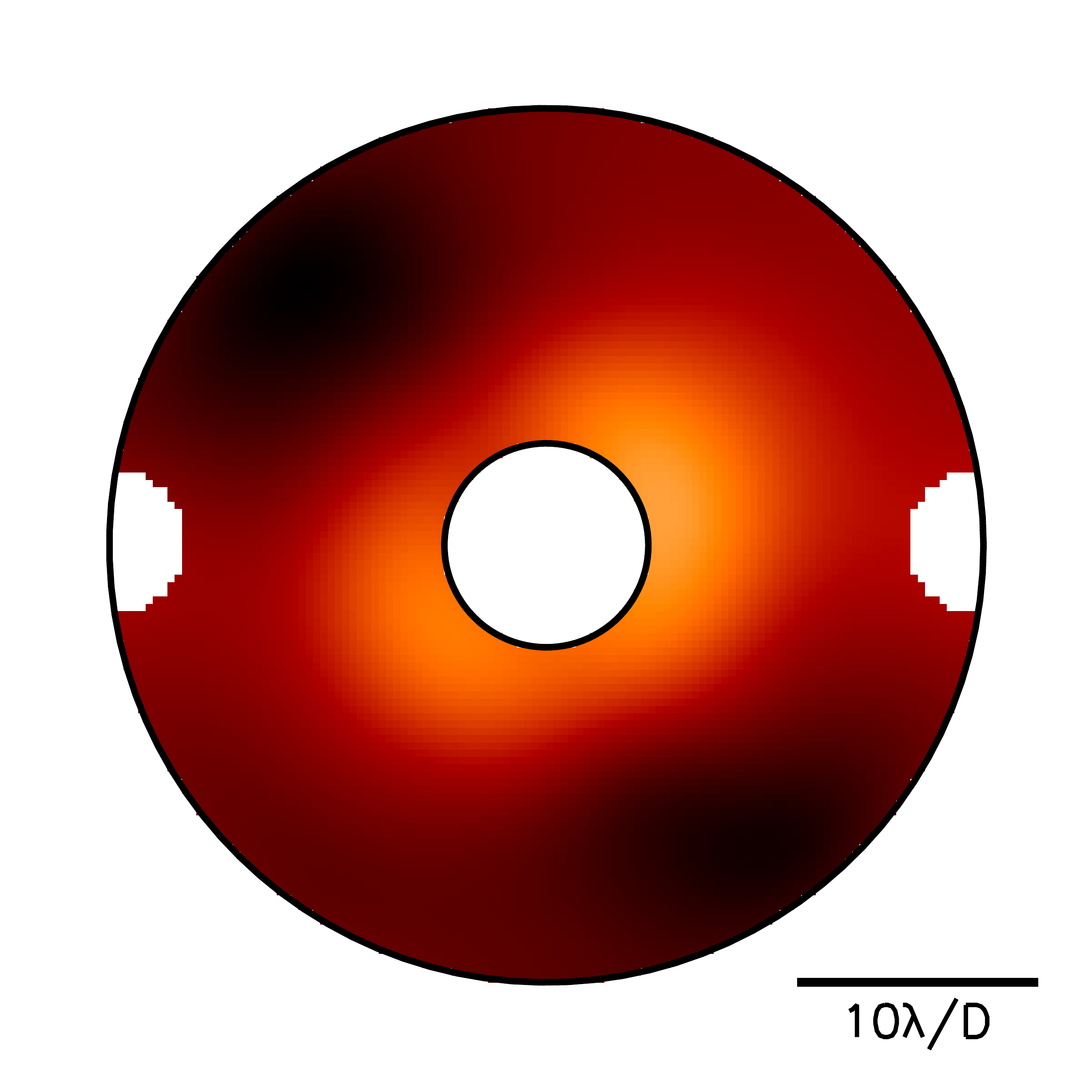}
\includegraphics{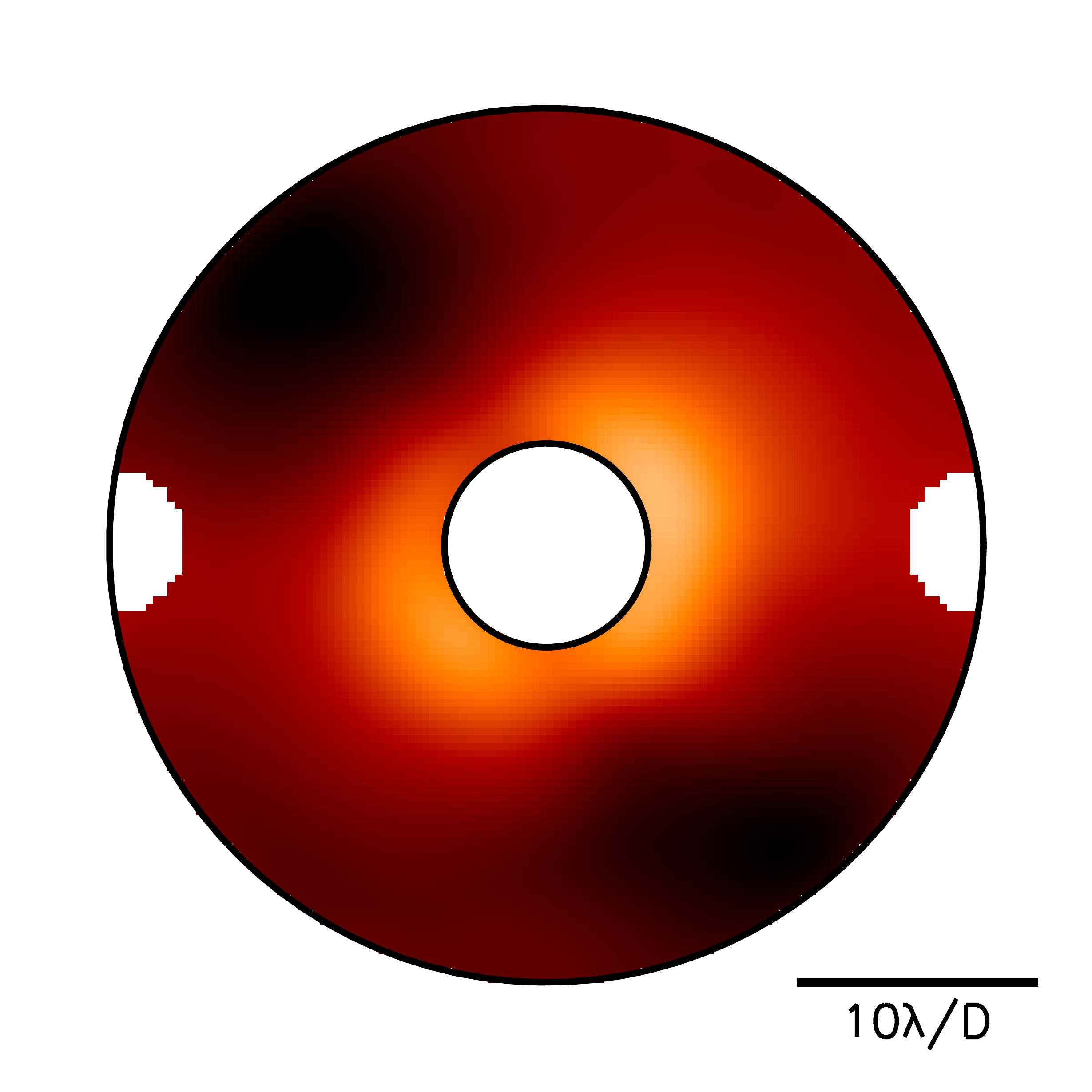}
\includegraphics{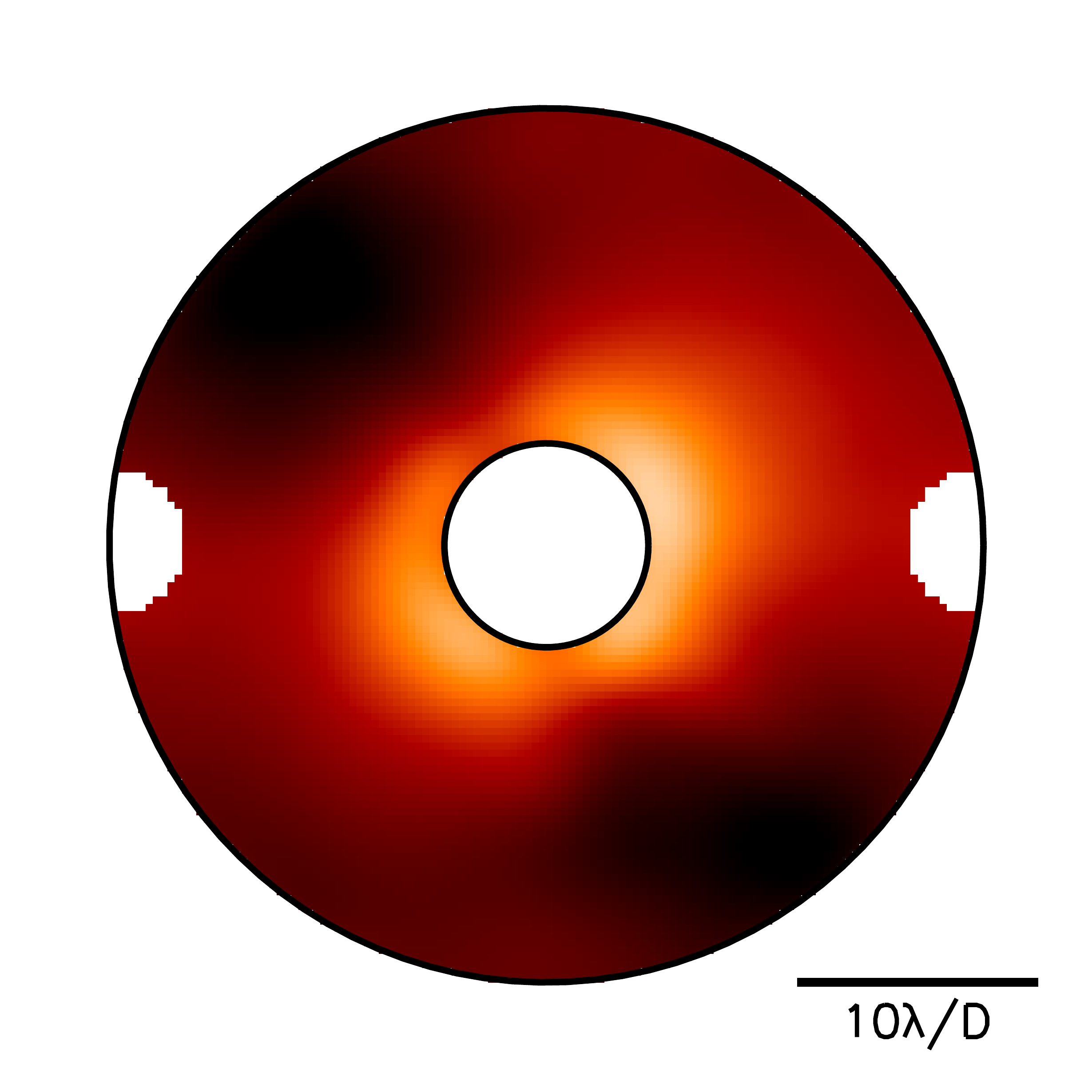}
\includegraphics{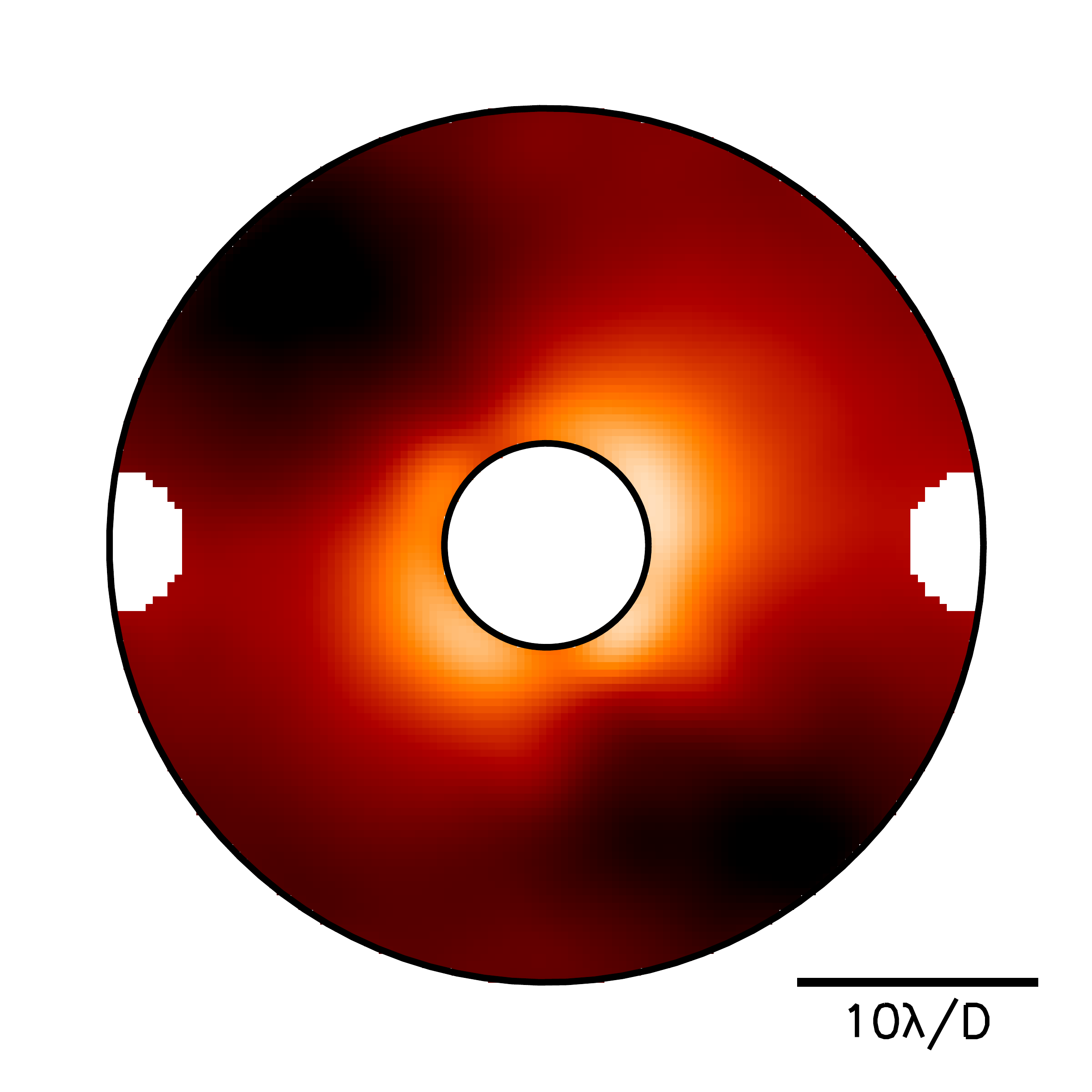}}
\resizebox{\hsize}{!}{\includegraphics{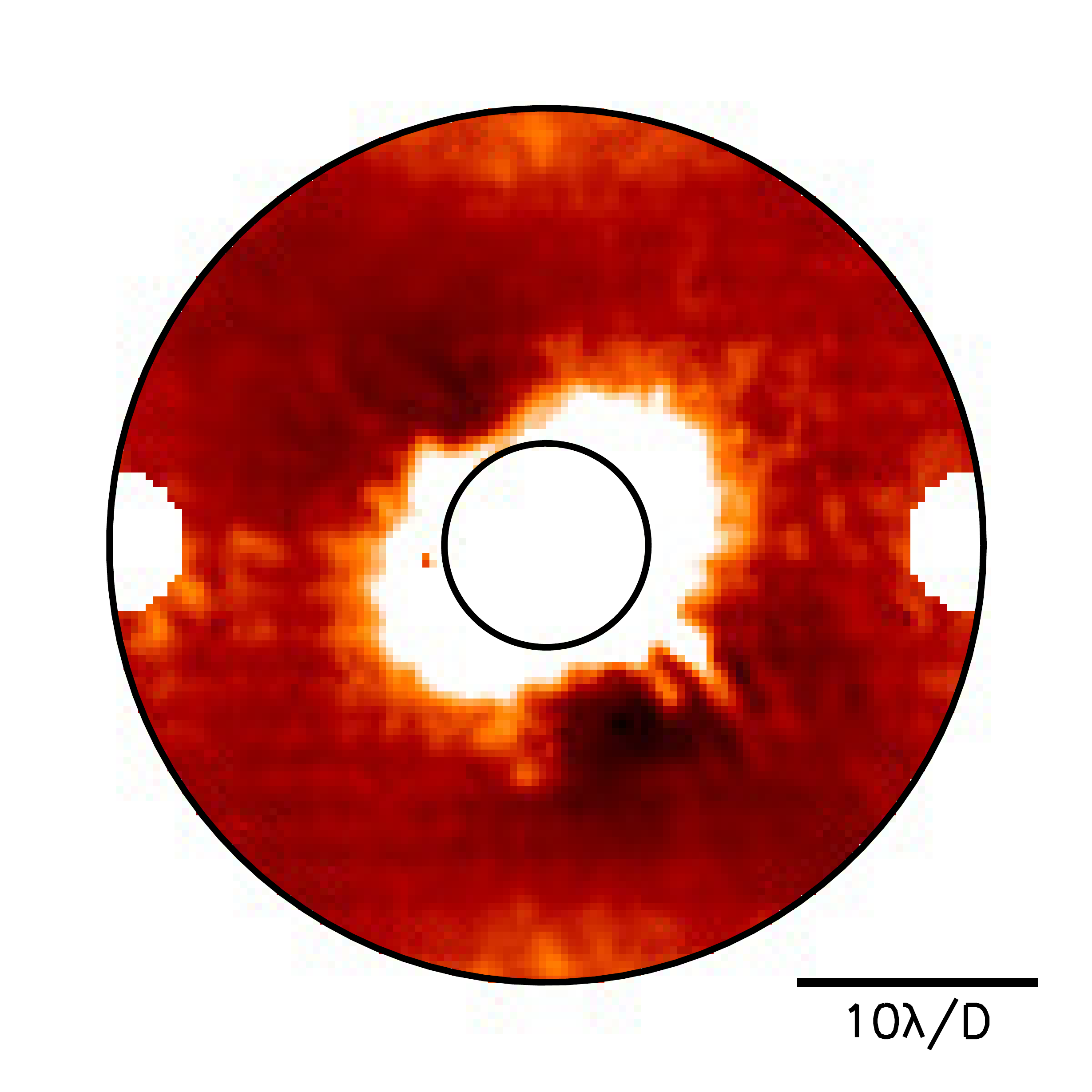}
\includegraphics{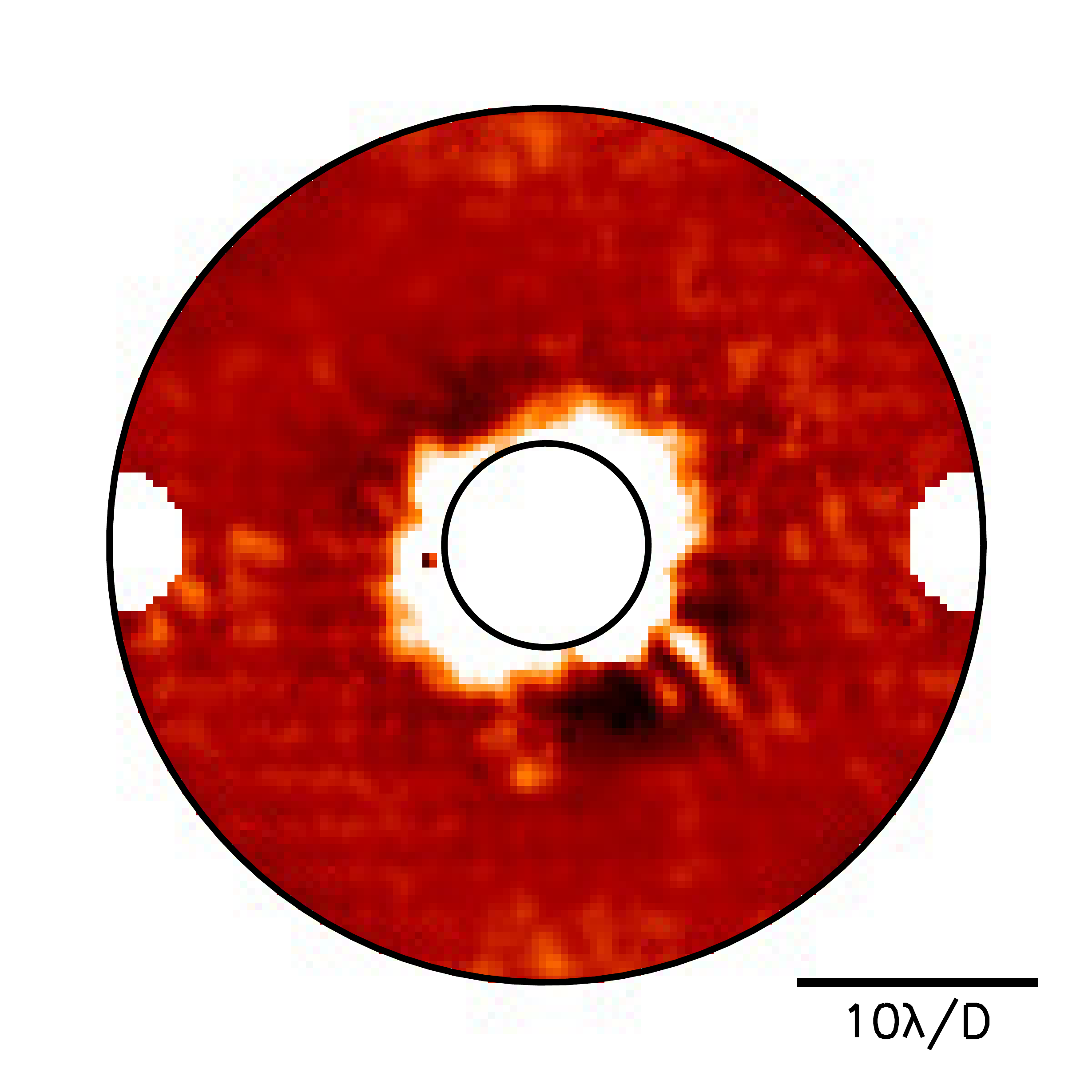}
\includegraphics{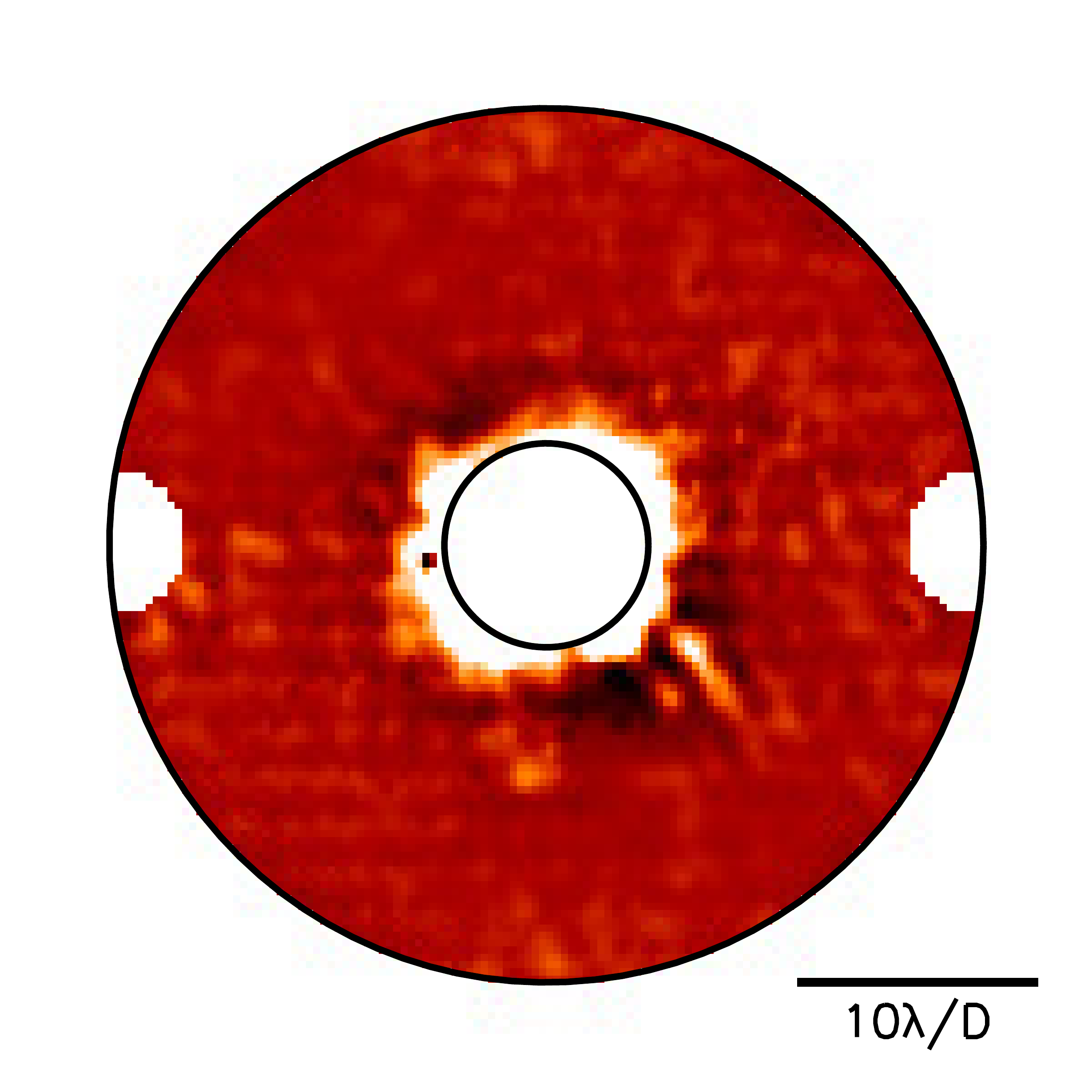}
\includegraphics{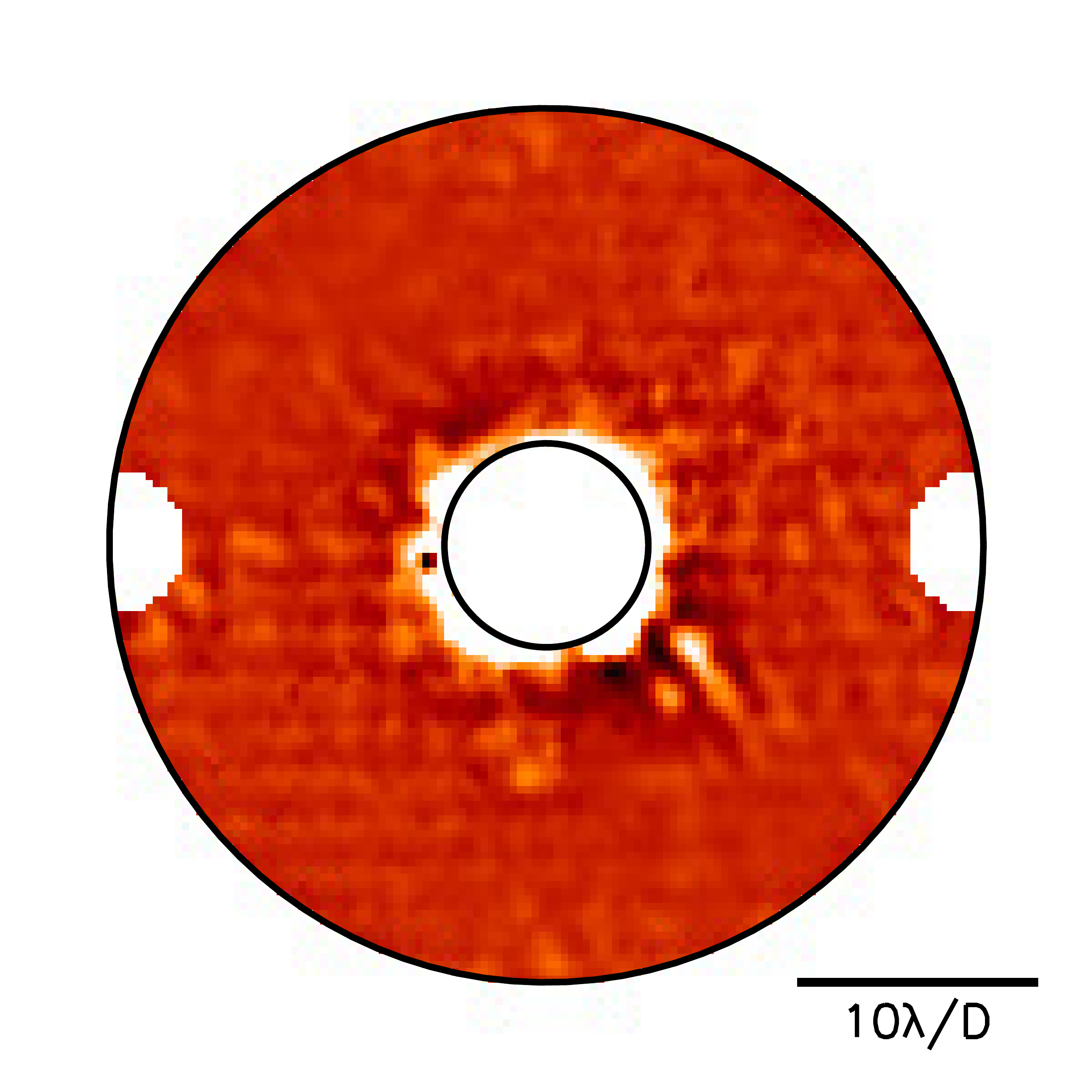}
\includegraphics{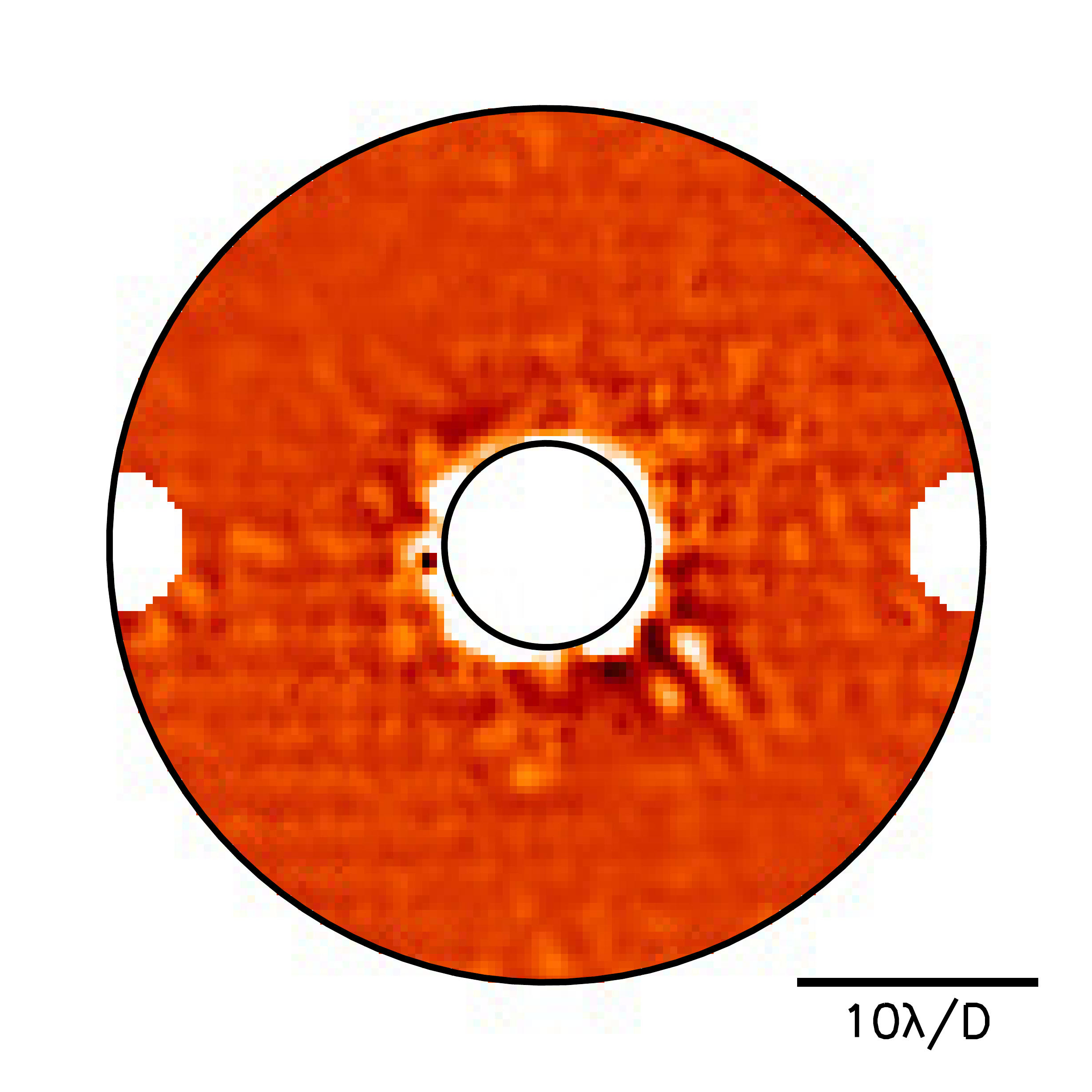}}
\caption{Spatially low-pass filtered images of the 51 Eri image shown in Fig.~\ref{Fig-imgall}-Right (top, logarithmic scale) and difference between the image and its low pass filtered version (bottom, linear scale). The images are masked to show only the AO-corrected area and hide the DM artefacts and the coronagraphic signature. From left to right with a filtering fraction (percentage of low frequencies kept in the image) of 5\%, 10\%, 15\%, 20\% and 25\%.}
\label{Fig-imgfilt}
\end{figure*}

To verify whether the high-pass filtering can indeed extract the WDH contribution, Fig.~\ref{Fig-imgfilt} shows the on-sky 51 Eri image (Fig.~\ref{Fig-imgall}, right) \postref{filtered} with different filtering fraction (percentage of low frequencies kept in the image)\postref{: the top row shows the low spatial frequencies and the bottom row shows the high spatial frequencies}. 
For a filtering fraction of 5\% (left column) \postref{the high-pass filtered image} (top row) results in a too smooth halo whereas from a filtering fraction of 15\% on, \postref{it shows sharper structures} due to the \postref{capture} of speckles. 
\postref{For a filtering fraction of 5\% the low-pass filtered image (bottom row) still shows a slight elongation along the wind speed direction at short separation, which completely disappears from 15\% on.} 
When the filtering fraction is too high, we can visualize very faint structures such as the grid of microlenses from the IFS design.
After testing different data set, a good trade-off based on visualization of the images \postref{(the low-pass filtered image shows a very smooth structure while its high-pass version shows no further directional elongation)}, is to use 15\% of the low spatial frequency content. \postref{This qualitative argument is confirmed by further analysis.}

\begin{figure}
\resizebox{\hsize}{!}{\includegraphics{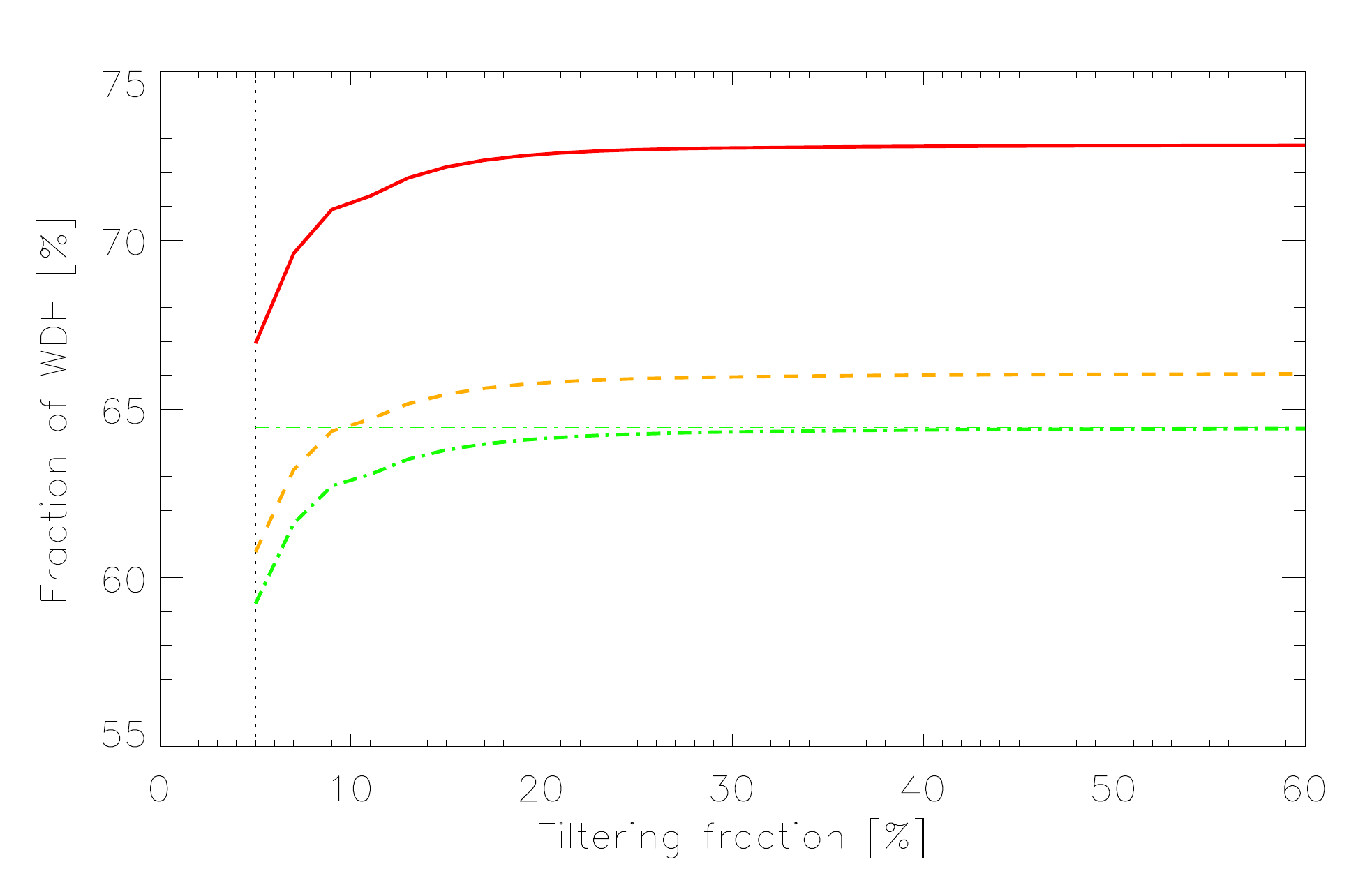}}
\caption{Amount of light within the WDH structure extracted by spatial filtering, as a function of the filtering fraction (starting from 5\%, at the vertical dotted black line). 
Red solid line is the first case containing only the WDH contribution, orange dashed line is the second case with NCPA and green dotted-dash line is the third case containing NCPA, LWE and LOR. 
\postref{The horizontal lines, with the same color code, are the real theoretical values extracted by simulating the exact same images but without the servolag error} (resp. $73\%$, $66\%$ and $64\%$).}
\label{Fig-ncpavsfilt}
\end{figure}

To check whether this filtering procedure indeed brings up mainly, if not only, the WDH component, we applied it on the three simulated data cases and compared with the exact same simulated cases but without the servolag error included. By comparing the images produced with and without WDH, we found that $73\%$ \postref{($66\%$ and  $64.5\%$)} of the light in the corrected area belongs to the WDH for the first case \postref{(second and third respectively)}.
To compare these absolute values to the ones extracted by low pass filtering the images, we computed the fraction of WDH \postref{(}the ratio between the total \postref{intensity} in the filtered masked image and the total \postref{intensity} in the \postref{non-filtered} masked image\postref{)} as a function of the filtering fraction (Fig.~\ref{Fig-ncpavsfilt}). 
The fractions of WDH extracted in the three cases are indeed lower than the absolute values \postref{(shown as horizontal lines in Fig.~\ref{Fig-ncpavsfilt})}, but from a filtering fraction of $22\%$ it reaches a plateau at the expected \postref{absolute} values. When adding NCPA and LWE or LOR, the starlight in the corrected area is scattered in other higher spatial frequencies that are not captured by the low pass filtering, hence the lower fraction of WDH for the case 2 and 3. As a conclusion, the WDH contribution can be extracted from the focal plane images by applying a low-pass filter with a filtering fraction of about $15\%$. \postref{Note that this rule of thumb is only valid for SPHERE data and the optimum fraction must be determined for a given instrument.}

\subsection{Direction of the WDH}
\label{ssec-dir}
In order to assess the direction of the WDH elongation, we developed the following procedure. 
In a first step, we apply a discrete Radon transform\footnote{The discrete Radon transform is a linear operator that transforms a given 2D image into a 2D map showing the intensity along lines over the image, as a function of the angle (for a square image of dimension $ N_{img} \times N_{img}$, the Radon map is of dimension $(\pi.N_{img}) \times (2.N_{img}-1)$).} \citep{Radon1917,Radon1917translation} to the estimate of the WDH, which provides the integrated intensity over one direction as a function of the angle. 
In a second step, to obtain the profile of the intensity as a function of the angle, $R_{WDH}(\theta)$, we average a few channels (typically a few pixels) of the Radon transform around its center (corresponding to the center of the image). 
In a third step, we perform a Gaussian fit around the maximum value of this Radon profile in order to extract the preferential direction of the WDH. Indeed, using directly the maximum value of the Radon profile is not the most robust option since, depending on the number of pixels within the masked image, the profile might be irregular. After testing different possibilities and data sets, we found out that performing a Gaussian fit around the maximum value of the Radon profile is the most robust method that does not require tuning any user-parameter. The uncertainties on the estimated direction are therefore the uncertainties on the Gaussian fit performed as, in an ideal case, the Radon profile is purely sinusoidal and the Gaussian fit around the maximum value is quite accurate. 
Thanks to this procedure, we extract the preferential direction of the starlight distribution \postref{in} the field of view. 

\begin{figure}
\resizebox{\hsize}{!}{\includegraphics{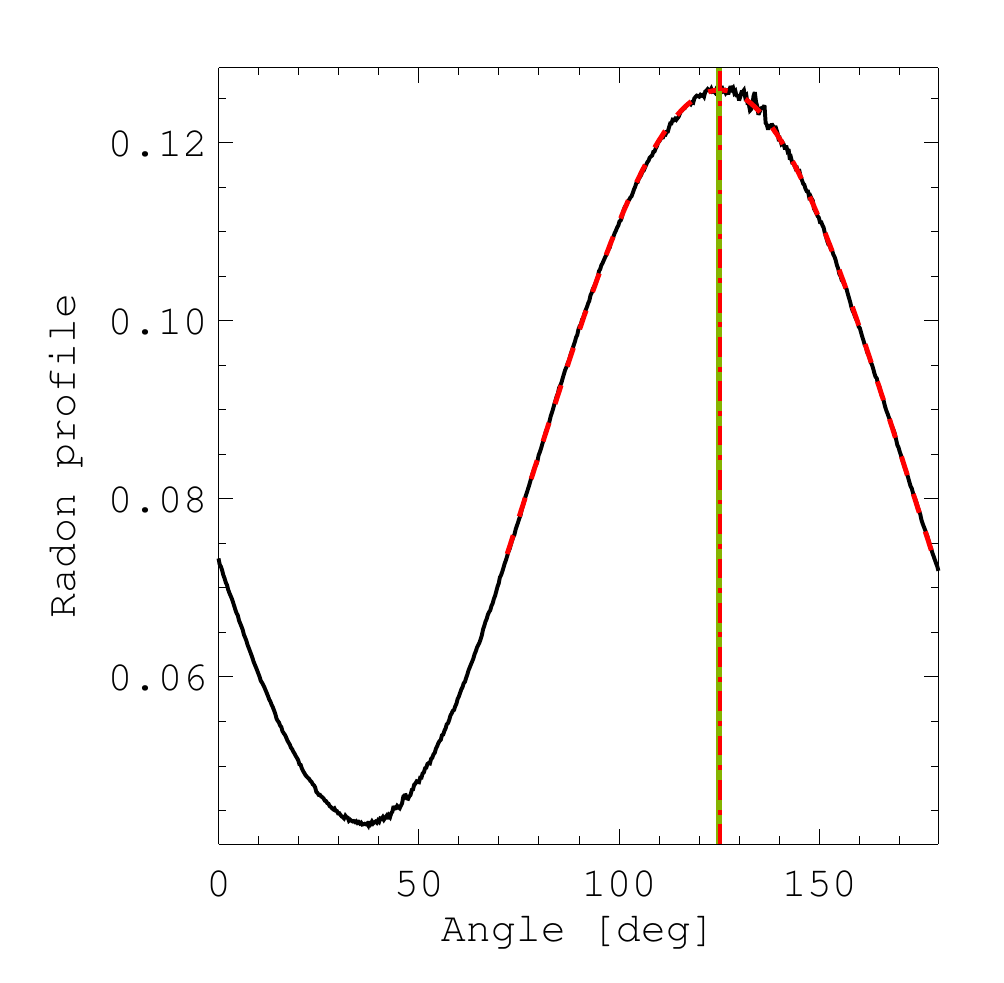}\includegraphics{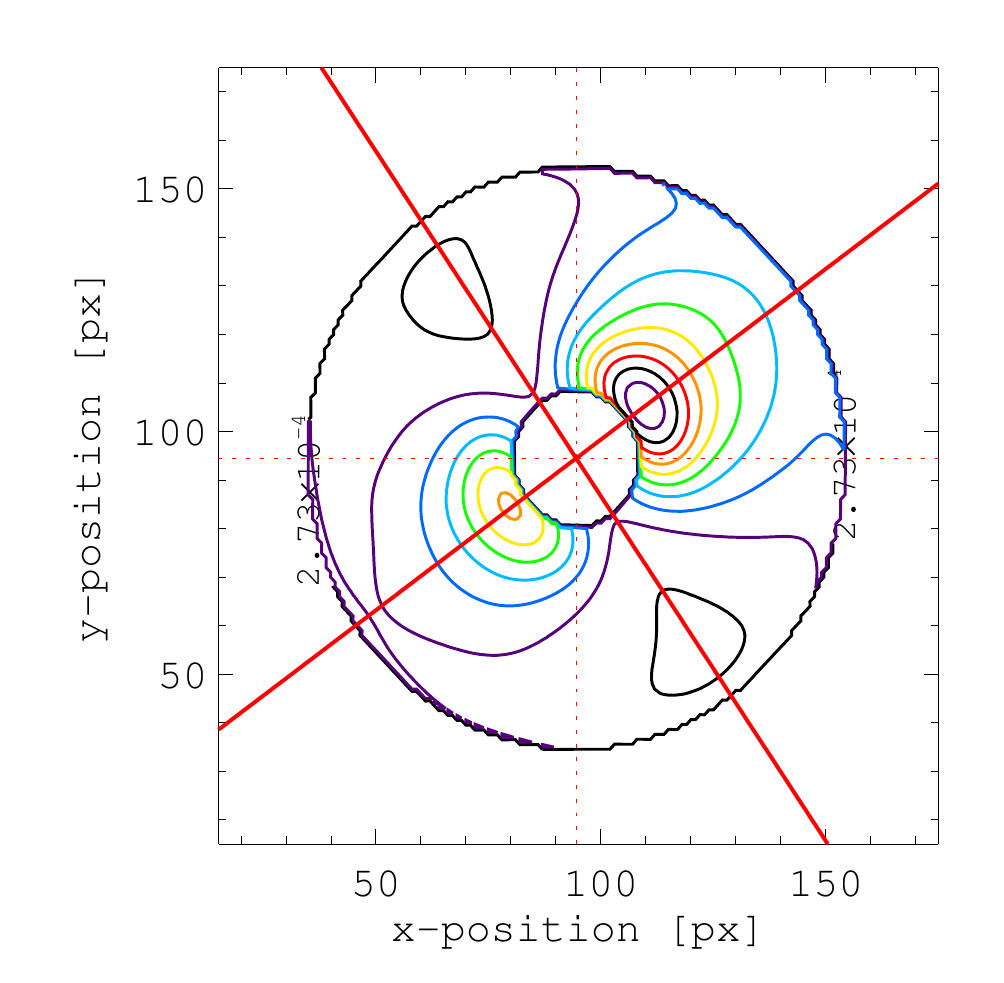}}
\resizebox{\hsize}{!}{\includegraphics{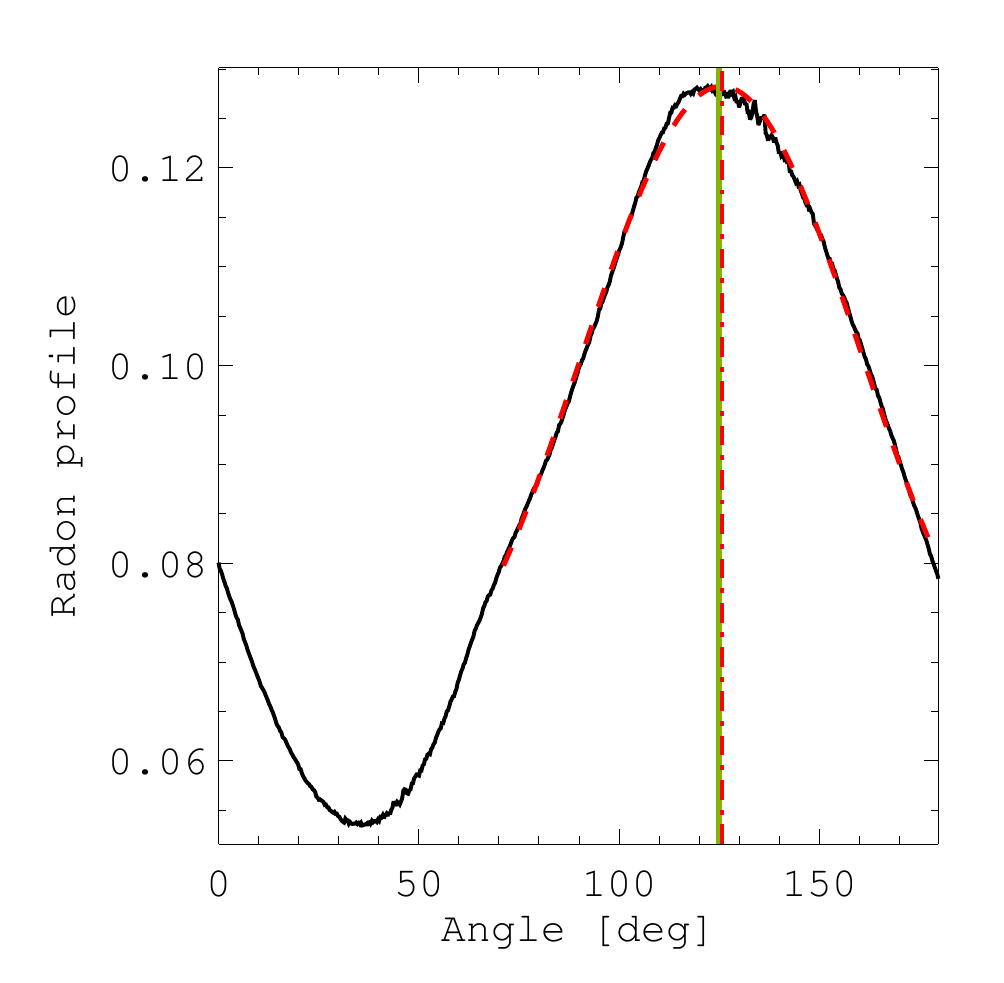}\includegraphics{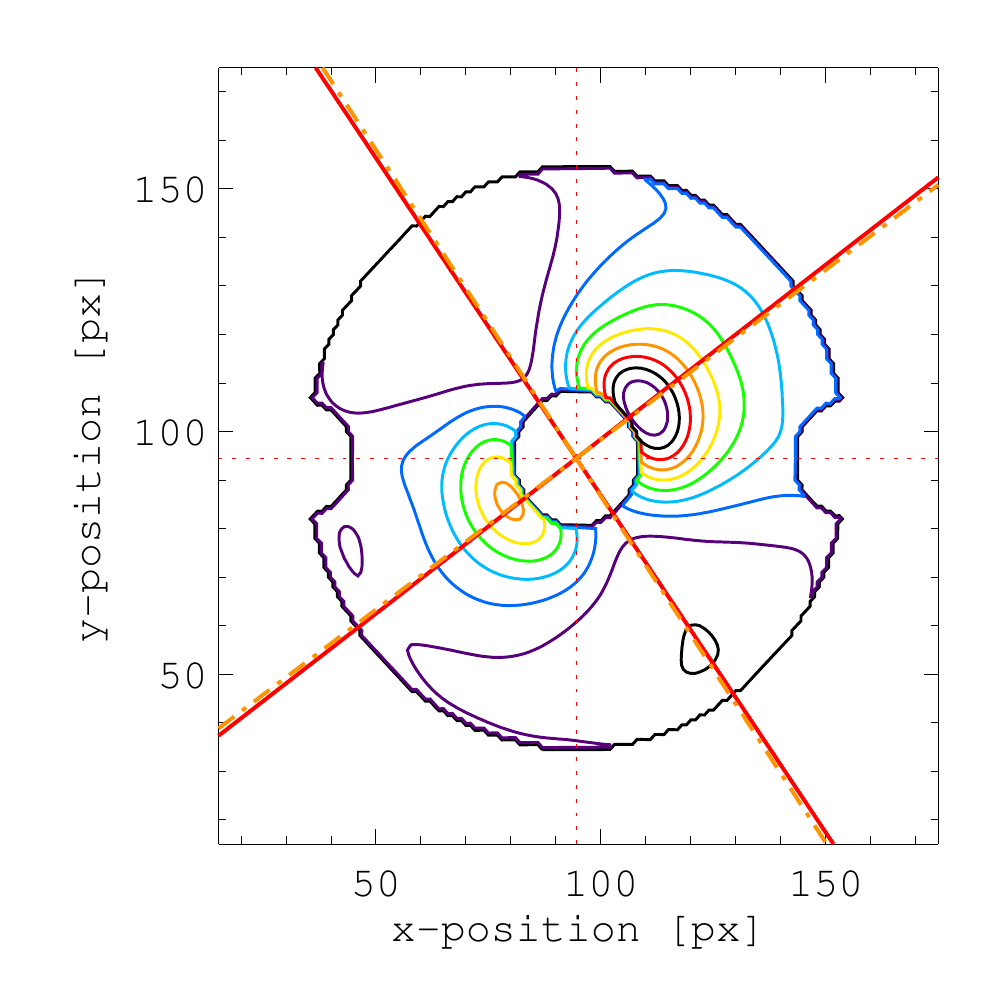}}
\resizebox{\hsize}{!}{\includegraphics{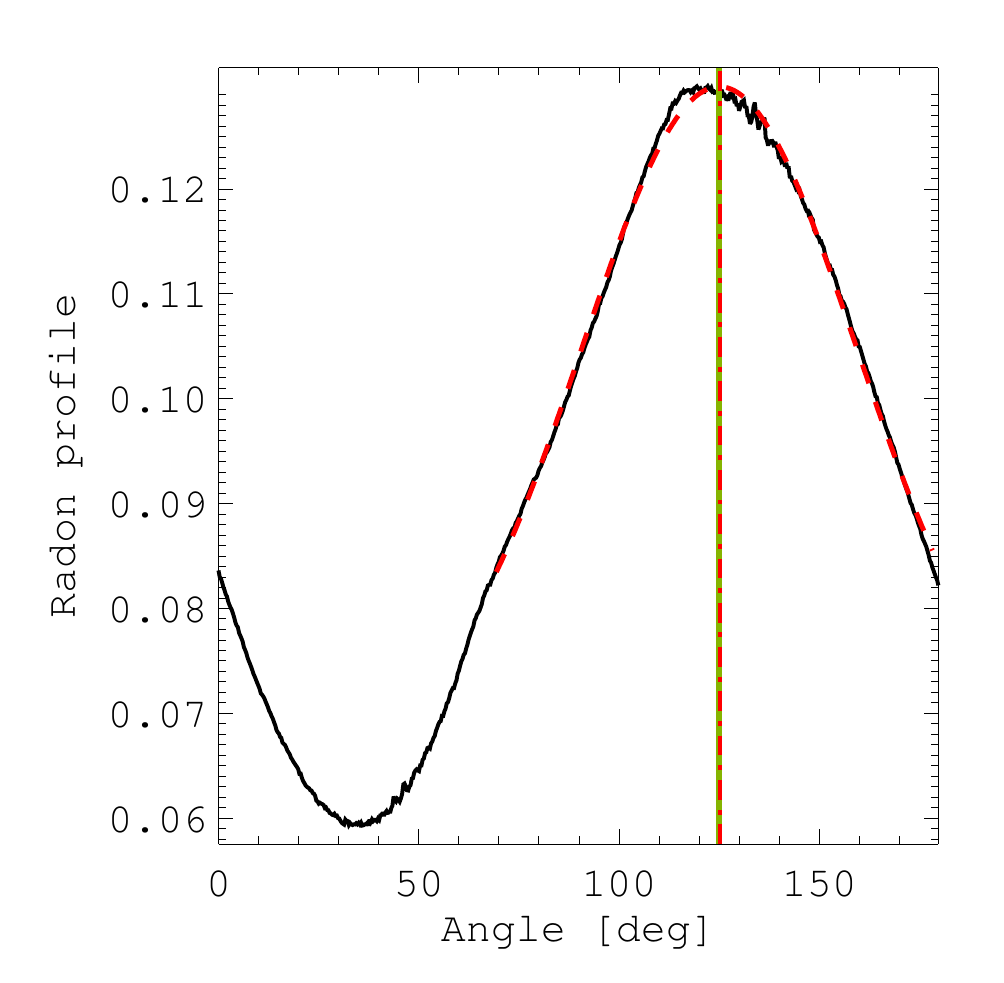}\includegraphics{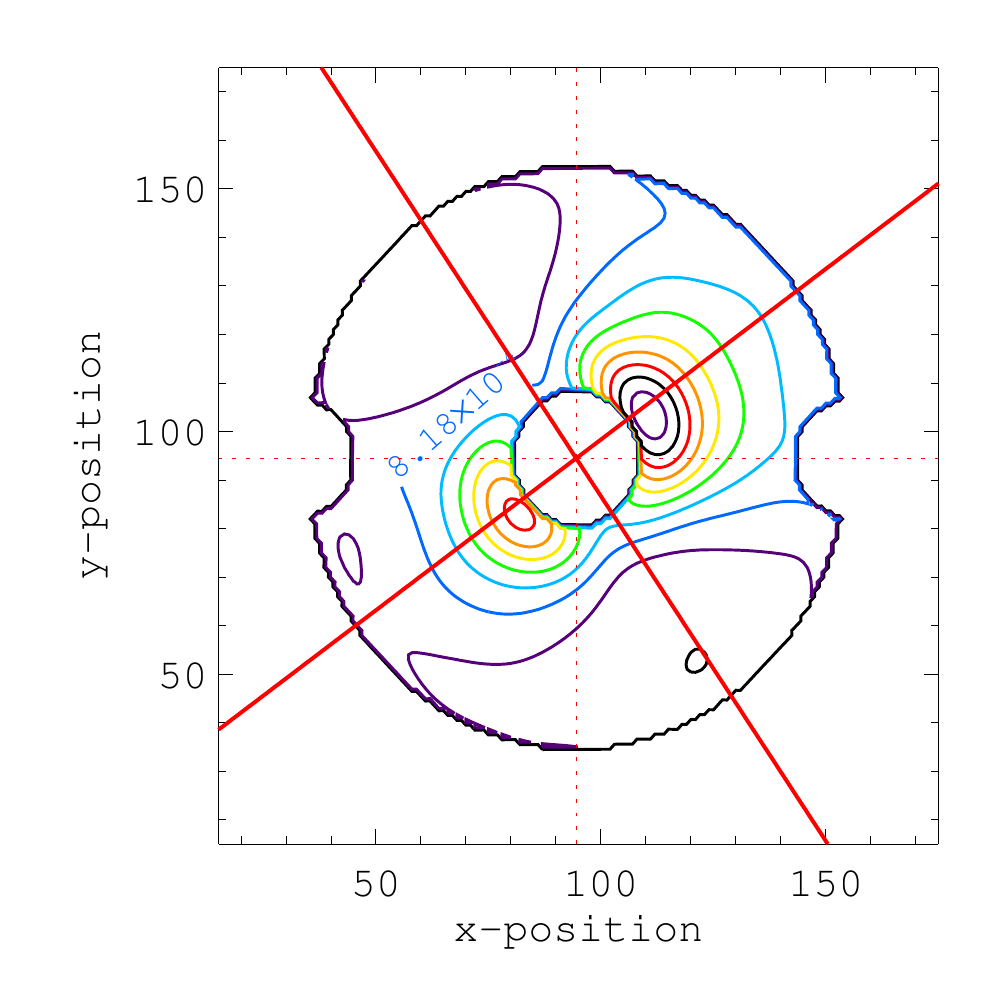}}
\resizebox{\hsize}{!}{\includegraphics{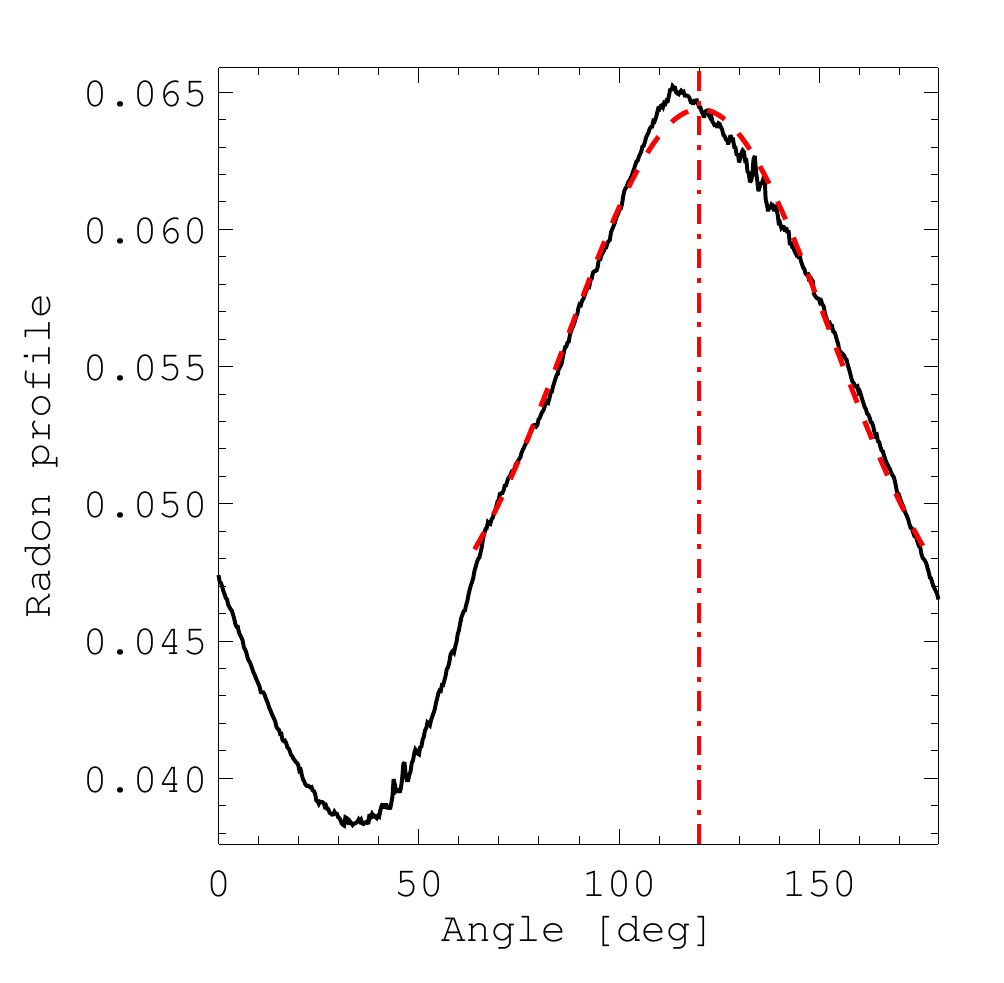}\includegraphics{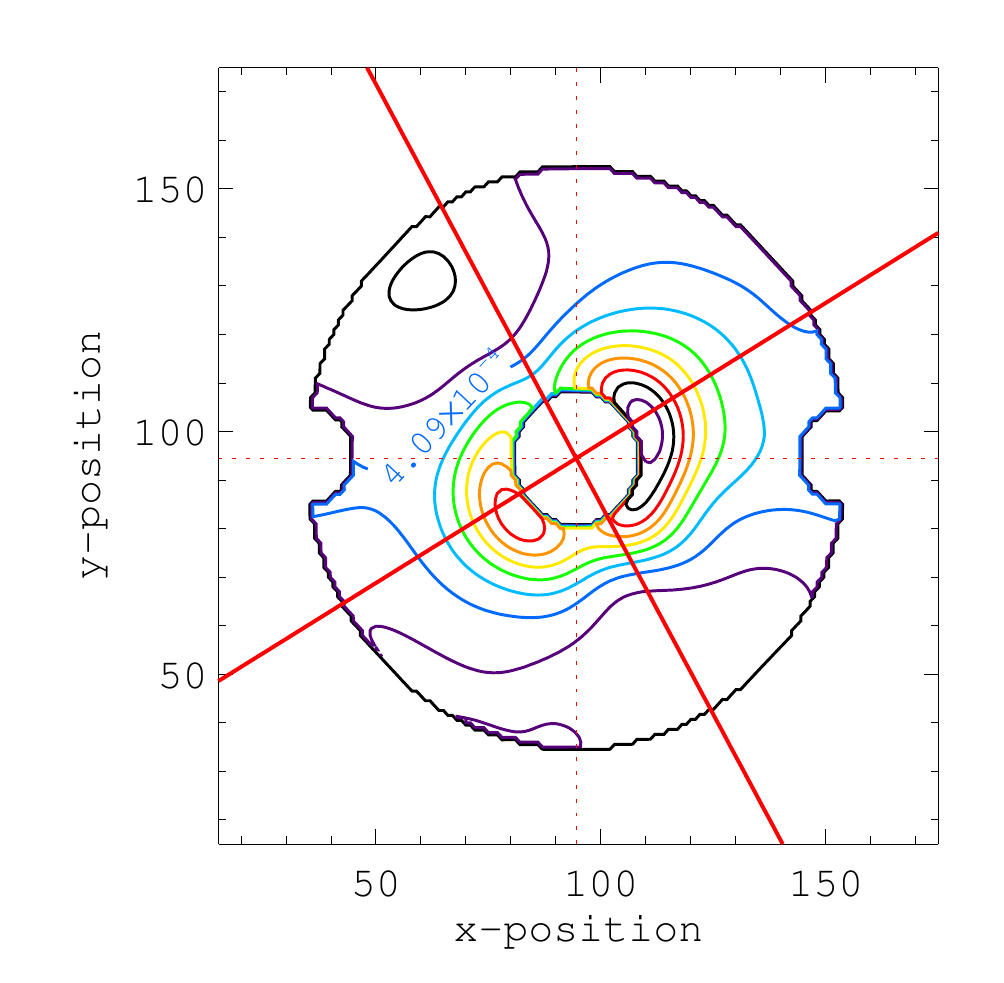}}
\caption{Estimation of the wind direction on the four cases: Extracted radon profile of the WDH and Gaussian fit (red dashed line) whose maximum (red dot-dashed line) shows the estimated preferential direction (left) and contour plot of the WDH showing the fitted wind direction (red cross) with the described procedure (right). 
From top to bottom: Case~1 (from AO simulations including only the fitting and servolag errors), case~2 (adding NCPA), case~3 (adding LOR and LWE) and case~4 (51~Eri on-sky data). Except for the case~1, a filtering fraction of $15\%$ has been used to isolate the WDH. For the first three cases simulated, the green line shows the simulated wind direction ($125$ degree).}
\label{Fig-radvssimu}
\end{figure}

In a first step, we checked that this approach is valid to extract the direction of the WDH on the first simulated data-set. 
Since in this data set only the WDH contribution is present, there is no need to perform the low-pass filtering. 
\postref{For this} case, the Radon profile (Fig.~\ref{Fig-radvssimu}, top-left) is quite smooth and the maximum of its Gaussian fit yields an estimated direction of $125.13$ degrees (see Fig.~\ref{Fig-radvssimu}, top-right). The uncertainty on the Gaussian fit is negligible which means systematic errors \postref{(}such as centering of the raw image and plate scale\postref{)} are \postref{dominating the error}. In the following we will therefore ignore the uncertainty on \postref{the} estimated \postref{direction}. 

As a second step, we \postref{apply this procedure on} the second and third simulated data set. 
After low-pass filtering the images with a filtering fraction of $15\%$, Fig.~\ref{Fig-radvssimu} (middle panels), show the obtained Radon profiles (left) and the extracted WDH contour plots with the fitted direction overlaid (right). 
On the simulations including NCPA \postref{(case~2)}, the estimated direction is $125.1$ degrees, as in the case without NCPA. The NCPA are mainly inducing high spatial frequencies in the focal plane \citep[see][]{Vigan2019zelda}, they do not affect greatly the extraction of the WDH and therefore of its direction. 
On the simulations including additionally LOR and LWE \postref{(case~3)}, the estimated direction is $124.5$ degrees. Indeed these low spatial frequencies slightly offset the WDH, as shown in the Radon profile that is not perfectly sinusoidal (Fig.~\ref{Fig-radvssimu}, middle-left). 

As a third step, we applied this approach to the on-sky data of 51 Eri for which the centering of the star behind the coronagraph is not perfect and some features due to LWE are visible (Fig.~\ref{Fig-imgall}, right). 
As expected, the obtained Radon profile deviates even more from a sinusoidal shape (Fig.~\ref{Fig-radvssimu}, bottom-left). The contour plot shows that the estimated direction is visually in line with the average halo (Fig.~\ref{Fig-radvssimu}, bottom-right). By construction, our approach does not fit perfectly the inner part or the outer part, \postref{but is a good fit overall}. 

\begin{figure}
\resizebox{\hsize}{!}{\includegraphics{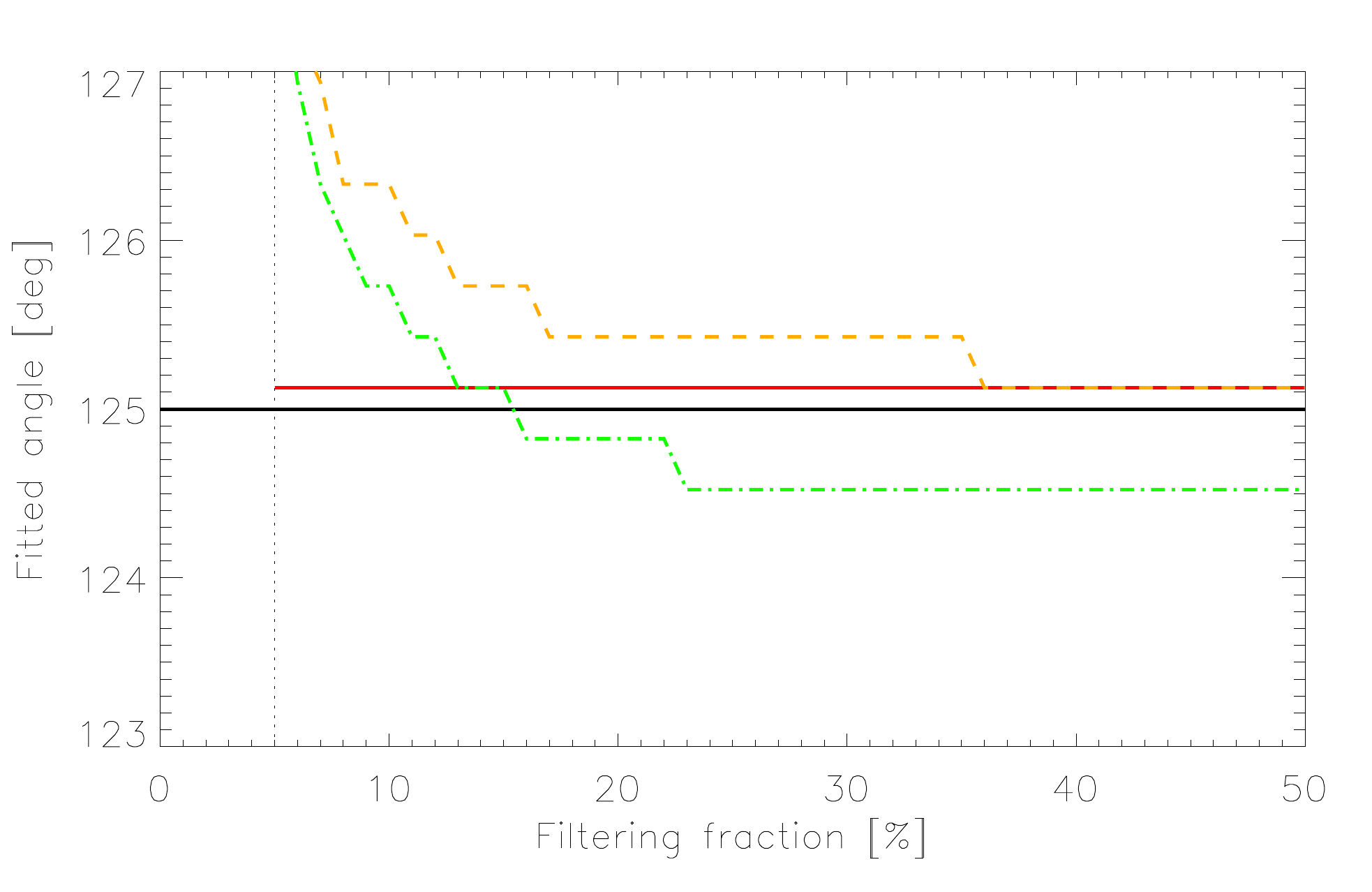}}
\caption{Fitted WDH direction in the simulated images as a function of the filtering fraction (starting from 5\%, at the vertical dotted black line). The horizontal black solid line is the injected value ($125$ degrees). Red solid line is the first case containing only the WDH contribution, orange dashed line is the second case with NCPA and green dotted-dash line is the third case containing NCPA, LWE and LOR.}
\label{Fig-angfitvsfilt}
\end{figure}

As a last step, we checked the robustness of the estimated direction with the filtering fraction used. For the three simulated data sets, Fig.~\ref{Fig-angfitvsfilt} shows the fitted angle of the WDH direction, as a function of the filtering fraction. From a filtering fraction of $15\%$, the estimated direction reaches a plateau around the injected value. In this specific case, even without filtering, the estimation is off from the injected value by only two degrees. We repeated this experiment on on-sky data (for which the real wind direction is not known) showing different amount of WDH. Except for the case of very low WDH contribution, the estimated direction is stable to $1$ degree for a filtering fraction between $5\%$ and $25\%$. In the case of very low WDH, the estimation is dominated by residual tip-tilt errors and varies of up to $7$ degrees between a filtering fraction of $5\%$ to $17\%$ but from $17\%$, a plateau is reached. \postref{Note that this experiment can be used to determined the optimal filtering fraction for a given HCI instrument.} 
\postref{In the context of the present study and our two future applications, we would like to estimate the wind direction within a few degrees accuracy.} 
The procedure to extract the direction of the WDH is therefore robust enough to further analyze the WDH structure. 

\subsection{Strength of the WDH}
\label{ssec-str}
When no WDH is present in the image, the latter procedure provides a \postref{random} direction. In order to check whether some WDH is present in the data, after the filtering and extraction of the Radon profile, we compute the standard deviation of the Radon profile. If this standard deviation is less or equal to the contrast expected without WDH, then there is no WDH. This threshold $C_{\tau}$ is obtained \postref{empirically} via the simulated PSD of the residual phase without servolag, from its minimum in the region where the profile of the Radon transform is computed (for the SPHERE data in H-band, this threshold is set to $C_\tau = 7.10^{-4}$). We checked this procedure on SPHERE on-sky data and it efficiently sorted the data without WDH from the data showing WDH. 

More generally, in order to assess the amount of starlight scattered into the WDH within the corrected area, we define its strength, $\mathcal{S}_{WDH}$, which is $100\%$ if all the starlight in the AO-corrected zone belongs to the WDH and $0\%$ if there is no WDH:
\begin{equation}
\mathcal{S}_{WDH} = \frac{\sigma(R_{WDH}(\theta)) - C_{\tau} }{\sigma(R_{WDH}(\theta)) + C_{\tau}} \mpostref{\times} 100,
\label{eq:Strength}
\end{equation}
where $\sigma(R_{WDH}(\theta))$ is the standard deviation of the Radon profile $R_{WDH}$ extracted previously. 
Note that the defined strength is not an exact estimate but provides a relative value with respect to the total intensity in the corrected area and serves as an indicator. 

When applying this metric to the three simulated data cases containing the same amount of WDH, the estimated strength as a function of the filtering fraction is about the same for all cases (to within $0.5\%$ of each other). From a filtering fraction of $10\%$ on, this metric reaches a plateau at a strength of $95\%$. \postref{On} simulated data containing only WDH, this metric provides a strength of $99.6\%$, \postref{and on simulated data without WDH, it provides a strength of $0.05\%$}. 

\begin{figure}
\resizebox{\hsize}{!}{\includegraphics{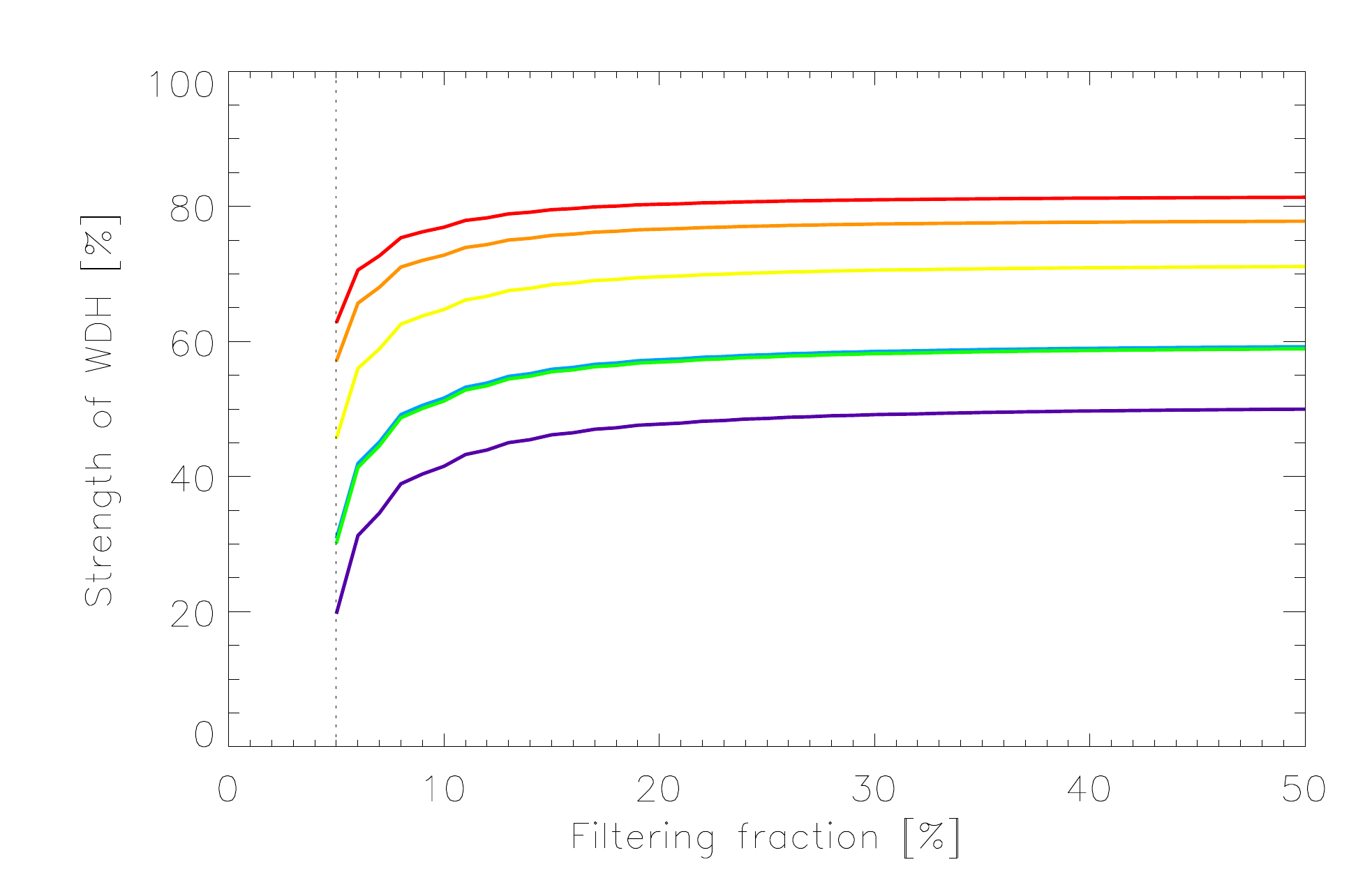}}
\resizebox{\hsize}{!}{\includegraphics[trim={0 0 0 1.1cm},clip]{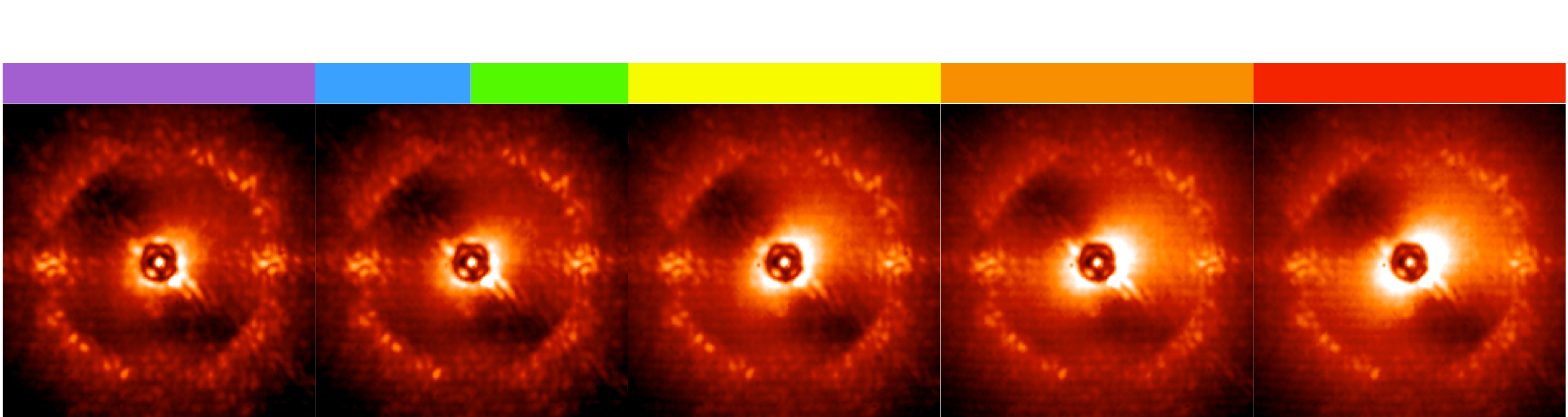}}
\caption{Evaluation of the strength of the WDH. Top: Strength of the WDH contribution in the AO-corrected area of the focal plane image, $\mathcal{S}_{WDH}$, defined at Eq.~\ref{eq:Strength}, as a function of the filtering fraction (starting at $5\%$, black dotted line). Bottom: Corresponding images from the 51~Eri data set showing low WDH residuals (left), average (middle) to high (right) WDH. The images corresponding to the green and blue lines look very similar so only one is shown here.}
\label{Fig-strengthvsfilt}
\end{figure}

In order to check that the defined metric makes sense on real data, we sorted out six images within the on-sky data cube of 51~Eri, showing visually more or less WDH (Fig.~\ref{Fig-strengthvsfilt}, bottom). The extracted strength of the WDH as a function of the filtering fraction is shown in Fig.~\ref{Fig-strengthvsfilt} (top), and ranges from $46\%$ in the weakest case of WDH (Fig.~\ref{Fig-strengthvsfilt}, bottom-left) to $80\%$ in the strongest case (Fig.~\ref{Fig-strengthvsfilt}, bottom-right). The variation of the strength obtained for the six cases \postref{is} consistent with what is observed in the image. 

\subsection{Asymmetry of the WDH}
\label{ssec-asy}
To quantitatively characterize the asymmetry of the WDH, we define its asymmetry factor, $\mathcal{A}_{WDH}$, which is 0\% when the WDH is perfectly symmetric and 100\% when the WDH shows only one wing:
\begin{equation}
\mathcal{A}_{WDH} = \frac{\int (\overline{I^+}(r,\theta) - \overline{I^-}(r,\theta) )\times M}{\int (\overline{I}(r,\theta) \times M)} \mpostref{\times} 100,
\label{eq:Asym}
\end{equation}
where $\overline{I}(r,\theta)=\overline{I^+}(r,\theta)+\overline{I^-}(r,\theta)$ is the total intensity contained in the WDH. The two wings, $I^+$ and $I^-$,  are obtained by cutting the image along the perpendicular direction to the WDH direction estimated previously, $I^+$ being the \postref{most} intense one. 

\begin{figure}
\resizebox{\hsize}{!}{\includegraphics{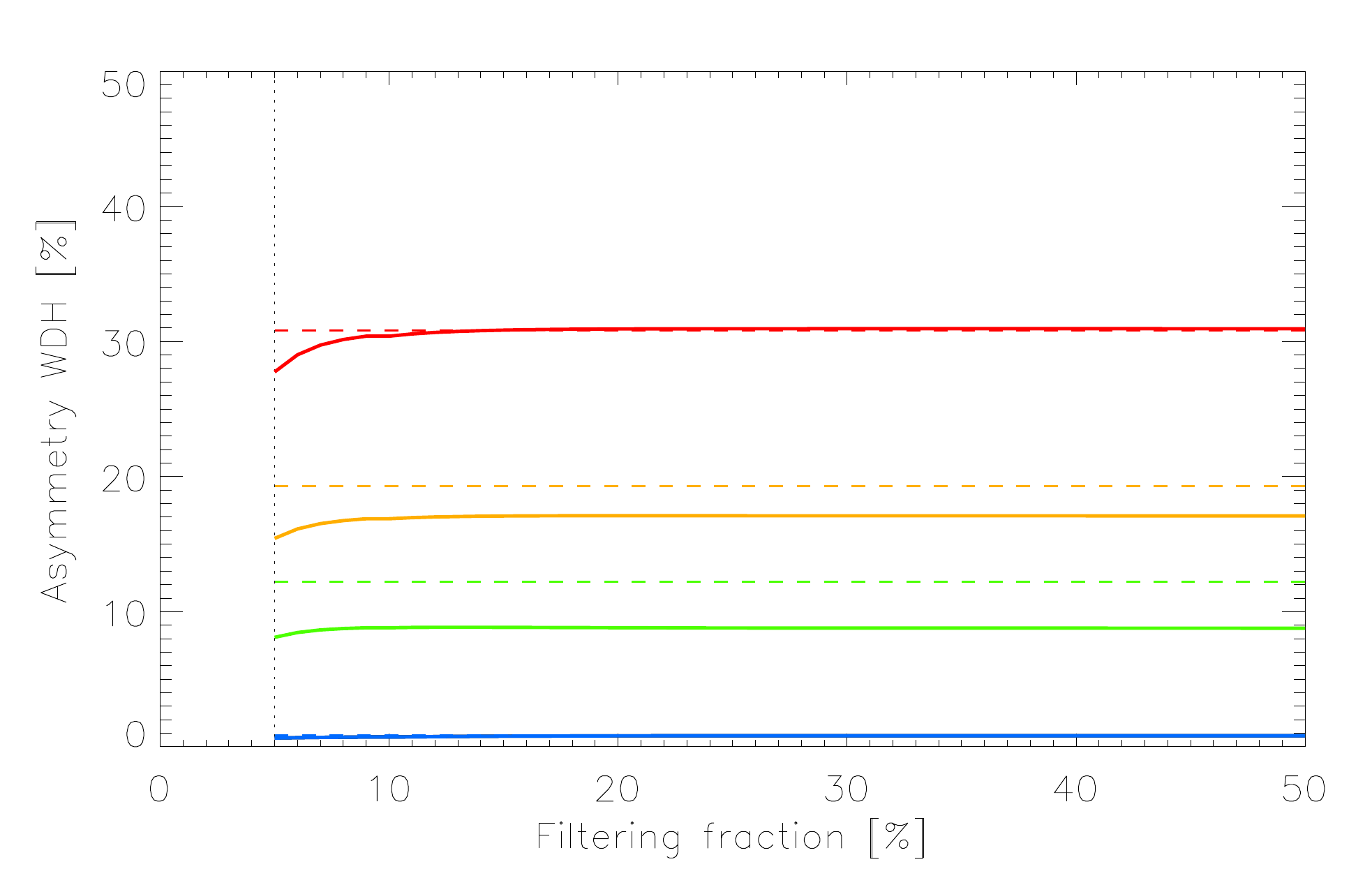}}
\resizebox{\hsize}{!}{\includegraphics[trim={0 0 0 1.1cm},clip]{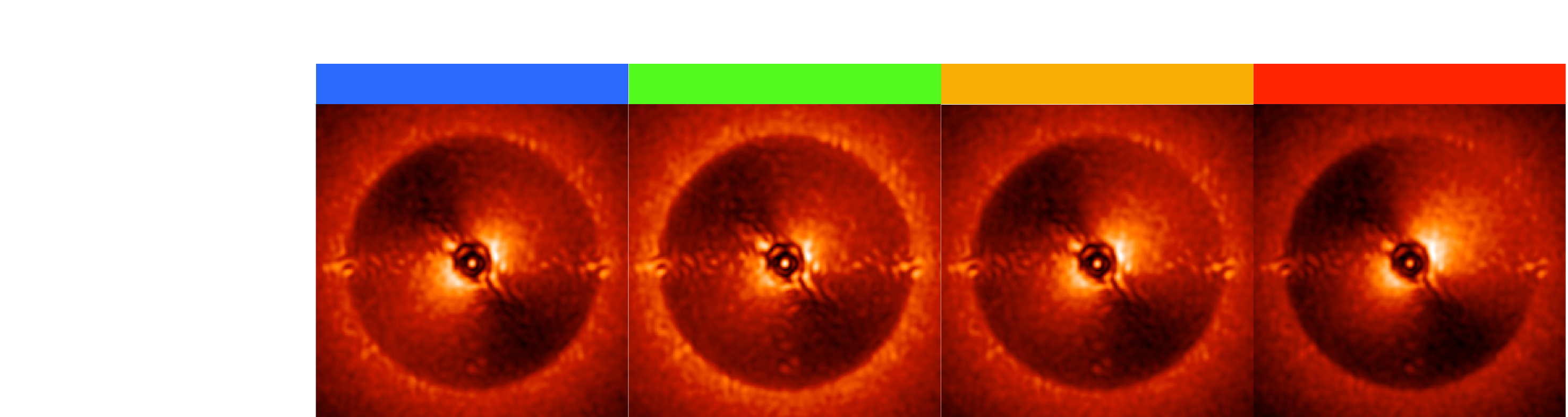}}
\caption{Evaluation of the asymmetry of the WDH. Top: Asymmetry of the WDH contribution in the AO-corrected area of the focal plane image, $\mathcal{A}_{WDH}$, defined at Eq.~\ref{eq:Asym}, as a function of the filtering fraction (starting at $5\%$, black dotted line). The blue solid line is for the image simulated without asymmetry. Bottom: Corresponding simulated images showing no (left) to large asymmetry of the WDH (right). The dashed lines correspond to the value extracted directly from the PSD.}
\label{Fig-asymvsfilt}
\end{figure}

For the three simulated cases, we obtain an asymmetry factor of $20\% \pm 2\%$, whatever the filtering fraction. To verify that this metric makes sense, we simulated images exactly in the same conditions as for the third case (including NCPA, LOR and LWE) but with a varying amount of asymmetry (Fig.~\ref{Fig-asymvsfilt}, bottom). We compare the extracted values from the images with the values obtained by analysing the PSD of the AO-residuals used to produce the simulated images (solid dashed lines \postref{in} Fig.~\ref{Fig-asymvsfilt}, top).
In the case without asymmetry the extracted value is indeed close to $0\%$ and the asymmetry factor gradually increases with the injected amount of asymmetry. 
In any of these cases, from a filtering fraction of $15\%$ the extracted asymmetry factor is fully stable. When the WDH is strong enough, the extracted value reaches exactly the value expected from the PSD\postref{. W}hen the WDH is fainter the extracted value is a few percent lower than the \postref{value expected} from the PSD.

We checked how our estimation of the direction is affected by the asymmetry of the WDH. Due to the method itself, the estimated direction is more accurate with a small amount of asymmetry but the error remains low (in the case with strong asymmetry, the error is less than $10~\mathrm{degrees}$). 

This section confirms that the described procedure \postref{is} valid to extract the WDH parameters directly from the focal plane image. In the following, we use the three metrics defined in this section to characterize the WDH (direction, strength and asymmetry), with a filtering fraction of $15\%$ (providing with stable results), in order to analyze the spatial, temporal and spectral behavior of the WDH.

\section{Effect of the WDH on the final contrast after post-processing}
\label{sec-PP}
Within the coronagraphic images sequence delivered by the latest generation of HCI instruments, the starlight residuals in the corrected area have a contrast level ranging from $10^{-3}$ to $10^{-5}$. To carve the starlight towards deeper contrast, advanced post-processing techniques have been developed, relying on observing strategies that provide a diversity between the starlight residuals and the potential circumstellar signals (disks or planets). Under good observing conditions (i.e. long coherence time), the main expected feature was quasi-statics speckles (QSS) that are neither stable enough to be calibrated nor varying fast enough to be smoothly averaged and removed by an appropriate filtering. Those QSS are originating from NCPA and, for short exposure time, atmospheric residuals \citep[e.g.][]{CantalloubeMsgr}. Consequently the post-processing techniques that have been developed are aiming at removing these QSS but are not tailored for the WDH, the LWE or other source of errors whose temporal, spectral or spatial behavior differ from the QSS. As a consequence the contrast reached after post-processing is usually worse than expected from simulations. After 5-years of SPHERE operations, the median contrast reached at $500~\mathrm{mas}$ is $2.10^{-5}$ (Langlois et al., in prep.), instead of the $10^{-6}$ expected from simulations before the commissioning of the instrument \citep{Vigan2010}.

The mainstream post-processing techniques used today rely on differential imaging that consist in: (1) estimating the QSS, (2) subtracting it from the image and (3) combining all the subtracted images available to increase the signal-to-noise-ratio of potential companions. 
In order to estimate the QSS, most HCI instruments working in near-infrared use the temporal diversity by carrying out the observations in pupil tracking mode to spatially fix the pupil (the \postref{instrumental aberrations remain static with time}) while the field of view (hence the circumstellar signals) rotates at a deterministic velocity: this is the Angular Differential Imaging \citep[ADI,][]{Marois2006} technique. If the instrument additionally provides simultaneous images at two or more wavelengths, one can use the spectral diversity as the QSS expands radially at increasing wavelength\postref{,} while the circumstellar signals remain \postref{at a fixed position}: this is the Spectral Differential Imaging \citep[SDI,][]{Racine1999} technique. At last, if a sufficient number of images on various targets are available from the instrument, one can apply Multiple Reference Differential Imaging \citep[MRDI,][with space-based instruments and with ground-based instruments resp.]{lafreniere2009hst,soummer2011rdisb,xuan2018rdigb,Ruane2019} using the correlation between the images to be processed and the library of images from the instrument.

For these three solutions, the estimated QSS is usually not perfectly estimated, which yields a high amount of speckle residuals, mostly at close separation to the star where the behavior of speckles is non-linear due to the coronagraphic device (for classical focal plane mask coronagraph). In addition, part of the circumstellar signal might be absorbed in the QSS model and removed from the image, yielding a lower signal-to-noise-ratio and, in the specific case of extended sources, a distorted shape or even a signal completely removed\footnote{For instance applying ADI on a centered circular face-on disk completely erases the signal, see \cite{Milli2012} for the effect of ADI on extended signals and \cite{Pairet2018disk} for the effect of an ADI-based PCA on extended signal.}. For point source detection, one can perform a high-pass spatial filtering to remove the light from the WDH, at a cost of a slight loss of signal that can be modeled and accounted for to characterize the candidate \citep{Cantalloube2015}. This is however not possible for extended sources, in which case most of the signal would be removed, in particular for low surface brightness disks seen face on. As a consequence, when comparing the number of disks detected in scattered light (about $40$) to the number of potentially detectable disks with an infrared excess greater than $10^{-4}$ from SPITZER data \citep{chen2014spitzer}, we find a detection rate of $\sim 25\%$ (Milli et al., in prep.).
It is still unclear if this is due to the actual disk configuration (too extended or too narrow to be seen by HCI) or due to the limited post-processing techniques available that absorb the disk signals and do not specifically account for other error terms than QSS.

In this section, we analyze the temporal, spectral and spatial variations of the WDH to see how it affects the post-processed contrast and to understand better how it could be removed by a different approach.


\subsection{Temporal variations of the WDH}
\label{sec-tempWHD}
\begin{figure*}
\resizebox{\hsize}{!}{\includegraphics{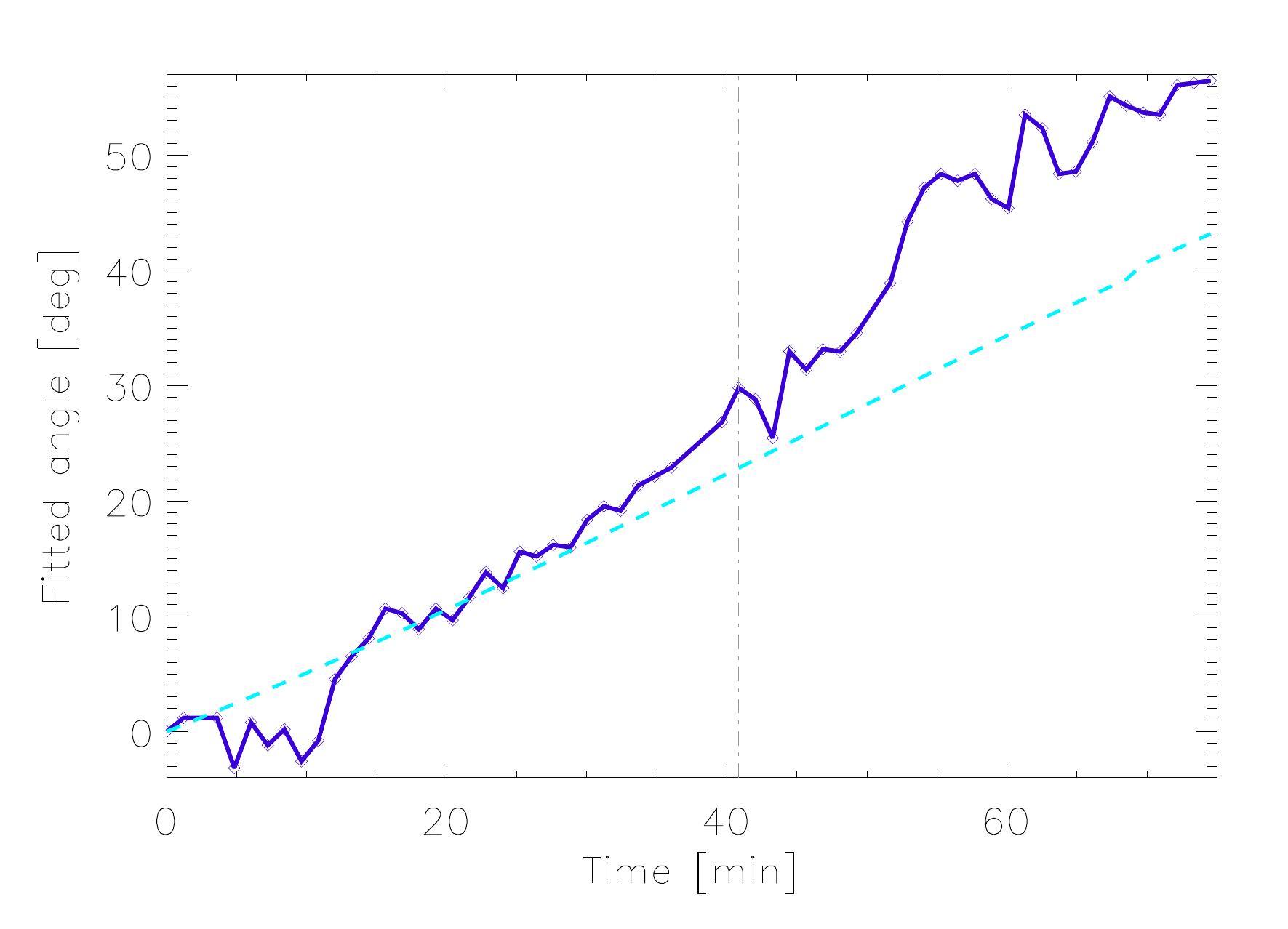}\includegraphics{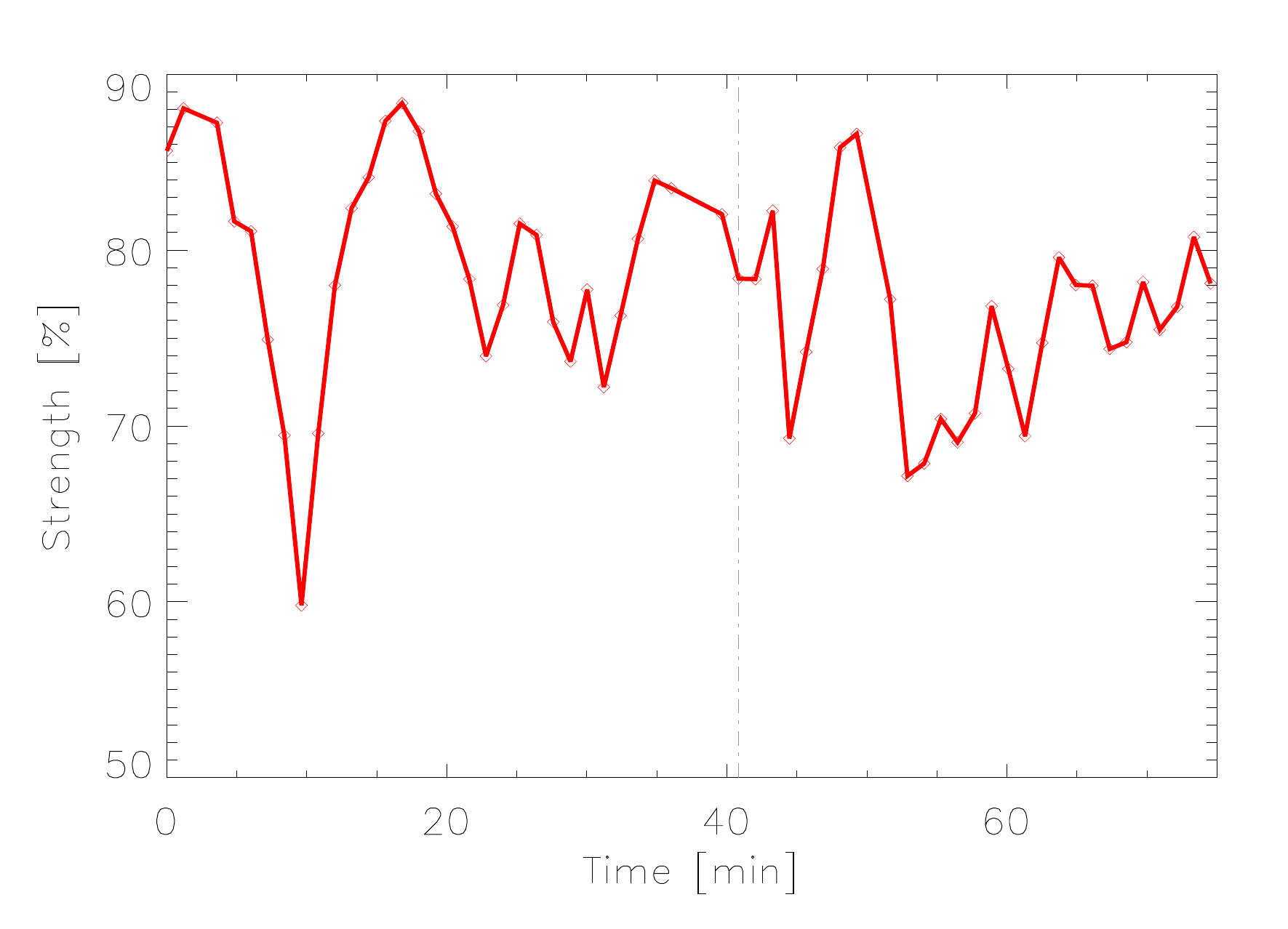}\includegraphics{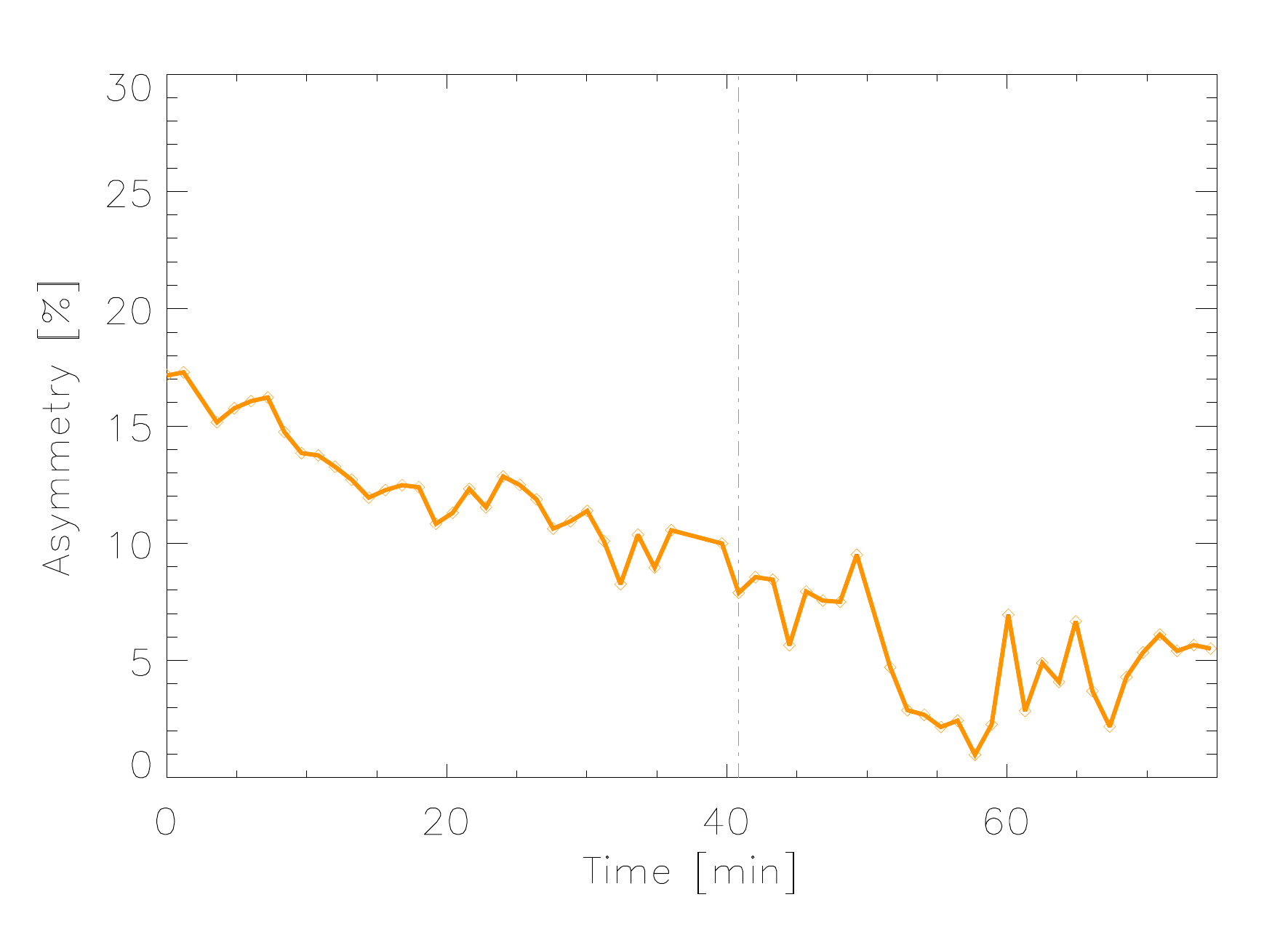}}
\caption{Temporal variation of the three WDH parameters in the case of the 51~Eri IFS data taken with SPHERE on September $25^{th}$, 2015. Left: Direction as a function of time, compared to the variation of the parallactic angle (blue dashed line). Middle: Strength as a function of time. Right: Asymmetry factor as a function of time. In the three cases, the gray dash-dotted line is when the target crossed the meridian (airmass of $1.08$).} 
\label{fig-wdhtime}
\end{figure*}

For the 51~Eri data set used in the previous section, Fig.~\ref{fig-wdhtime} shows the variation of the direction (left), the strength (middle) and the asymmetry (right) of the WDH, as a function of the time between the first image to the next ones. The whole observation sequence lasts $75~\mathrm{min}$ and the reduced cube consist of $64$ frames that are each made of $4$ binned images of $16~\mathrm{sec}$ integration time (the total number of raw images is therefore $256$). Four frames have been automatically rejected by the SPHERE reduction pipeline \citep{Delorme2017sphdc} due to their bad quality. 

The direction of the WDH is rotating linearly with time and seems to follow the parallactic angle, as expected if the WDH is induced by the upper atmospheric jet stream layer at $12~\mathrm{km}$ with a stable direction. By fitting the slope of the WDH direction as a function of the parallactic angle variation, using a robust affine fit (to avoid taking into account outlier data points for which the WDH is too low to extract accurately its direction), we find that the slope is not perfectly equal to one but rather $1.3$. This slight discrepancy is expected if the jet stream layer wind direction changes over the timescale of the observing sequence, as shown in \postref{Fig.~\ref{Fig-evol} (bottom)}. 
The strength is varying erratically, from $60\%$ to $88\%$, which corresponds to what we can observe by visualizing the data cube\footnote{We compared the temporal variation of the estimated strength of the WDH from this data set with both the SPHERE AO telemetry and the MASS-DIMM turbulence profiler measurements of the turbulence coherence time. As shown in Fig.~\ref{Fig-telemetry}, the variation of the extracted WDH strength from the SPHERE images is visually consistent with the variation of the coherence time via these two external measurements.}. 
The asymmetry factor is decreasing slowly from $17\%$ to almost no asymmetry, without link to the strength or the airmass variation (which, at first order, increases and decreases around the meridian crossing). 

\begin{figure*}
\resizebox{\hsize}{!}{\includegraphics{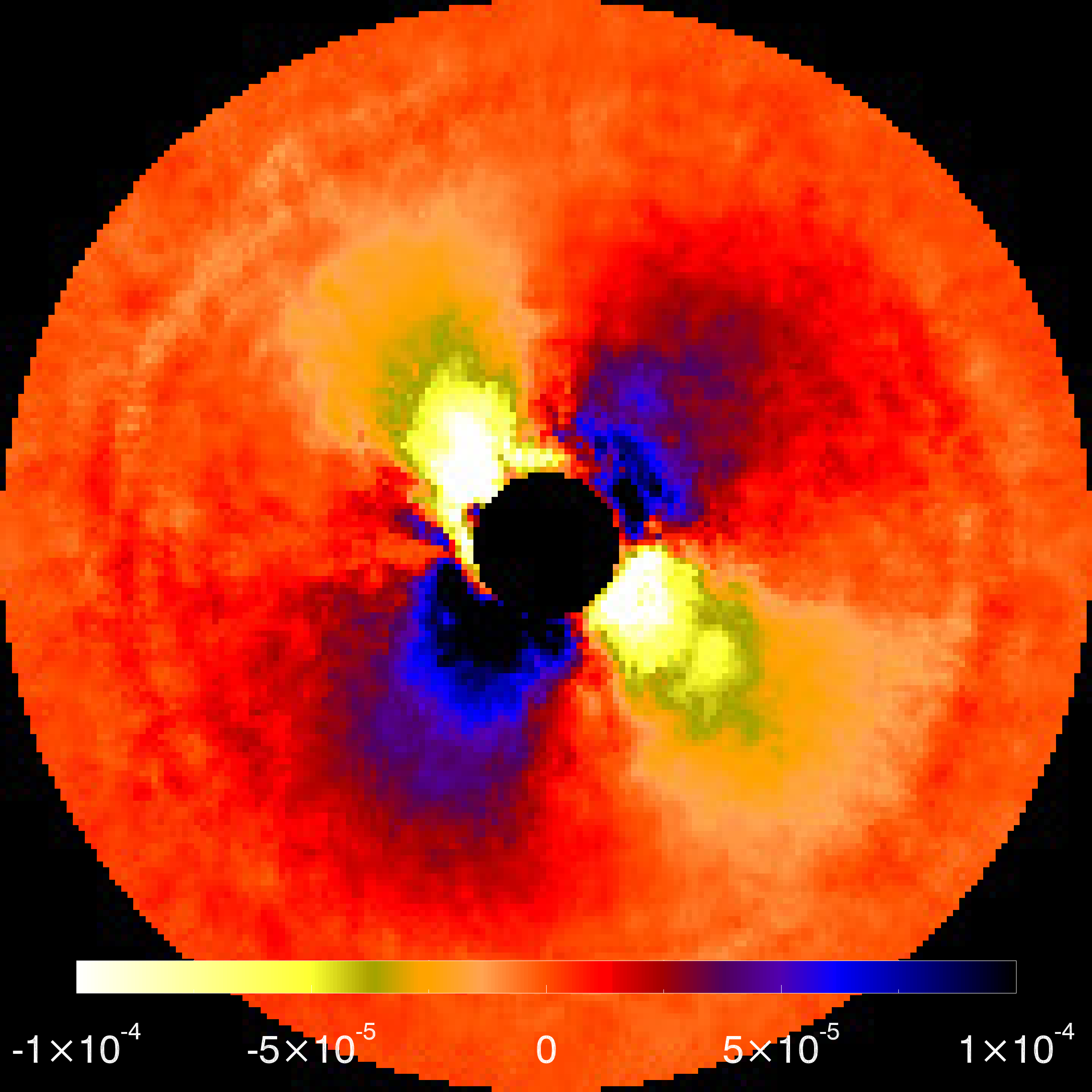}\includegraphics{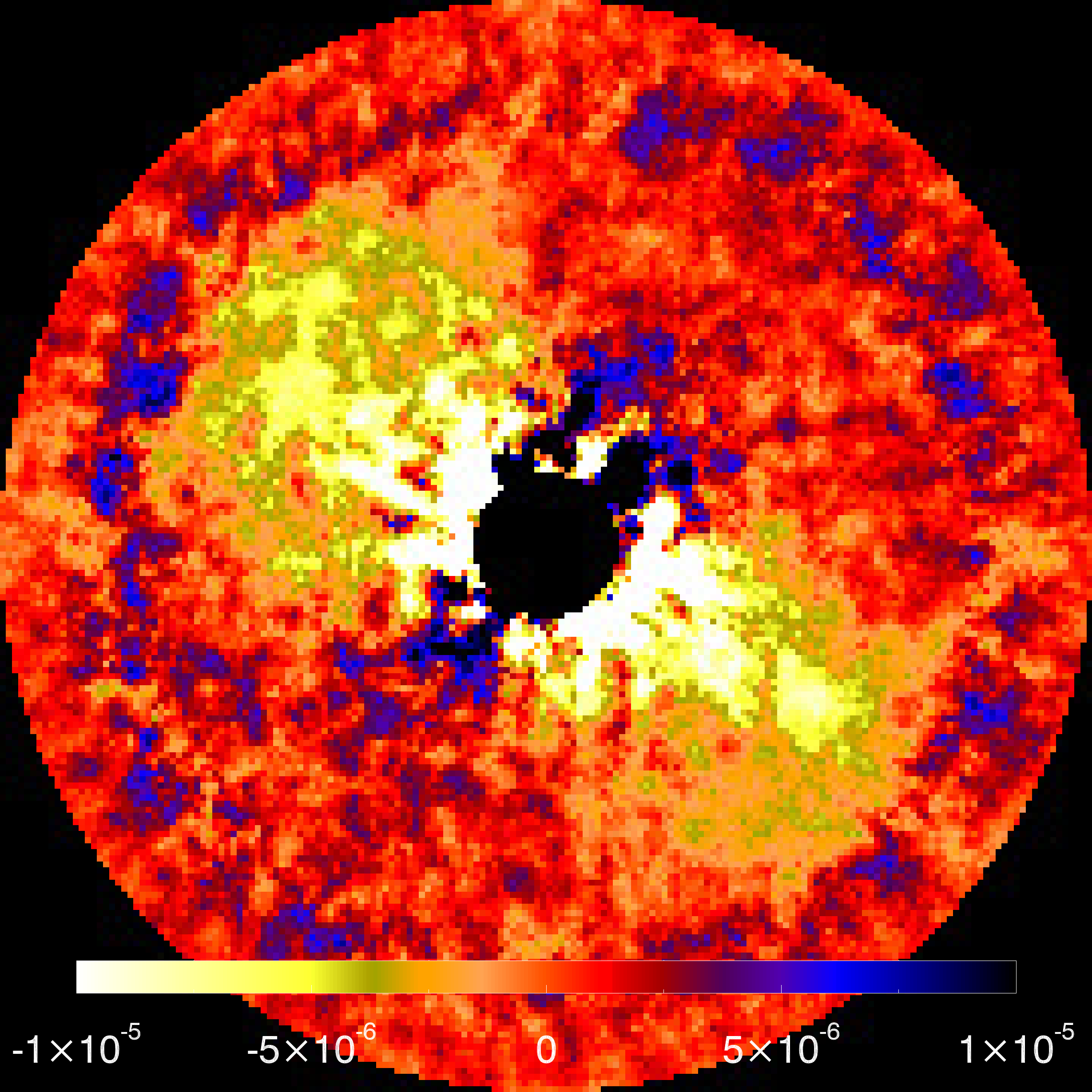}\includegraphics{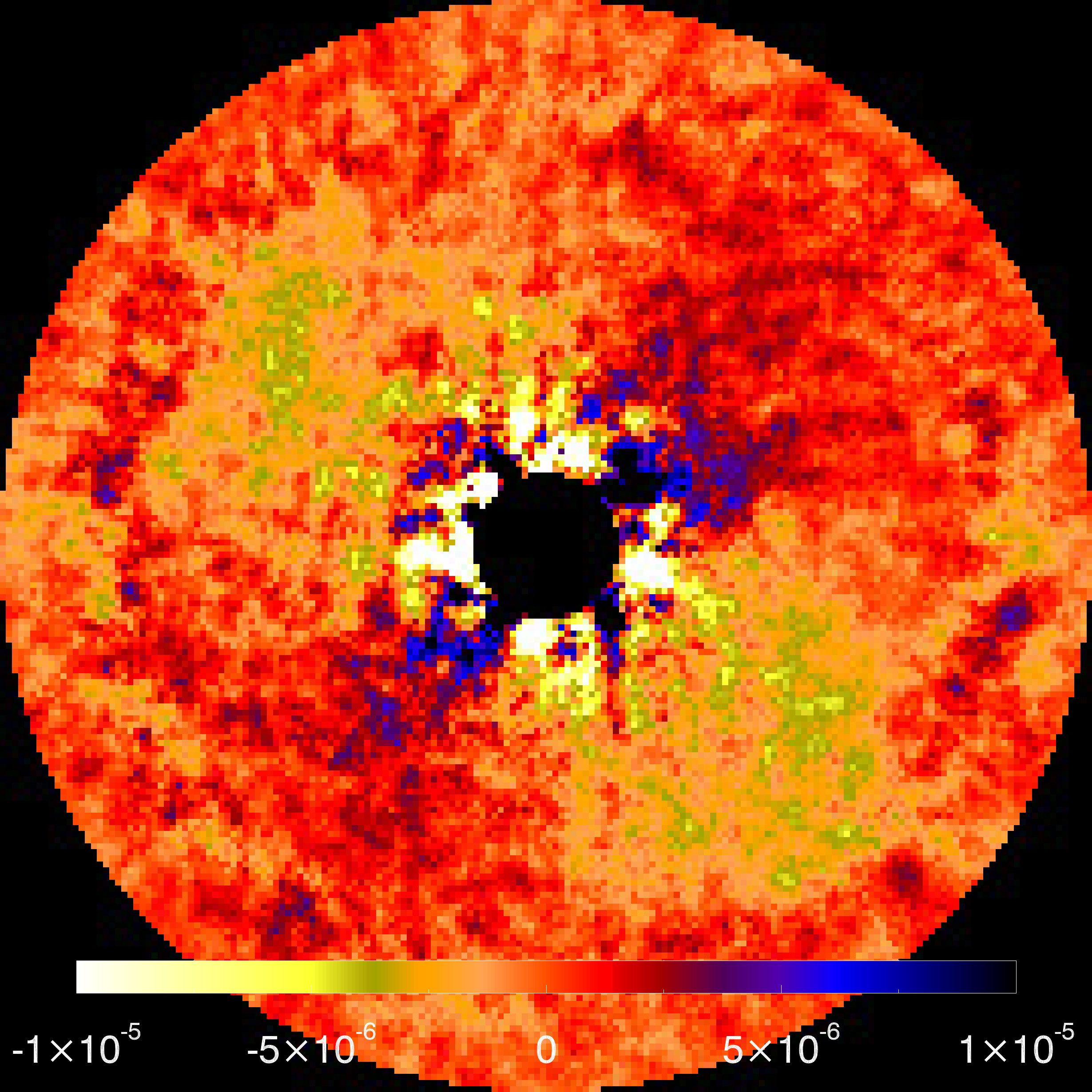}\includegraphics{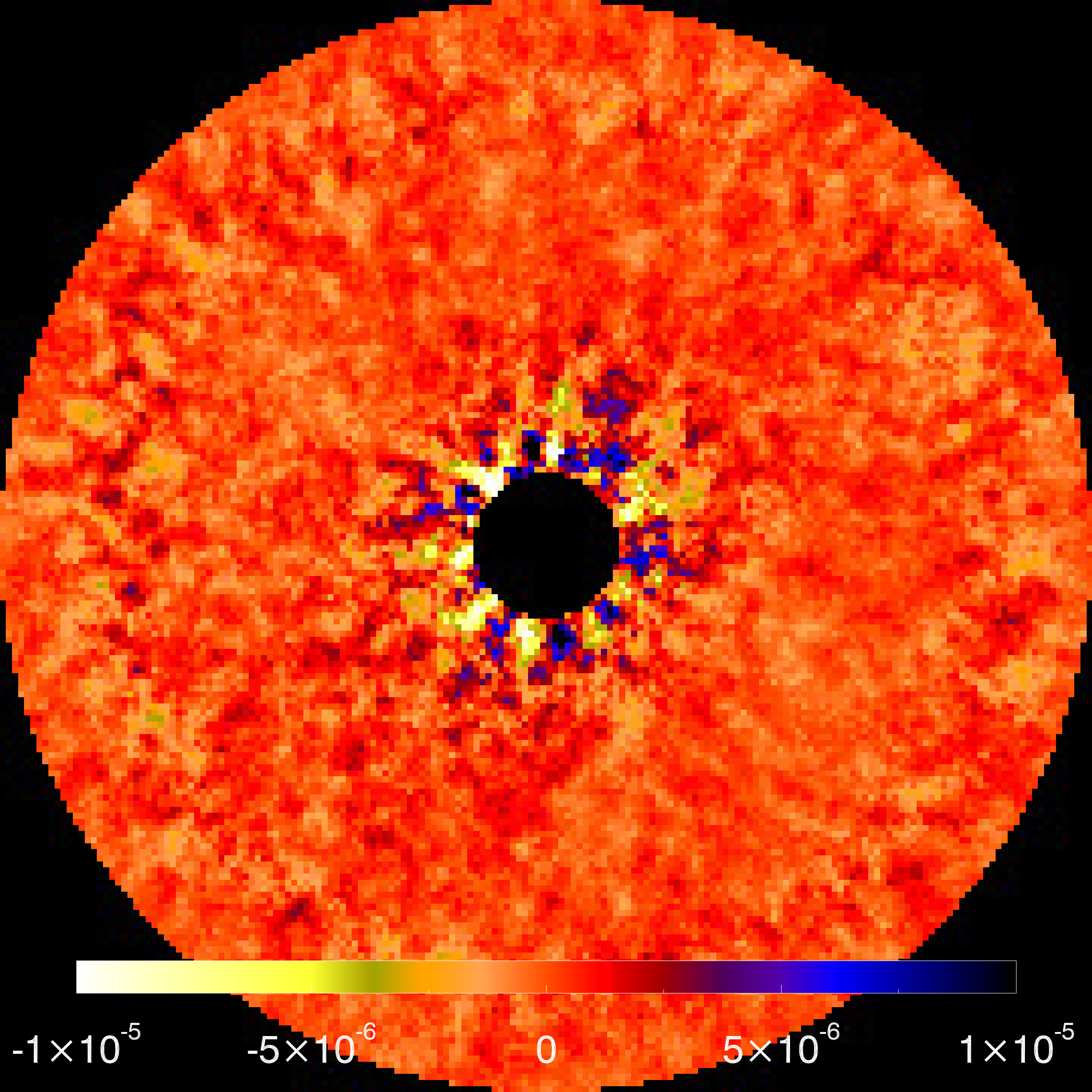}}
\caption{Post-processed image of the 51~Eri SPHERE data cube \postref{(as in Fig.~\ref{Fig-imgall}, left)} using ADI-based algorithms. 
Left: classical-ADI \citep[cADI][]{Marois2006} for which the temporal median is used as a QSS field model.
Middle-left: Principal component analysis \citep[PCA][]{amara2012pynpoint,Soummer2012}, for which a linear combination of the 3 first principal components is used as a QSS field model.
Middle-right: PCA using 5 components. 
Right: PCA using 10 components. \postref{About one order of magnitude in contrast is lost due to the WDH residuals}.} 
\label{Fig-adi}
\end{figure*}

For ADI-based algorithm, the important aspect is that the starlight residuals are spatially stable in time to be efficiently removed. We repeated this analysis on various data sets from SPHERE and the strength of the WDH is always varying \postref{significantly} from one frame to another. This prevents the ADI-based algorithm from capturing correctly the level of this feature in the model of the starlight residuals, resulting in the presence of a strong asymmetric WDH residuals in the final post-processed images, as shown in Fig.~\ref{Fig-adi}. 
In addition, the WDH direction follows the parallactic angles, that is to say the trajectory of an object\postref{. Thus,} if one simply rotates and median combines the temporal data cube, the WDH signature co-aligns and adds-up. 
As a consequence, the temporal median of the data cube does not capture the WDH whose strong signature remains in the combined subtracted images (c-ADI, Fig.~\ref{Fig-adi}, left). When applying a principal component analysis \citep[PCA\postref{,}][]{Soummer2012,amara2012pynpoint}, the intensity of the WDH residuals depends on the number of principal components kept to \postref{build} the model. As shown in Fig.~\ref{Fig-adi} (right), a more aggressive PCA removes low spatial frequencies but also considerably reduce the signal sought, mostly for extended features\postref{,} and still leaves high residuals at close separation to the star. We additionally processed the same data set with locally optimized combination of images\footnote{As implemented in the SpeCal pipeline dedicated to post-process SPHERE infrared data, described in \cite{Galicher2018specal}.} \citep[LOCI\postref{,}][]{Lafreniere2007} and Non-negative matrix factorization\footnote{As implemented in the VIP open-source pipeline gathering various state-of-the-art  post-processing methods, described in \cite{gonzalez2017vip}.} \citep[NMF\postref{,}][]{ren2018nmf}\postref{, which}  also leave strong WDH residuals in the final processed frame. In all these cases, the companion 51~Eri~b \citep{macintosh2015discovery} is hardly visible in any of the final processed frames. \postref{After cADI, the typical contrast of the WDH residual is $10^{-3}$ at $300~mas$, compared to $5.10^{-6}$ when the WDH is filtered out.} 

\subsection{Spectral variations of the WDH}
\label{sec-specWHD}
To investigate the spectral variation of the WDH, we run our analysis procedure on the three simulated IFS data cubes and on the on-sky data cube of 51~Eri described in Sect.~\ref{sec-WDH} (whose image\postref{s} at \postref{the} shortest wavelength is shown in Fig.~\ref{Fig-imgall}). Each cube is constituted of $39$ images at wavelength $\lambda$ ranging from $0.96~\mathrm{\mu m}$ to $1.64~\mathrm{\mu m}$. 

\begin{figure*}
\resizebox{\hsize}{!}{\includegraphics{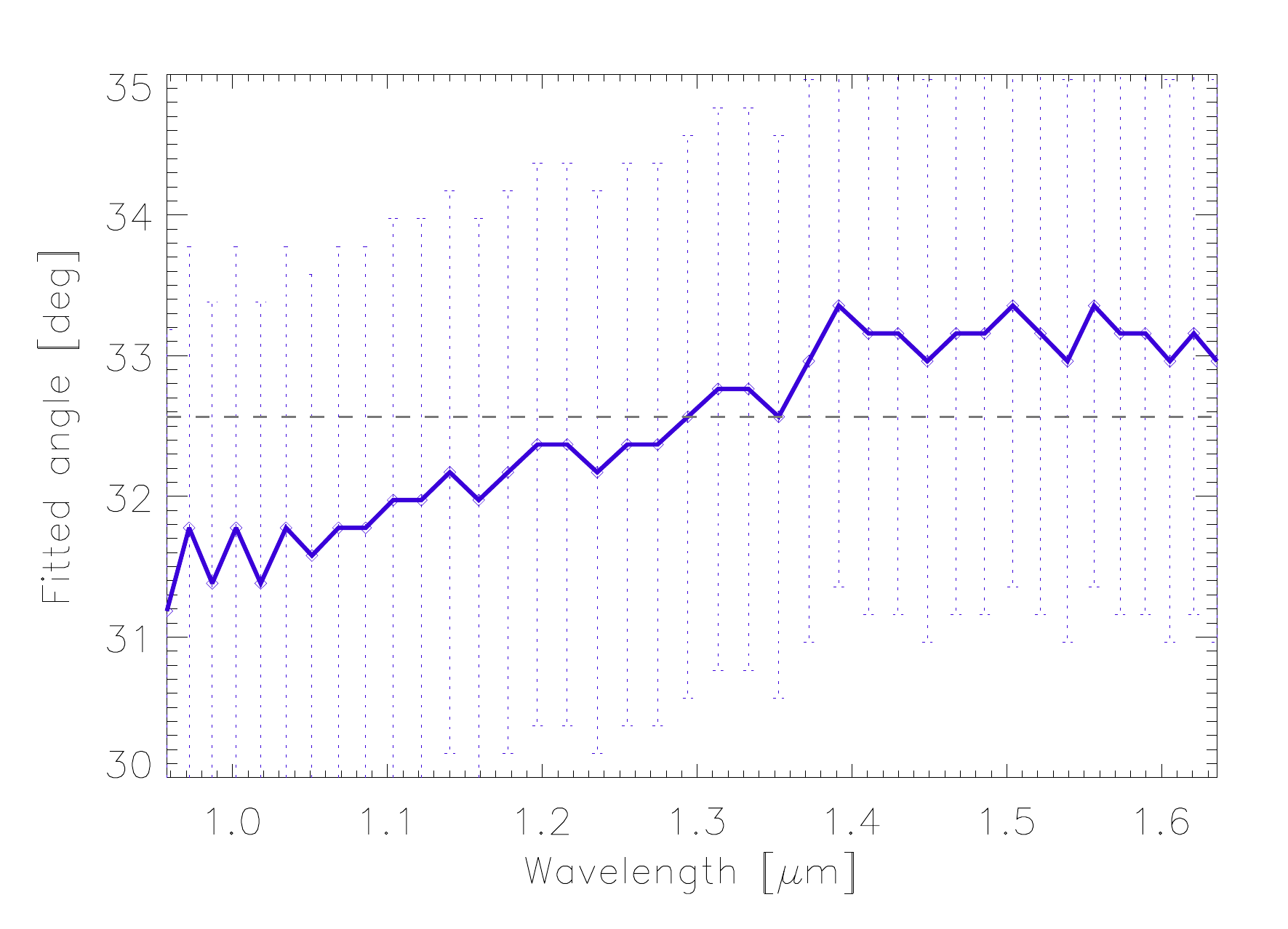}\includegraphics{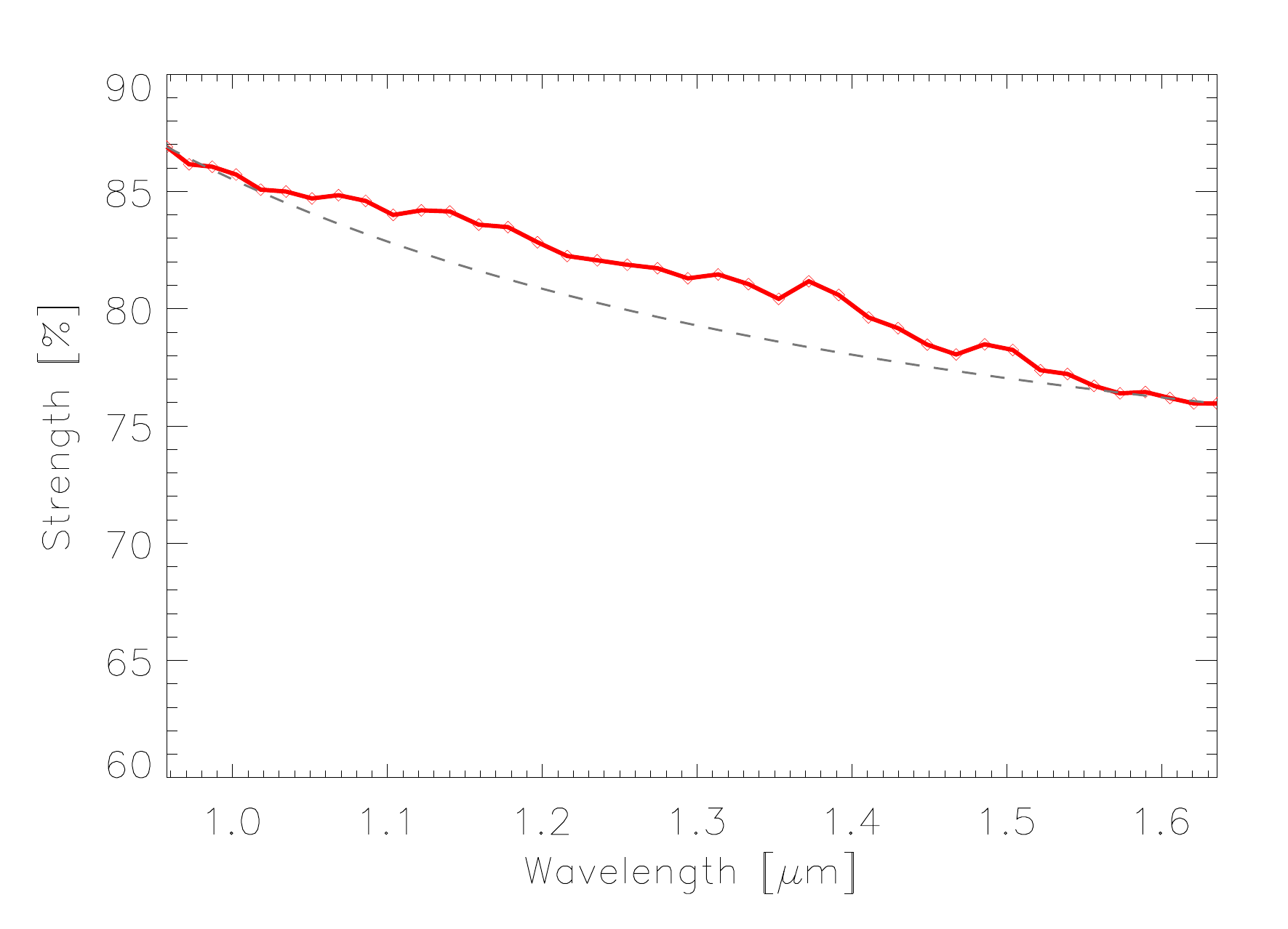}\includegraphics{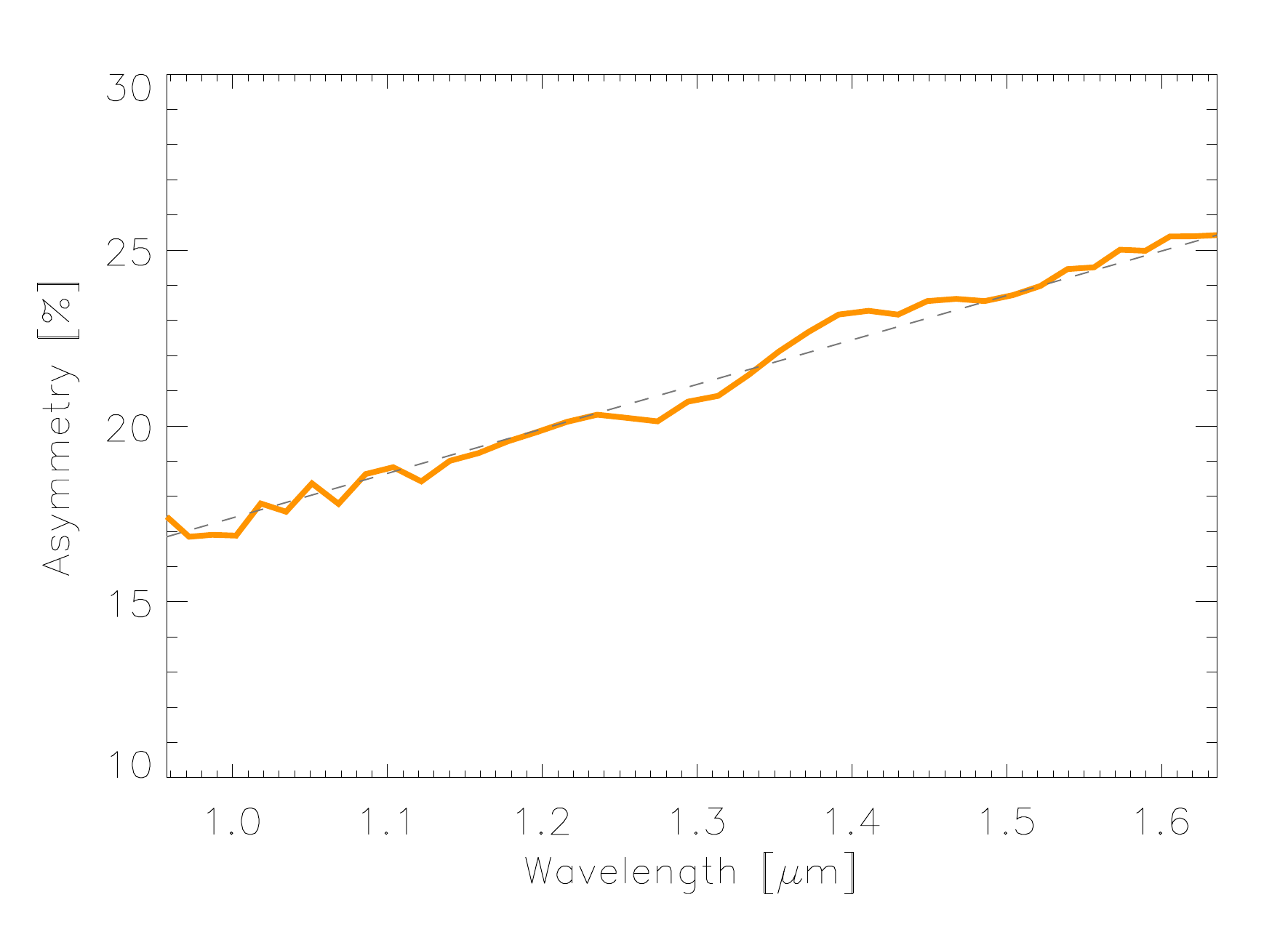}}
\caption{Spectral variation of the three WDH parameters in the case of the 51~Eri IFS data taken with SPHERE on September $25^{th}$, 2015. Left: Direction as a function of wavelength. Middle: Strength of the WDH as a function of the wavelength. Right: Asymmetry of the WDH as a function of the wavelength. In each plot, the dashed grey line shows the expected trend.} 
\label{fig-wdhspec}
\end{figure*}

The direction of the WDH is obviously constant with the wavelength, which is indeed verified as shown in Fig.~\ref{fig-wdhspec} (left). Our method provides a constant direction varying \postref{at most by} $2$~degrees for all of the four tested cases. 
As \postref{mentioned} in Sect.~\ref{sec-WDH}, the turbulence coherence time $\tau_0$, through Eq.~\ref{eq:tau0}, scales with the wavelength as $r_0$, that is to say as $\lambda^{6/5}$. The variance of the AO residual phase due to the servolag error, through Eq.~\ref{eq:servolag}, scales as $\tau_0^{-5/3}$. Therefore, the WDH \postref{strength} scales with the wavelength as $\lambda^{-2}$, which is indeed what we observe in Fig.~\ref{fig-wdhspec} (middle). The asymmetry factor is expected to steadily increase with the wavelength, as demonstrated in \citep{Cantalloube2018}. This is indeed what is observed in Fig.~\ref{fig-wdhspec} (right) where, in the four cases, the asymmetry \postref{of the WDH} follows the \postref{wavelength} variation.

\begin{figure}
\resizebox{\hsize}{!}{\includegraphics[scale=0.48]{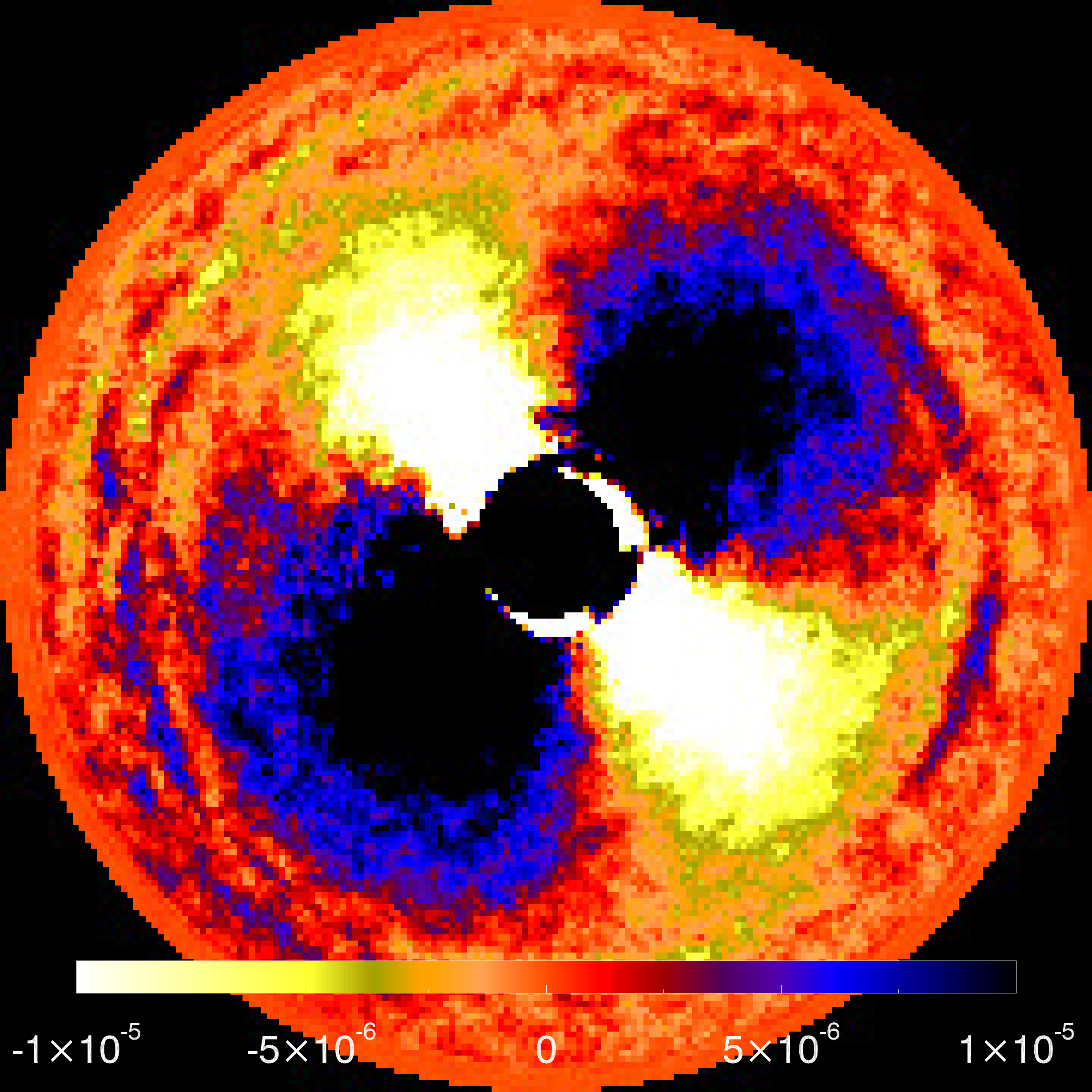}\includegraphics[scale=0.48]{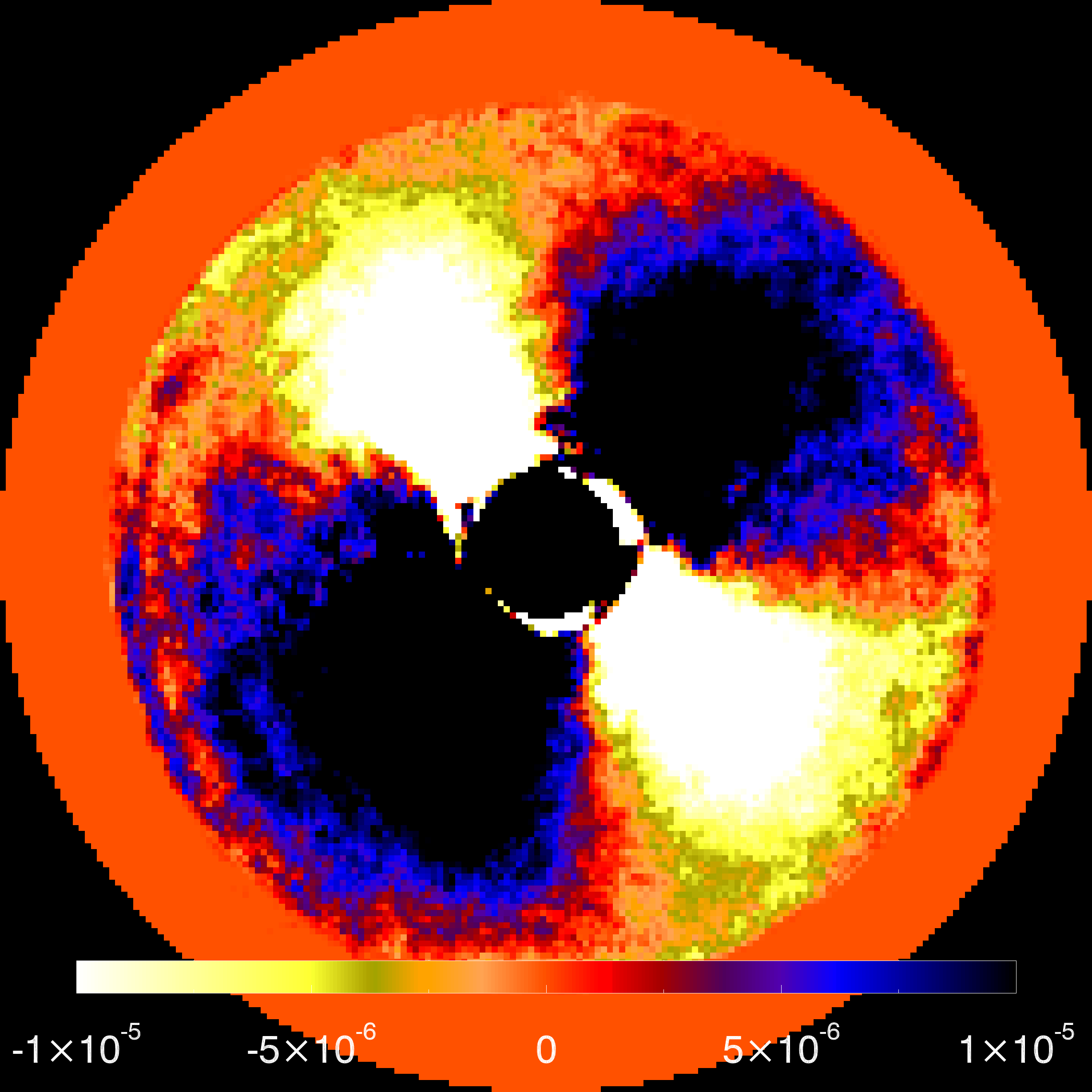}}
\caption{Post-processed images of the 51~Eri data using SDI. Left: the reference is the $10^{th}$ channel ($1.12~\mathrm{\mu m}$), which is rescaled and removed from the frame in the first channel {$1.00~\mathrm{\mu m}$}. Right: the reference is this time the $20^{th}$ channel ($1.31~\mathrm{\mu m}$). \postref{About half an order of magnitude in contrast is lost due to the WDH residuals.}}
\label{fig-wdhsdi}
\end{figure}

The widely accepted model to perform SDI is, assuming that the aberrations are achromatic and that the small phase approximation is valid, that the speckle field at a given wavelength $s_{\lambda_1}(\mathbf{r})$ can be rescaled to a second speckle field at a different (but close) wavelength, $s_{\lambda_2}(\mathbf{r})$, via the square of the ratio between the two wavelengths: $s_{\lambda_2}(\frac{\lambda_2}{\lambda_1} \mathbf{r}) = (\frac{\lambda_1}{\lambda_2})^2 \times s_{\lambda_1}(\mathbf{r})$ \citep{Racine1999,Pueyo2007}. After rescaling one image (the reference channel), SDI simply subtracts it from the other (the science channel). Since SDI assumes that the QSS varies as $\lambda^2$ whereas the WDH varies as $\lambda^{-2}$ and its asymmetry as $\lambda$, a SDI post-processing\postref{\footnote{The data have been processed using the pipeline described in \cite{Vigan2016gj758}.}} leaves a strong asymmetric WDH imprint in the residual map, as shown in Fig.~\ref{fig-wdhsdi}. This imprint gets worse as the reference channel is \postref{further} from the science channel since the rescaling square law deviates more from the actual WDH \postref{spectral variation}. \postref{After classical SDI, the typical contrast of the WDH residual (as in Fig.~\ref{fig-wdhsdi} left) is $10^{-4}$ at $300~mas$, compared to $6.10^{-5}$ when the WDH is filtered out.} 

\subsection{Spatial variations of the WDH}
\label{sec-spatWHD}
\begin{figure*}
\resizebox{\hsize}{!}{\includegraphics{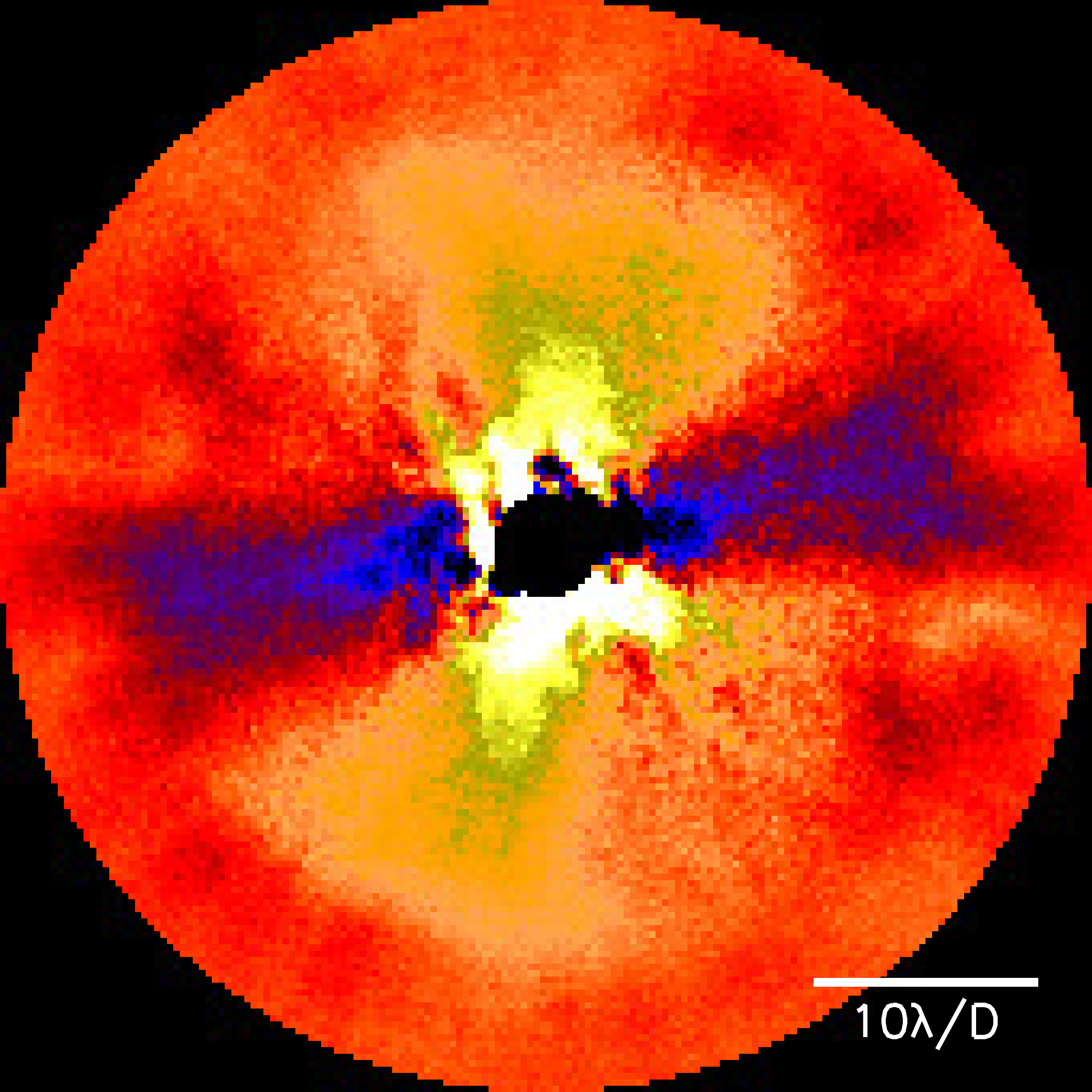}\includegraphics{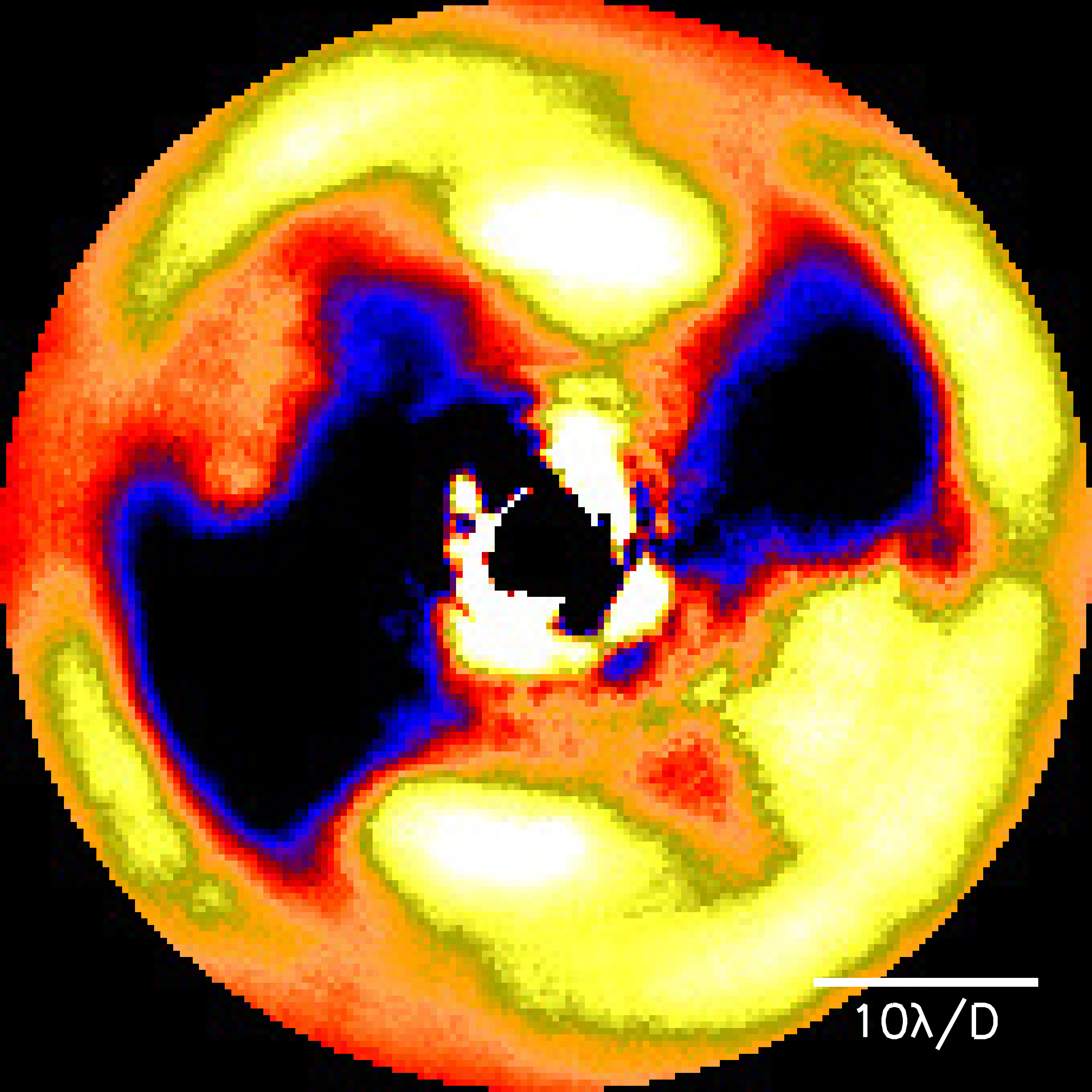}\includegraphics{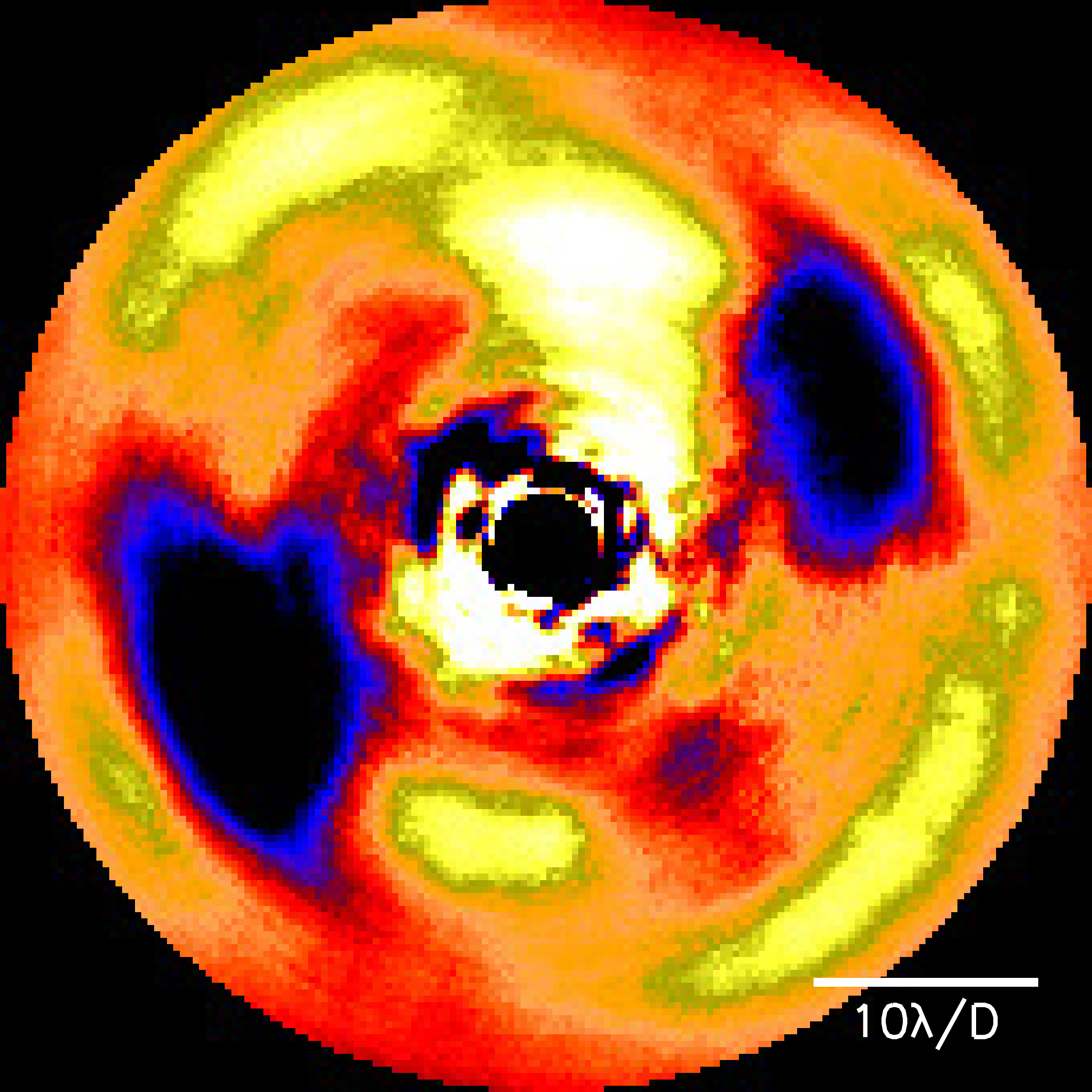}\includegraphics{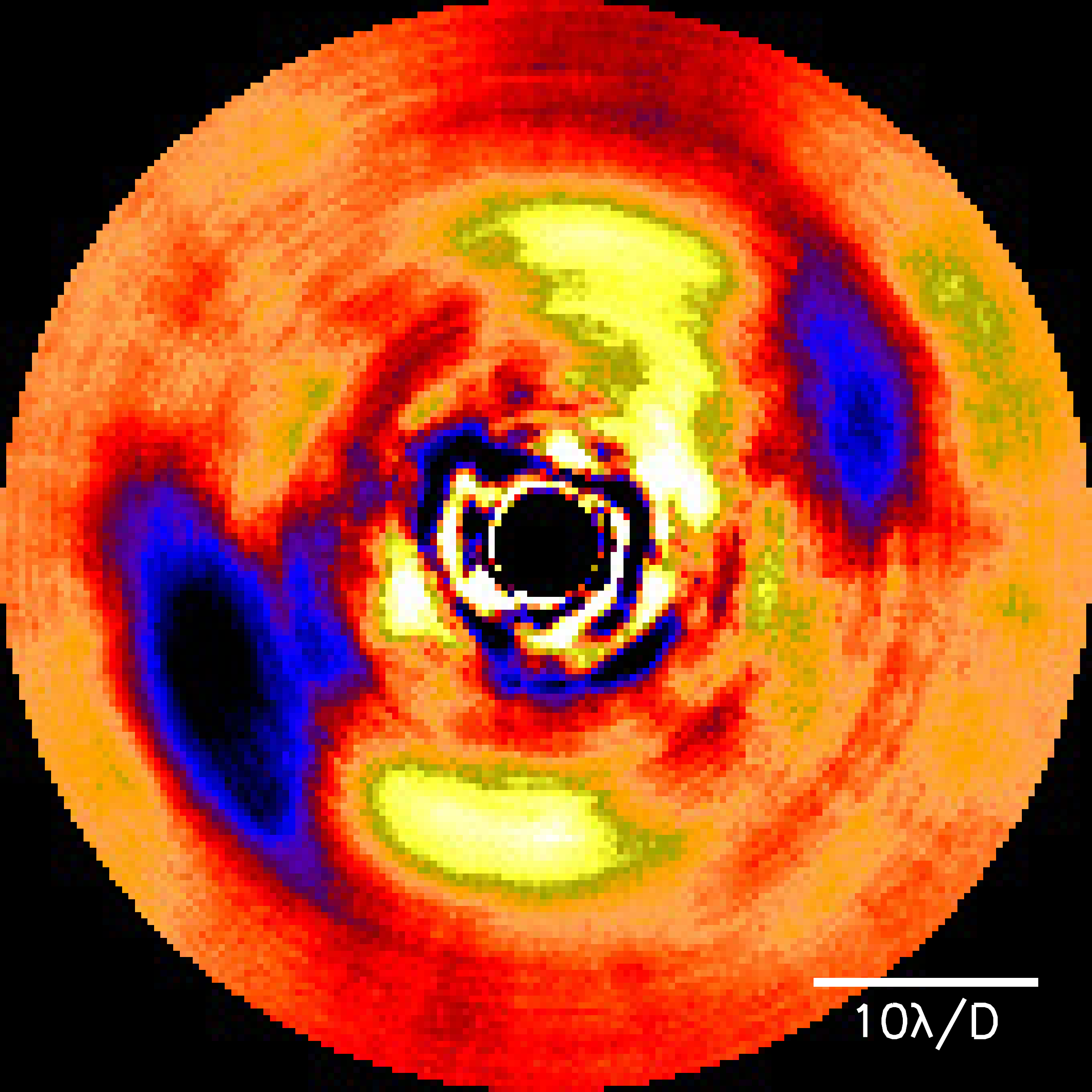}}
\caption{Residual WDH signature in RDI post-processed images from SHARDDS SPHERE-IRDIS data taken in broadband H ($1.625~\mathrm{\mu m} \pm 0.15$).
Left: ADI-PCA post-processed image, using $2$ principal components that are subtracted, shown for comparison. 
Right: RDI-PCA post-processed images, using $10$ (left), $25$ (middle) and $50$ principal components. All frames share the same color scale.}
\label{fig-rdi}
\end{figure*}

As discussed in Sect.~\ref{sec-WDH}, the shape, direction and intensity of the WDH signature depends on various external atmospheric turbulence parameters, on the observed target and on the observation set-up. In the context of RDI, mainly developed for extended source extraction, this means that it is unlikely that a set of reference stars of the same spectral type as the science target exhibits a similar WDH signature. 
As a consequence, we observe again strong WDH residuals in RDI post-processed images, as illustrated \postref{in} Fig.~\ref{fig-rdi}. 
The images shown were processed using an algorithm implementing the PCA, as described in \cite{soummer2012detection}. In ADI (left), the principal components were computed from the science frames themselves while in RDI (right) the principal components were computed from a library of frames which did not include the science target. This library was build from targets part of the SHARDDS program survey for debris disk using SPHERE-IRDIS \citep{Wahhaj2016,Choquet201749cet,Marshall2018hd105} and the frames were selected based on their correlation with the science target (Milli et al. in prep). We see that WDH residuals still remain in the RDI-PCA post-processed images, mostly at large distance where it is known that RDI reaches a turnover \citep{Ruane2019} and shows an equal or higher level of residuals compared to ADI-PCA. The effect of the asymmetry is even more emphasized in this example where the upper-right residual wing is smaller than the other. These considerations depend on how the reference to be subtracted is picked within the library of images. Usually the highest correlation is set on the QSS and not the WDH. Also using larger library should leave more chances to get a reference that also contains a similar WDH signature. \postref{Using a basic RDI implementation, the contrast loss due to the WDH residual signature is of about an order of magnitude in the affected regions.}

\postref{In a next paper we will propose a method to reduce the impact of the WDH on the contrast after differential imaging, without affecting close-in and/or extended signals. The idea is to use the analysis procedure described in Sect.~\ref{sec-anaWDH} to estimate a model of the WDH and subtract it so as to recover a better contrast, intrinsically limited by the NCPA.}


\section{Conclusions and perspectives}
\label{sec-ccl}
In this paper, we conducted a study of the wind driven halo visible in VLT/SPHERE images, a specific signature that significantly affects the contrast performance of the instrument.  
We provided a detailed examination of the different parameters that are playing a role in building up this WDH, from the instrument design to its interaction with the observing conditions and operations. \postref{W}hen the turbulence coherence time is below $3~\mathrm{ms}$, the WDH appears in SPHERE coronagraphic images \postref{(for its AO system running at $1380~\mathrm{Hz}$)}, yielding an occurrence rate of about  $30\%$, according to turbulence profiling measurements. \postref{This occurrence rate will be refined by an upcoming detailed analysis of the SPHERE-SHINE survey data sample.}

\postref{To that end, we} proposed a procedure to analyze the WDH contribution directly from the focal plane images\postref{. We} checked, thanks to various simulations, that this procedure is able to extract the relevant parameters to describe the WDH: its direction, its strength and its asymmetry. \postref{In the future, we intend to use this procedure for a statistical analysis on the full SPHERE-SHINE sample, and correlate the results to the atmospheric turbulence parameters, essentially the turbulence coherence time and the scintillation measured by the MASS-DIMM and Stereo-SCIDAR at Paranal observatory, and by the AO telemetry data.} Such a study will bring a deeper understanding of this specific limitation, towards proposing a way to alleviate this frequent contrast limitation on the data already acquired with SPHERE, but also to give important outlooks for the design of upgrades and future instruments equipped with an HCI mode such as SPHERE+, GPI2.0 \citep{chilcote2018gpi2} and the three ELT first light instruments METIS \citep{brandl2018metis,kenworthy2018metishci}, HARMONI \citep{thatte2016harmoni,carlotti2018harmonihci} and MICADO \citep{davies2016micado,baudoz2014micadohci}.

\postref{Using the established procedure to analyze the WDH contribution}, we highlighted \postref{its} effect on the final contrast \postref{after post-processing}. \postref{C}urrent post-processing techniques, \postref{based on differential imaging,} fail to cope with this signature and strong WDH residuals appear in the post-processed images, hindering the detection of close planets and/or circumstellar disks. \postref{As such, the WDH is responsible for a contrast loss of about an order of magnitude (about a factor of 10).} In a \postref{forthcoming} paper, we will develop a parametric model of the WDH, function of the three parameters discussed in the present paper, to fit the WDH signature directly from the images. We will then establish a way to subtract this model from the images without altering extended signals or affecting the signal-to-noise ratio of point sources (Cantalloube et al., in prep). 



In addition, some post-processing methods, such as MEDUSAE \citep{Ygouf2013,Cantalloube2017} or COFFEE \citep{Paul2013,Herscovici2019} are aiming at estimating the phase responsible for the observed coronagraphic \postref{image} (using respectively the spectral diversity of an IFU or induced phase diversity). The AO residuals are taken into account in the model via the turbulence phase structure function. \postref{In that context,} studying and understanding the WDH is crucial to obtain a correct model of \postref{the} AO residuals for \postref{such} algorithms to be \postref{operational} on sky. These type of algorithms are relying on a full model of the HCI instrument and reach theoretical contrasts close to the photon noise limit, making them potentially the next big step to post-processing techniques.

\begin{acknowledgements}
\postref{We would like to thank the anonymous referee for her/his beneficial comments on the present paper. 
F.C.: acknowledges J. Antichi for fruitful discussions about the Talbot length in the context of SPHERE.} 
O.J.D.F: STFC studentship (ST/N50404X/1). 
T.H.: T.H. acknowledges support from the European Research Council under the Horizon 2020 Framework Program via the ERC Advanced Grant Origins No. 832428. 
J.O. and N.B.: This work was supported by the Science and Technology Funding Council (UK) (ST/P000541/1), UKRI Future Leaders Fellowship (UK) (MR/S035338/1) and Horizon 2020; this project has received funding from the European Union’s Horizon 2020 research and innovation programme under grant agreement No. 730890. 
A.V.: This project has received funding from the European Research Council (ERC) under the European Union’s Horizon 2020 research and innovation programme, grant agreement No. 757561 (HiRISE).
\end{acknowledgements}


\bibliographystyle{aa} 
\bibliography{my_biblio}

\begin{appendix}

\section{MASS-DIMM measurements at Paranal observatory}
\begin{figure}
\centering
\includegraphics[scale = 0.5]{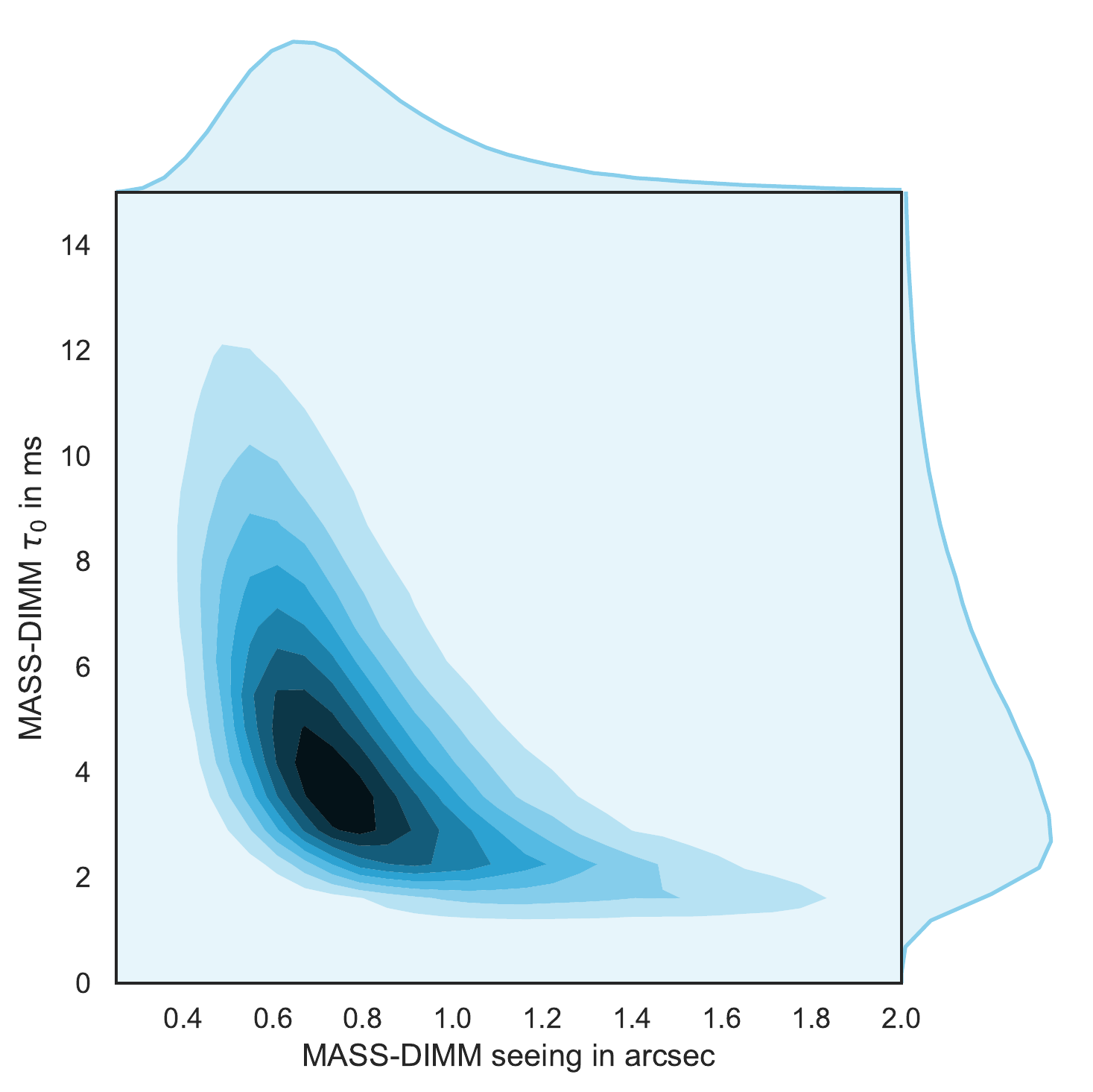}
\caption{Joint probability distribution of the seeing and coherence time, as measured by the Paranal MASS-DIMM located $7~\mathrm{m}$ above the platform, using data from April 2016 to April 2019.}
\label{Fig-coude}
\end{figure}

Figure \ref{Fig-coude} shows the  turbulence coherence time \postref{vs.} the seeing, both measured by the MASS-DIMM \citep{Kornilov2007massdimm} turbulence profiler installed at Paranal observatory. 

\section{MASS-DIMM and SPARTA measurements compared to the WDH strength}
\label{app-comp}
\begin{figure}
\centering
\resizebox{\hsize}{!}{\includegraphics{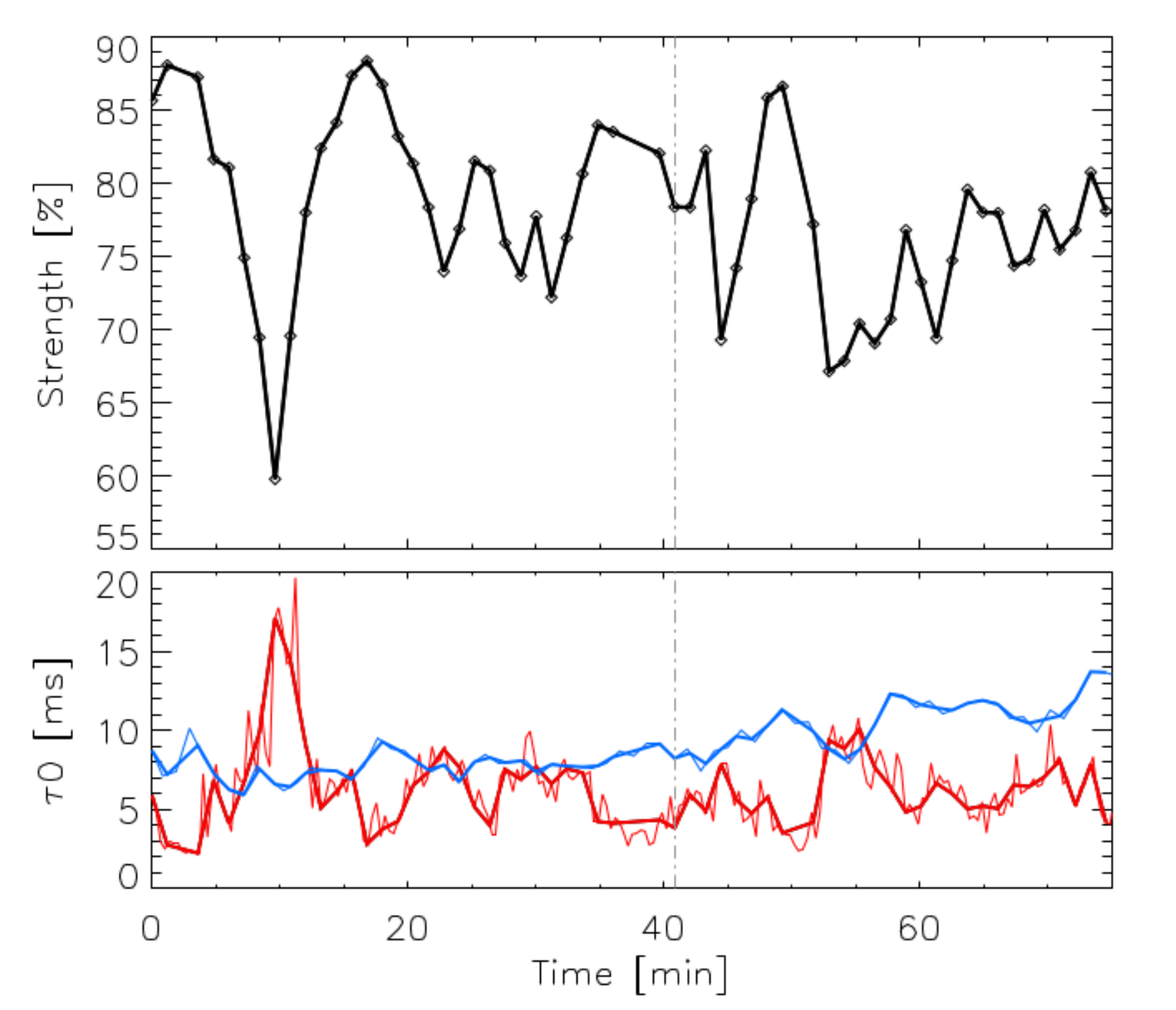}}
\caption{Upper panel: Strength of the WDH as a function of time extracted from the 51~Eri data taken with SPHERE (at $1~\mathrm{\mu m}$) on September $25^{th}$ 2015, starting at 8h19m40s and finishing at 9h35m25s (UTC). Lower panel: Measured turbulence coherence time $\tau_0$ as a function of time during the 51~Eri observation sequence. The AO-telemetry data are from SPHERE-SPARTA (red solid lines) and the atmospheric profiling data are from the MASS-DIMM (blue solid lines). In both cases, the full data are shown by the thin line while the thick line is the data interpolated at the image rate (black diamonds in upper panel). In both panels, the meridian crossing of the target star is indicated by the grey dotted-dash line.}
\label{Fig-telemetry}
\end{figure}

The RTC of any AO-equipped instruments at Paranal observatory, SPARTA \citep{fedrigo2006sparta}, provides AO-telemetry data such as Strehl, seeing, Fried parameter and coherence time along the line of sight. The coherence time provided by the SPHERE-SAXO telemetry is not an absolute value but its temporal variation matches the actual temporal variations of the coherence time. \postref{In} Fig.~\ref{Fig-telemetry}, the upper panel shows the strength of the WDH from the SPHERE 51~Eri data (as shown in Fig.~\ref{fig-wdhtime}, left), while the lower pannel shows the coherence time  $\tau_0$ measured by the SPARTA SPHERE AO telemetry (red lines) and the $\tau_0$ measured by the DIMM-MASS that we corrected for the airmass and the wavelength (blue lines), since the DIMM-MASS value are \postref{given} at zenith and at $500~\mathrm{nm}$. The variation of the SPARTA estimated $\tau_0$ follows quite well the strength of the WDH extracted directly from the SPHERE images: the highest $\tau_0$ indeed corresponds to the lowest WDH strength and vice-versa. However, the variation of the MASS-DIMM measured $\tau_0$ is not showing the WDH trend, which can be expected due to its non-local measurement.

\end{appendix}

\end{document}